\pdfoutput=1

\documentclass[acmsmall, nonacm]{acmart}
\AtBeginDocument{%
  }

\usepackage[ruled,vlined,linesnumbered]{algorithm2e}
\SetKwInOut{KwInput}{Input}
\SetKwInOut{KwOutput}{Output}
\SetKw{And}{and}

\usepackage{subcaption}
\usepackage{cleveref}
\usepackage{enumitem}
\usepackage{multirow}

\crefname{definition}{Definition}{Definitions}
\Crefname{definition}{Definition}{Definitions}
\crefname{equation}{Equation}{Equations}
\Crefname{equation}{Equation}{Equations}
\crefname{figure}{Figure}{Figures}
\Crefname{figure}{Figure}{Figures}
\crefname{table}{Table}{Tables}
\Crefname{table}{Table}{Tables}
\crefname{section}{Section}{Sections}
\Crefname{section}{Section}{Sections}
\crefname{algorithm}{Algorithm}{Algorithms}
\Crefname{algorithm}{Algorithm}{Algorithms}
\crefname{appendix}{Appendix}{Appendices}
\Crefname{appendix}{Appendix}{Appendices}
\crefname{theorem}{Theorem}{Theorems}
\Crefname{theorem}{Theorem}{Theorems}

\newcommand{\best}[1]{\textbf{#1}}
\newcommand{\secondbest}[1]{\underline{#1}}

\setcopyright{cc}
\setcctype{by}
\acmJournal{PACMMOD}
\acmYear{2026} \acmVolume{4} \acmNumber{6 (SIGMOD)} \acmArticle{382}
\acmMonth{12} \acmDOI{10.1145/3856377}

\begin{document}

\title{TraHist: Fast and Accurate Cardinality Estimation for~Trajectory Data (Extended Version)}
\titlenote{This is the extended version of the paper ``TraHist: Fast and Accurate Cardinality Estimation for Trajectory Data,'' to appear in Proc. ACM Manag. Data 4, 6 (SIGMOD), Article 382 (December 2026), https://doi.org/10.1145/3856377. This version additionally includes the full appendices. The source code is available at \url{https://github.com/YJ960/TraHist}.
\vspace{2pt}}

\author{Yujeong Jo}
\orcid{0009-0005-8029-4073}
\email{joyujeong960@korea.ac.kr}
\affiliation{%
  \institution{Korea University}
  \city{Seoul}
  \country{Republic of Korea}
}

\author{Incheol Baek}
\orcid{0000-0002-6857-5666}
\email{inch307@korea.ac.kr}
\affiliation{%
  \institution{Korea University}
  \city{Seoul}
  \country{Republic of Korea}
}

\author{Changbeom Shim}
\authornote{Changbeom Shim and Yon Dohn Chung are the corresponding authors.}
\orcid{0000-0001-6604-0855}
\email{changbeom.shim@curtin.edu.au}
\affiliation{%
  \institution{Curtin University}
  \city{Bentley}
  \country{Australia}
}

\author{Yon Dohn Chung}
\authornotemark[2]
\orcid{0000-0003-2070-5123}
\email{ydchung@korea.ac.kr}
\affiliation{%
  \institution{Korea University}
  \city{Seoul}
  \country{Republic of Korea}
}

% \renewcommand{\shortauthors}{Jo et al.}

\begin{abstract}
  Cardinality estimation is fundamental for query optimization in database systems. Histogram-based estimators are widely used for their simplicity and efficiency but rely on point-centric partitioning that assumes record independence. This assumption breaks in trajectory data, where records from the same moving object show strong spatiotemporal dependencies. Conventional axis-aligned histograms therefore fragment trajectories across partition boundaries, leading to overestimation through duplicate counting. This error reflects a structural limitation of partition boundaries that are misaligned with trajectory-induced density patterns. To this end, we propose \textit{\textbf{TraHist}}, a histogram-based estimator approximating density-aligned partitioning via efficient grid-based propagation. TraHist builds density-adaptive rectilinear buckets to preserve trajectory continuity and local density coherence while keeping histogram efficiency. Experiments on large-scale real-world datasets show that TraHist reduces estimation error by over 20$\times$ relative to the strongest competing estimator in query regimes where trajectory fragmentation is a dominant source of estimation error, while maintaining sub-millisecond query latency and competitive construction cost. 
\end{abstract}




% \received{17 April 2026}
% \received[revised]{19 August 2026}
% \received[accepted]{12 September 2026}

\maketitle

\begin{figure}[t]
    \centering

    \begin{subfigure}{0.31\linewidth}
        \centering
        \includegraphics[
            width=0.92\linewidth,
            trim=0 10 0 0,
            clip
        ]{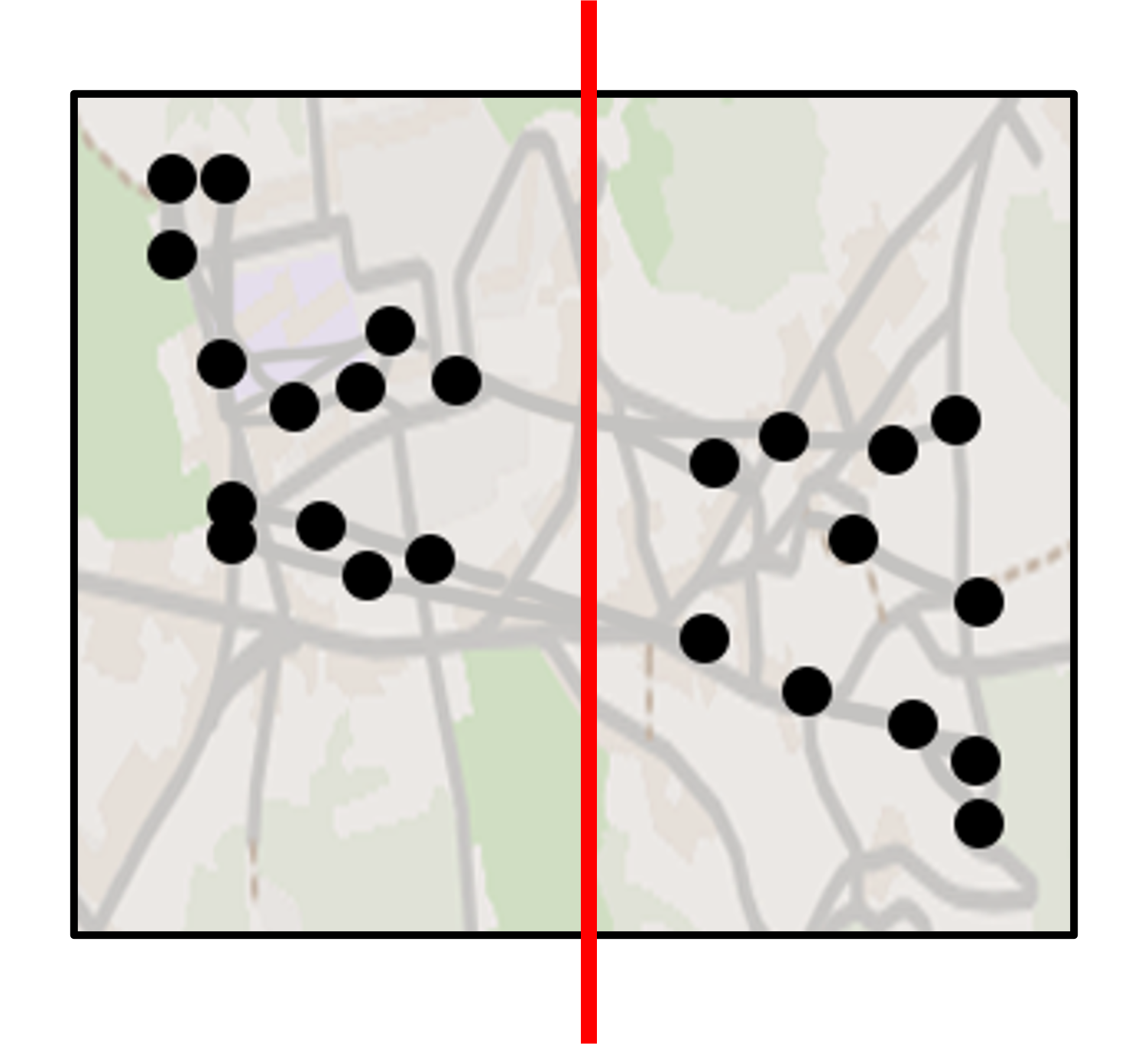}
        \caption{\mbox{Point-centric Partitioning}}
    \end{subfigure}%
    \hspace{0.01\linewidth}%
    \begin{subfigure}{0.31\linewidth}
        \centering
        \includegraphics[
            width=0.92\linewidth,
            trim=0 10 0 0,
            clip
        ]{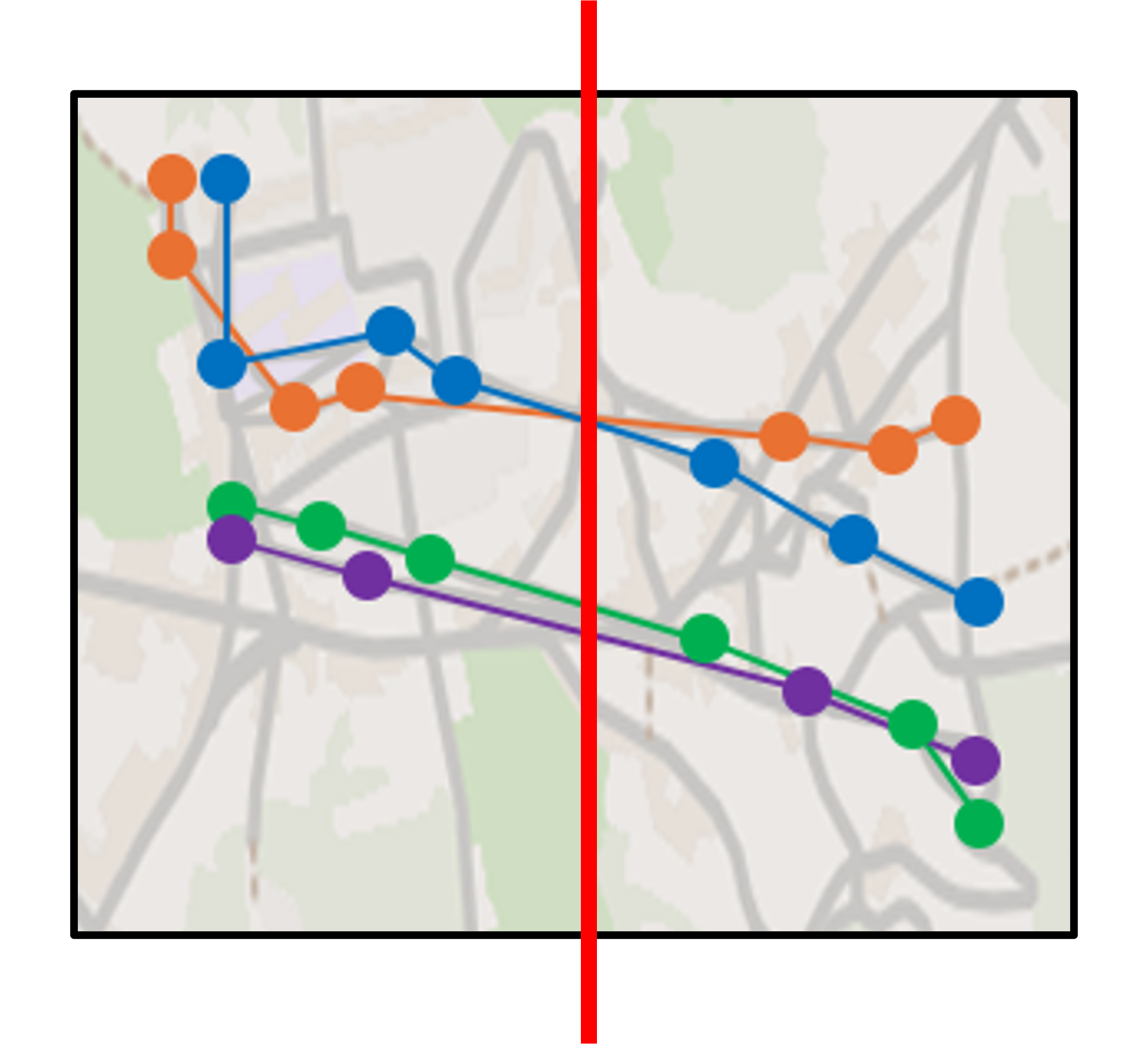}
        \caption{\mbox{Trajectory Fragmentation}}
    \end{subfigure}%
    \hspace{0.01\linewidth}%
    \begin{subfigure}{0.31\linewidth}
        \centering
        \includegraphics[
            width=0.92\linewidth,
            trim=0 10 0 0,
            clip
        ]{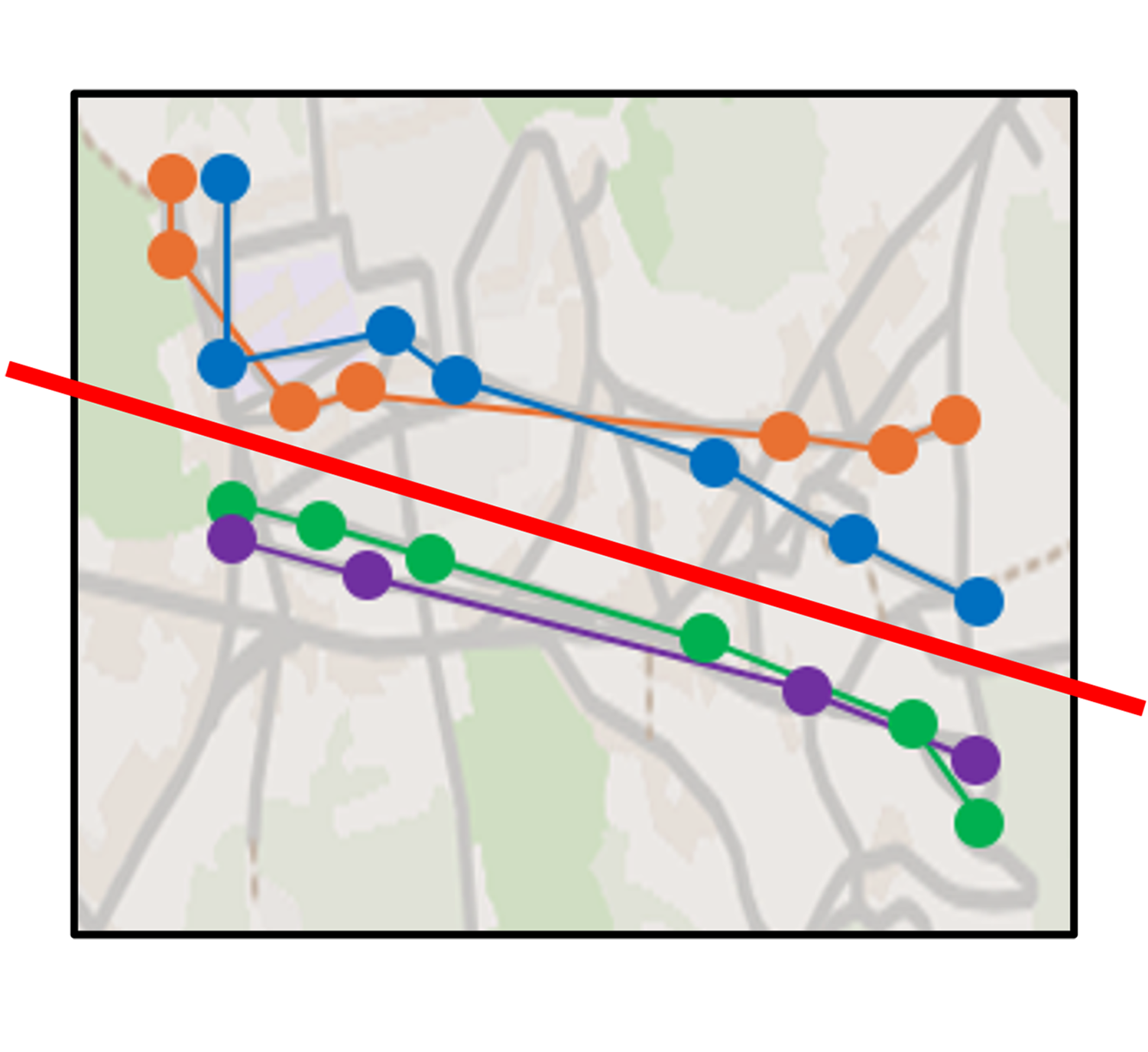}
        \caption{\mbox{Trajectory-aware Partitioning}}
    \end{subfigure}

    \caption{Trajectory fragmentation in point-centric partitioning and the motivation for trajectory-aware partitioning. (a) Point-centric histograms treat trajectory records as independent points, while (b) axis-aligned boundaries (red lines) cut trajectories, causing fragmentation. (c) Trajectory-aware partitioning instead adapts boundaries to data patterns, preserving object-level continuity.}
    \Description{Three schematic panels compare point-centric and trajectory-aware partitioning. Panel (a) shows trajectories represented as independent points under axis-aligned partitioning. Panel (b) illustrates trajectories being split by partition boundaries, resulting in fragmentation. Panel (c) shows boundaries adapted to the data pattern to better preserve trajectory continuity.}
    \label{fig:introduction}
\end{figure}

%%%%%%%%%%%%%%%%%%%%%%%%%%%%%%%%%%%%%%%%%%%%%%%%%%%%%%%%%%%%%%%%%

\section{Introduction}
In database systems, cardinality estimation is the task of estimating the number of records that satisfy a query predicate~\cite{selinger1979access}. It is a fundamental component of query optimization, as inaccurate estimates can lead to suboptimal execution plans and unpredictable query performance~\cite{leis2015good}. This problem becomes particularly challenging when query predicates are evaluated over datasets with strong correlations among records. Such correlations are inherent in trajectory data, where sequences of records generated by the same moving object form strong spatiotemporal dependencies~\cite{zheng2015trajectory, wang2021survey}. Consequently, range predicates select multiple records originating from the same object, challenging conventional estimators.
\par Most cardinality estimators in modern database query optimizers rely on histograms as compact synopses to summarize data distributions~\cite{leis2015good}. These methods partition the data space using axis-aligned boundaries and estimate query results by aggregating counts from buckets intersecting the query predicate~\cite{muralikrishna1988meh, ioannidis1995voptimal, poosala1996maxdiff, poosala1997mhist, jagadish1998voptimal2, bruno2001stholes, liu2021lhist}. This abstraction implicitly treats records as independent points, assuming that partition boundaries preserve their semantic interpretation across buckets. While such an assumption is reasonable for uncorrelated point data, it breaks down in trajectory workloads. As shown in~\cref{fig:introduction}, a single trajectory may intersect multiple buckets as it traverses the data space, causing records from the same object to be assigned to different buckets. Under bucket-wise aggregation, these records are treated as separate contributions, leading to duplicate counting of the same object~\cite{orlando2007spatio, braz2007approximate, fanaeepour2015case}. Crucially, this overestimation does not arise from overly coarse partitioning or insufficient bucket-level accuracy. Even when bucket summaries are accurate, object-level continuity remains fragmented at partition boundaries.
\par Although existing approaches to cardinality estimation improve accuracy through more expressive summaries or models, they primarily focus on capturing data distributions of individual records. Traditional multi-dimensional histograms~\cite{muralikrishna1988meh, ioannidis1995voptimal, poosala1996maxdiff, poosala1997mhist, jagadish1998voptimal2, aboulnaga1999stgrid, bruno2001stholes} enhance estimation accuracy by refining bucket boundaries to better capture local data distributions or adapt to query feedback. Learned multi-dimensional histogram (LHist)~\cite{liu2021lhist} further improves accuracy by learning more expressive models within buckets, while still relying on axis-aligned partitioning. In contrast, model-based learned cardinality estimators~\cite{kipf2019mscn, yang2019naru, hilprecht2020deepdb} do not depend on histogram-style space partitioning, but instead use trained models to estimate cardinalities from query predicates by leveraging learned statistical relationships in the data. While effective for capturing correlations in relational data, these approaches do not capture inter-record dependencies arising from spatiotemporal continuity in trajectories. Moreover, such methods typically require model training and inference at estimation time, which departs from the lightweight, training-free nature of histogram-based estimators. As a result, prior work leaves open the question of \textit{how to design partition boundaries that respect trajectory continuity while retaining histogram efficiency}.
\par To address this gap, we propose \textit{\textbf{TraHist}}, a histogram-based cardinality estimator designed for trajectory data. TraHist revisits histogram partitioning from the perspective of object-level continuity, constructing partitions that adapt to trajectory traversal patterns. By aligning partition boundaries with trajectory-induced density structures, our method alleviates fragmentation of individual trajectories across buckets and improves the representativeness of data grouped within each bucket. This design is guided by an idealized partitioning criterion we term \textit{density-aligned partitioning}, which aims to align partition boundaries with trajectory-induced density variations. In practice, TraHist approximates this criterion in a scalable manner by constructing density-adaptive rectilinear buckets compatible with efficient histogram-based estimation.

\par In summary, we make the following contributions.
\begin{itemize}
    \item We analyze a fundamental limitation of point-centric, axis-aligned histogram partitioning for trajectory data, which fragments trajectories and degrades bucket representativeness.
    \item We establish \textit{density-aligned partitioning} as a design principle, providing an idealized criterion for addressing both structural and statistical deficiencies of existing histograms.
    \item We propose \textit{TraHist}, a trajectory-aware histogram-based cardinality estimator that approximates density-aligned partitioning by constructing density-adaptive rectilinear buckets.
    \item We conduct extensive experiments on large-scale real-world trajectory datasets, showing that TraHist achieves significantly higher estimation accuracy than existing methods while maintaining fast query performance and competitive construction cost.
\end{itemize}

\Cref{tab:notation} summarizes the notation used throughout the paper.

\begin{table}[t]
    \centering
    \caption{Summary of frequent notation.}
    \label{tab:notation}
    {
    \renewcommand{\arraystretch}{1}
    \setlength{\aboverulesep}{0pt}
    \setlength{\belowrulesep}{0pt}
    \setlength{\arrayrulewidth}{0.8pt}
    \setlength{\tabcolsep}{3pt}
    \newcommand{\topgap}{\rule{0pt}{2.7ex}}
    \newcommand{\botgap}{\rule[-1.5ex]{0pt}{0pt}}
    \ifdefined\qcatw\else\newlength{\qcatw}\fi
    \settowidth{\qcatw}{\textbf{Partitioning \&}}
    \newcommand{\cat}[2]{\multirow{#1}{\qcatw}{\raggedright\textbf{#2}}}
    \resizebox{0.78\textwidth}{!}{%
    \begin{tabular}{@{}p{\qcatw}@{\hspace{9pt}}|@{\hspace{9pt}}c@{\hspace{9pt}}l@{}}
        \toprule[0.8pt]
        \topgap\botgap\textbf{Category} & \textbf{Notation} & \textbf{Description} \\
        \midrule[0.8pt]
        \cat{5}{Spatial \& trajectory basics}
        & \topgap$\Omega$ & Spatial domain \\
        & $\mathcal{D}$ & Trajectory dataset \\
        & $T$ & A trajectory, $T=\langle p_1,\dots,p_n\rangle$ \\
        & $p$ & A trajectory point, $p\in\Omega$ \\
        & \botgap$L(T)$ & Piecewise-linear spatial representation of $T$ \\
        \midrule
        \cat{6}{Grid \& \\density metrics}
        & \topgap$\mathcal{G}$ & Regular spatial grid (CountGrid) \\
        & $x$ & A grid cell in $\mathcal{G}$ \\
        & $C(x)$ & Raw point count in grid cell $x$ \\
        & $\rho(x)$ & Continuous trajectory-induced point density at location $x$ \\
        & $\rho_p(x)$ & Log-transformed density value at grid cell $x$ \\
        & \botgap$\mu(x)$ & Local mean of $\rho_p$ around $x$ \\
        \midrule
        \cat{4}{Partitioning \& bucket formation}
        & \topgap$\Pi$ & Partition of the spatial domain \\
        & $B_k$ & The $k$-th histogram bucket \\
        & $\mathrm{LM}(x)$ & LabelMap assigning grid cell $x$ to bucket $B_k$ \\
        & \botgap$\mathcal{N}(x)$ & Neighboring grid cells of $x$ \\
        \midrule
        \cat{4}{Query \& cardinality estimation}
        & \topgap$Q$ & Spatial range query region \\
        & $RQ(Q)$ & Set of trajectories whose induced polylines intersect $Q$ \\
        & $y_q$ & True trajectory-level cardinality of query $q$ \\
        & \botgap$\hat{y}_q$ & Estimated trajectory-level cardinality of query $q$ \\
        \bottomrule[0.8pt]
    \end{tabular}%
    }
    }
\end{table}

%%%%%%%%%%%%%%%%%%%%%%%%%%%%%%%%%%%%%%%%%%%%%%%%%%%%%%%%%%%%%%%%%

\section{Background} \label{sec:background}

\subsection{Trajectory Data and Query} \label{subsec:definitions}
\par We consider trajectory datasets consisting of movement records generated by distinct objects over a spatial domain $\Omega \subset \mathbb{R}^2$. Each object produces a sequence of spatial samples as it moves through the domain. Most existing DBMSs store and index trajectory data using point-based abstractions, treating individual samples as records, while many spatial queries are defined with respect to entire trajectories rather than individual points~\cite{graser2018evaluating, wang2021survey}.

\begin{definition}[Trajectory]\label{dfn:trajectory}
    A trajectory $T$ is defined as an ordered sequence of spatial points sampled from the movement of a single object, i.e., $T = \langle p_1, p_2, \ldots, p_n\rangle$, where each point $p_i\in \Omega$ represents a spatial location~\cite{wang2021survey}. We write $L(T)=\bigcup_{i\text{=}1}^{n\text{-}1}\overline{p_ip_{i\text{+}1}}$ for the polyline induced by $T$, i.e., the piecewise-linear spatial representation obtained by connecting consecutive samples in order.
\end{definition}

\begin{definition}[Trajectory Range Query]\label{dfn:trajectory-range-query}
    Given a trajectory dataset $\mathcal{D}$ and a spatial range $Q\subset\Omega$, a trajectory range query returns the set of trajectories whose polylines intersect $Q$: 
    \begin{equation}\label{eq:query}
        RQ(Q)=\{T\in\mathcal{D}\mid L(T) \cap Q \neq \emptyset\}.
    \end{equation} 
    The trajectory-level cardinality is defined as the number of distinct trajectories returned by the query, i.e., $|RQ(Q)|$.
\end{definition}

\par The query predicate is defined over $L(T)$, whereas a point-centric synopsis is constructed from sampled points in $\mathcal{D}$. The directly observable sample-hit condition $\exists p \in T \ such\ that\ p\in Q$ identifies only a subset of $RQ(Q)$: it misses intersections that occur along a segment between consecutive samples. This gap widens as trajectories are sampled sparsely relative to $Q$. Moreover, bucket-wise aggregation of sampled points can treat contributions from the same trajectory independently and count that trajectory multiple times across buckets. TraHist targets $|RQ(Q)|$ and estimates it from sample-derived point-density and distinct-trajectory statistics.
\par In this work, we focus on \textit{spatial} trajectory range CE: timestamps determine the ordering of the sampled points in $T$, and hence $L(T)$, but do not appear in the query predicate. Spatiotemporal range predicates additionally constrain time and therefore define a distinct CE problem with separate temporal semantics. Within this spatial setting, we target the standard \textsc{Intersects} predicate, defined in terms of the DE-9IM~\cite{ogc_sfa_2011}. More specific topological relations for trajectories~\cite{bornholdt2025trajectory} define cardinality functions beyond our target; \cref{app:topological-predicates} discusses the additional statistics required for their cardinality estimation and how the \textsc{Intersects} estimate can support their query processing.

\subsection{Cardinality Estimation Problem}\label{subsec:cardinality-estimation-problem}
\par Cardinality estimation typically summarizes a dataset with a compact synopsis that supports efficient query optimization. Histograms are widely used synopses: they partition the data domain into a finite set of disjoint regions, called \textit{buckets}, and store a representative statistic for each bucket.
\par Given a query workload $\mathcal{W}$, the objective of a histogram $H$ is to minimize the estimation error, subject to a constraint on the number $\vert B\vert$ of buckets, which corresponds to a memory budget. The estimation error is defined as
\begin{equation} \label{eq:cardest}
    \mathrm{Err}(H)=\sum_{q\in \mathcal{W}}\mathrm{Err}(y_q, \hat{y}_{q}),
\end{equation}
where $y_q$ is the true cardinality for a query $q$, $\hat{y}_q=f_H(q)$ is the cardinality estimated by the estimator function $f_H$ associated with the histogram $H$. The function $\mathrm{Err}(\cdot)$ denotes an error metric.
\par In this work, we focus on histogram-based cardinality estimation for trajectory range queries. Specifically, for a query $q$ with spatial range $Q$, given a partitioning $\Pi = \{B_1, B_2, \ldots, B_{\vert B\vert}\}$ and an estimator function $f_H$, the goal is to estimate the true trajectory-level cardinality $y_q=|RQ(Q)|$ using $\hat{y}_q=f_H(Q)$.

%%%%%%%%%%%%%%%%%%%%%%%%%%%%%%%%%%%%%%%%%%%%%%%%%%%%%%%%%%%%%%%%%

\begin{figure}[t]
    \centering
    \begin{subfigure}{0.38\linewidth}
        \centering
        \includegraphics[
            width=\linewidth,
            trim=0 10 0 0,
            clip
        ]{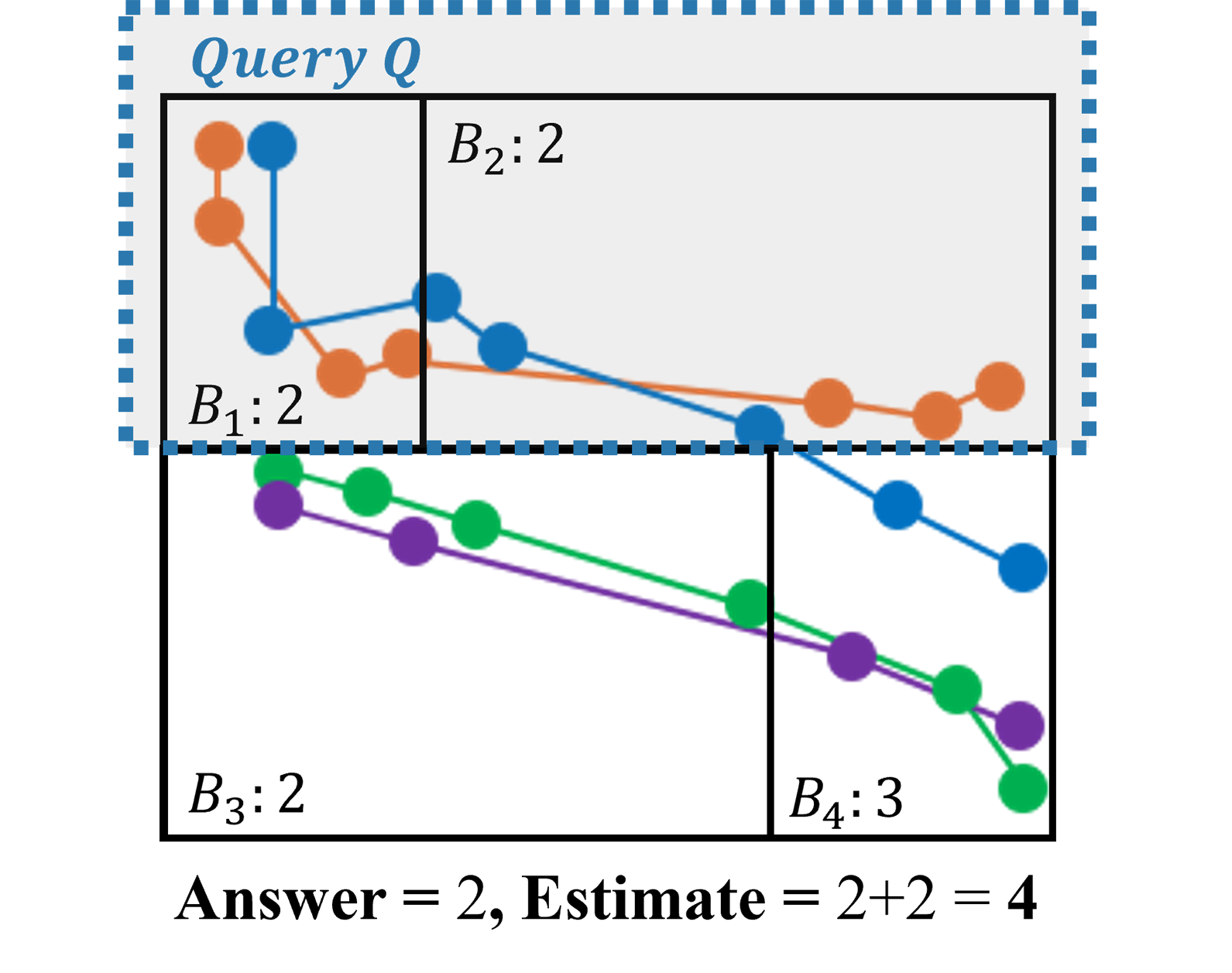}
        \caption{Trajectory Fragmentation}
        \label{fig:motiv_trajfrag}
    \end{subfigure}%
    \hspace{0.03\linewidth}%
    \begin{subfigure}{0.38\linewidth}
        \centering
        \includegraphics[
            width=\linewidth,
            trim=0 10 0 0,
            clip
        ]{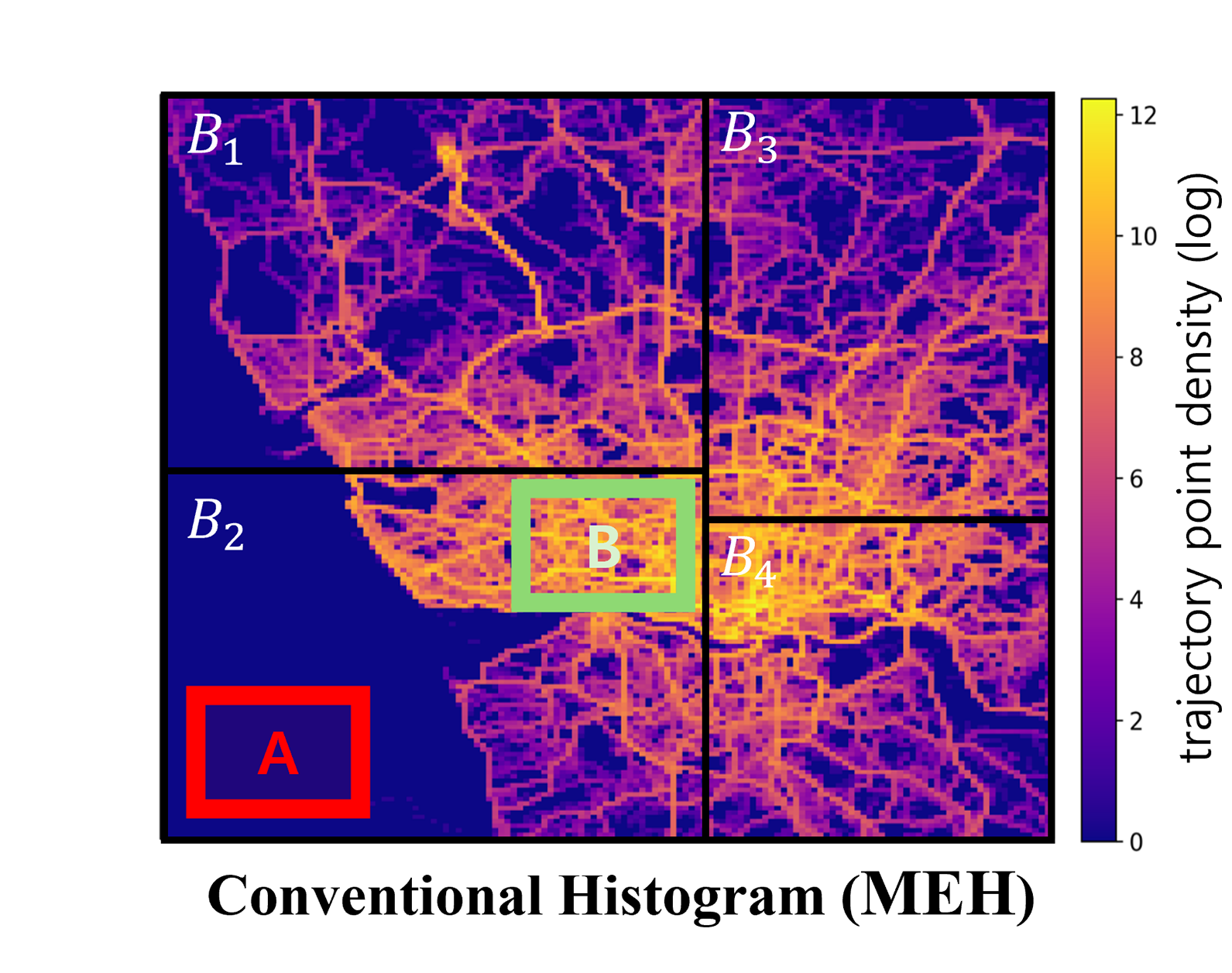}
        \caption{Loss of Density Coherence}
        \label{fig:motiv_density}
    \end{subfigure}

    \caption{Limitations of conventional histogram-based estimators for trajectory data.
    (a) \textit{Fragmented Trajectories}: Every trajectory split across multiple buckets can be counted multiple times under independent bucket aggregation.
    (b) \textit{Loss of density coherence}: On real trajectory data (\textit{Porto}), axis-aligned buckets aggregate regions with substantially different densities (regions \textsf{A} and \textsf{B}), undermining bucket-level summaries.}
    \label{fig:motivation}
    \Description{Two panels illustrate limitations of conventional histogram partitioning for trajectory data. Panel (a) shows a trajectory spanning multiple buckets, causing duplicate counting under independent bucket aggregation. Panel (b) shows axis-aligned buckets on the Porto dataset, where a single bucket contains regions with substantially different trajectory point densities.}
\end{figure}

\section{Motivations} \label{sec:motivations}
\par Conventional histogram-based cardinality estimators summarize data distributions using compact partitions under assumptions of record independence and local uniformity. In trajectory workloads, however, applying such axis-aligned partitioning to sampled trajectory points gives rise to two structural factors that conventional histogram partitioning cannot address simultaneously. One factor originates from ignoring dependencies among samples from the same trajectory, resulting in trajectory fragmentation that leads to duplicate counting under bucket-wise aggregation (\cref{subsec:frag_traj}). The other is caused by biased and spatially correlated density patterns of trajectories, which undermine the formation of locally coherent partitions (\cref{subsec:density_coherence_loss}). These two factors are illustrated in~\cref{fig:motivation}. Constructing effective histograms for trajectory data therefore requires jointly accounting for both object-level continuity and density heterogeneity. 

\subsection{Fragmented Trajectories} \label{subsec:frag_traj}
A defining characteristic of trajectory data is that sampled points generated by the same object are dependent and jointly induce the polyline $L(T)$. However, histogram-based estimators adopt a point-centric abstraction, treating these dependent samples as independent points during both partitioning and estimation. Under axis-aligned rectangular partitioning, $L(T)$ may cross multiple bucket boundaries, causing sampled points from the same trajectory to be assigned to different buckets. We refer to this loss of object-level continuity as \textit{trajectory fragmentation}.
\par The origin of trajectory fragmentation lies in the geometric mismatch between axis-aligned bucket boundaries and the dominant directions of trajectory movement. Consider a regular axis-aligned grid with cell width $h$ and a segment of $L(T)$ with length $d$ oriented at an angle $\theta$ with respect to the x-axis. The expected number of grid boundary crossings is proportional to $\frac{d}{h}\left(|\cos \theta| + |\sin \theta|\right)$. This quantity is minimized when the segment is aligned with a coordinate axis and maximized when it is diagonal to the grid axes.
\par Trajectory fragmentation directly impacts cardinality estimation under point-centric aggregation. Consider a trajectory $T\in RQ(Q)$ whose sampled points are assigned to $k$ buckets that overlap $Q$. Although $T$ should contribute exactly once to $|RQ(Q)|$, histogram-based estimators aggregate its bucket-level contributions independently, yielding up to $k$ contributions instead of one. \cref{fig:motiv_trajfrag} shows a trajectory whose points fall into several buckets overlapping $Q$, each of which contributes a separate count for the same trajectory to the aggregated estimate. This error persists even when individual buckets accurately summarize local point statistics, as it arises from fragmented object-level aggregation rather than estimation noise within buckets.
\par A natural intuition is to mitigate this error by grouping regions traversed by the same trajectory into a single bucket. However, enforcing continuity alone fundamentally alters the role of histogram partitions. Buckets formed solely to preserve trajectory continuity may span spatially heterogeneous regions, aggregating records from distinct areas of the domain and degrading the reliability of bucket-level summaries. This analysis indicates that trajectory fragmentation cannot be resolved by na\"ive continuity-preserving partitioning without regard to local density coherence.

\subsection{Loss of Density Coherence in Buckets}
\label{subsec:density_coherence_loss}
A core objective of histogram-based estimators is to construct partitions in which bucket contents can be accurately summarized by representative statistics during query estimation. This objective fundamentally depends on the homogeneity of data distributions within individual buckets and directly influences the global estimation error defined in~\cref{subsec:cardinality-estimation-problem}. Let $\rho(x)$ denote the local trajectory point density (points per unit area) at location $x$. For a bucket $B$, the degree of density heterogeneity can be quantified by the \textit{intra-bucket variance} $Var(B)$, defined by:
\begin{equation} \label{eq:intra-bucket_variance}
    {Var}(B)=\frac{1}{A(B)}\int_{x\in B}(\rho(x)-\mu_{B})^2dx,
\end{equation}
where $\mu_{B}=\frac{1}{A(B)}\int_{B}\rho(x)dx$ represents the average density within bucket $B$, and $A(B)$ denotes the area of $B$. A core design goal of histogram-based estimators is therefore to construct partitions that keep $Var(B)$ low, ensuring that bucket-level summaries provide accurate representations of local data distributions during query estimation.
\par This objective is particularly difficult in trajectory datasets due to the biased and spatially correlated nature of trajectory-induced point densities. Individual trajectories vary in length, speed, and orientation, and collectively induce highly anisotropic spatial density patterns. Rather than forming compact and isotropic clusters, trajectory point density tends to concentrate along elongated paths corresponding to dominant movement directions. As a result, density coherence is largely preserved along trajectory-aligned directions, while substantial variation arises across them. \cref{fig:motiv_density} shows axis-aligned buckets aggregating regions with substantially different trajectory point densities into a single bucket.
\par Independent of trajectory fragmentation, this directional density structure poses a fundamental challenge for histogram construction. Conventional axis-aligned rectangular partitions are generally misaligned with dominant density directions and therefore tend to mix regions of substantially different density within the same bucket. As a result, intra-bucket variance becomes inflated, undermining the effectiveness of representative bucket-level summaries and degrading estimation accuracy. Refining bucket granularity or employing more expressive bucket-level models can mitigate this effect only to a limited extent, because the underlying geometric mismatch persists unless the partition boundaries themselves are adapted to the density structure.

%%%%%%%%%%%%%%%%%%%%%%%%%%%%%%%%%%%%%%%%%%%%%%%%%%%%%%%%%%%%%%%%%

\section{Proposed Method: TraHist}
This section presents TraHist, a fast and accurate cardinality estimator for trajectory data. We first introduce density-aligned partitioning as a design principle that characterizes desirable partition boundaries for trajectory data (\cref{subsec:DAP_principle}). Since directly optimizing this principle in continuous space is computationally infeasible, TraHist adopts a density-adaptive approximation based on grid-based propagation to construct rectilinear buckets (\cref{subsec:implementation}). Finally, we describe how the resulting histogram is used to estimate trajectory query cardinality (\cref{subsec:estimation}).

\subsection{Density-aligned Partitioning} \label{subsec:DAP_principle}
\par The fundamental challenge in trajectory cardinality estimation lies in designing a partitioning scheme that preserves trajectory continuity and maintains local density coherence. Conventional histograms struggle to satisfy both objectives simultaneously, as they restrict boundaries to fixed coordinate axes that are often misaligned with the directional structure of trajectory-induced density fields. These limitations motivate the design principle that we term \textit{density-aligned partitioning}.
\par Intuitively, a partition is density-aligned if its boundaries follow directions along which the trajectory-induced density varies minimally, instead of being tied to fixed coordinate axes. Let $\Pi$ denote a partition of the spatial domain into $|B| $ regions, $\delta \Pi$ its boundary set, and $t(x)$ the unit tangent vector to the boundary at location $x$. The local density geometry is described by the gradient $\nabla \rho(x)$. The rate of density change along a boundary can be quantified by the magnitude of the inner product $\lvert \nabla \rho(x) \cdot t(x) \rvert$.  Integrating this quantity over the partition boundaries with respect to arc length yields the following idealized formulation:
\begin{equation}  \label{eq:ideal_density_partitioning}
    \Pi^* = \underset{\Pi}{\arg\min}\int_{\delta\Pi}\vert\nabla\rho(x)\cdot t(x)\vert\ ds.
\end{equation}
\par This objective characterizes an ideal scenario in which partition boundaries are aligned with local density geometry and follow directions of minimal density change. However, directly optimizing this formulation is computationally impractical and incompatible with the efficiency requirements of histogram-based estimators. It induces a generally non-convex, combinatorial optimization over continuous boundary configurations under global consistency constraints, with a search space that grows rapidly as spatial resolution increases.

\begin{figure}[t]
    \centering

    \begin{subfigure}[b]{0.245\linewidth}
        \centering
        \includegraphics[width=\linewidth]{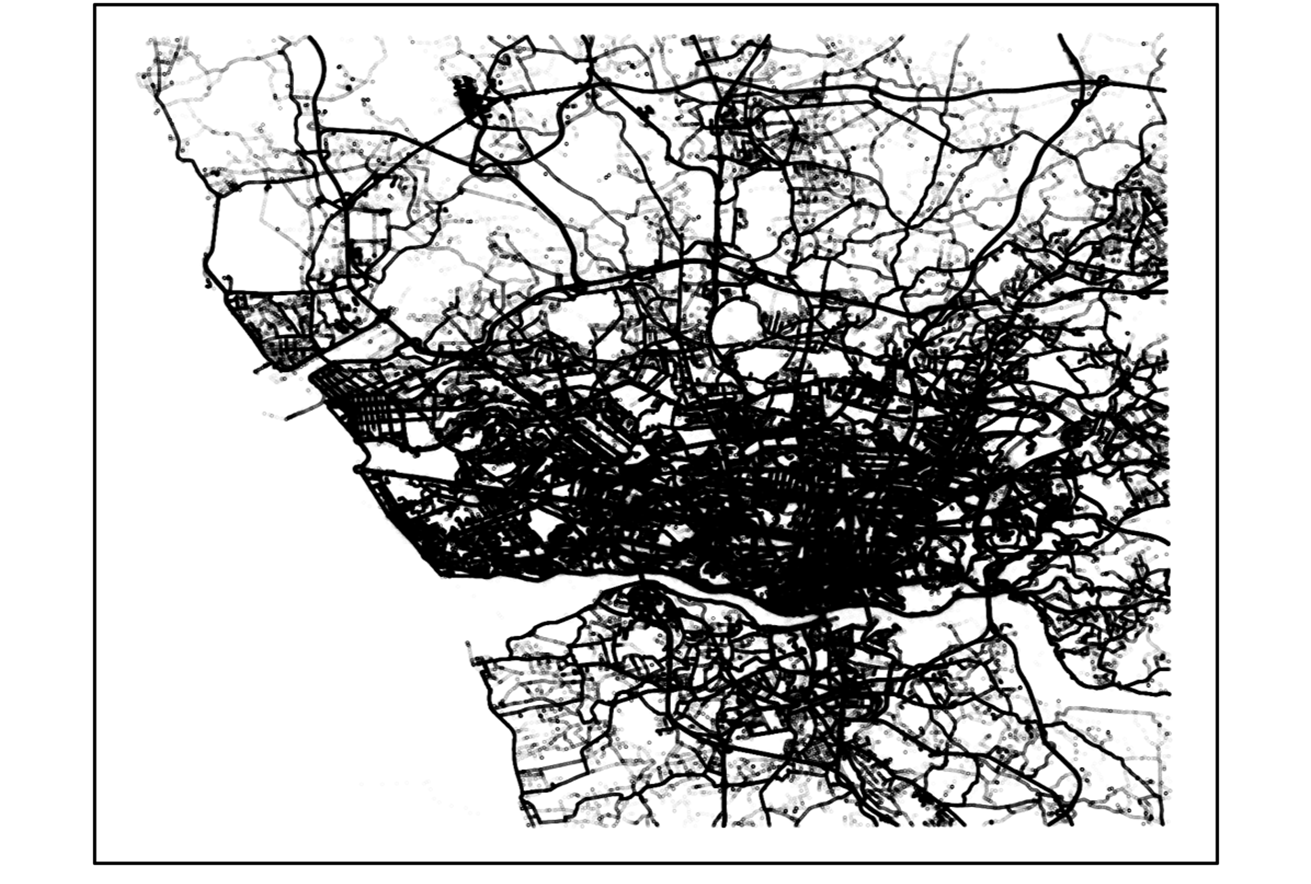}
        \caption{Dataset}
    \end{subfigure}
    \hfill
    \begin{subfigure}[b]{0.245\linewidth}
        \centering
        \includegraphics[width=\linewidth]{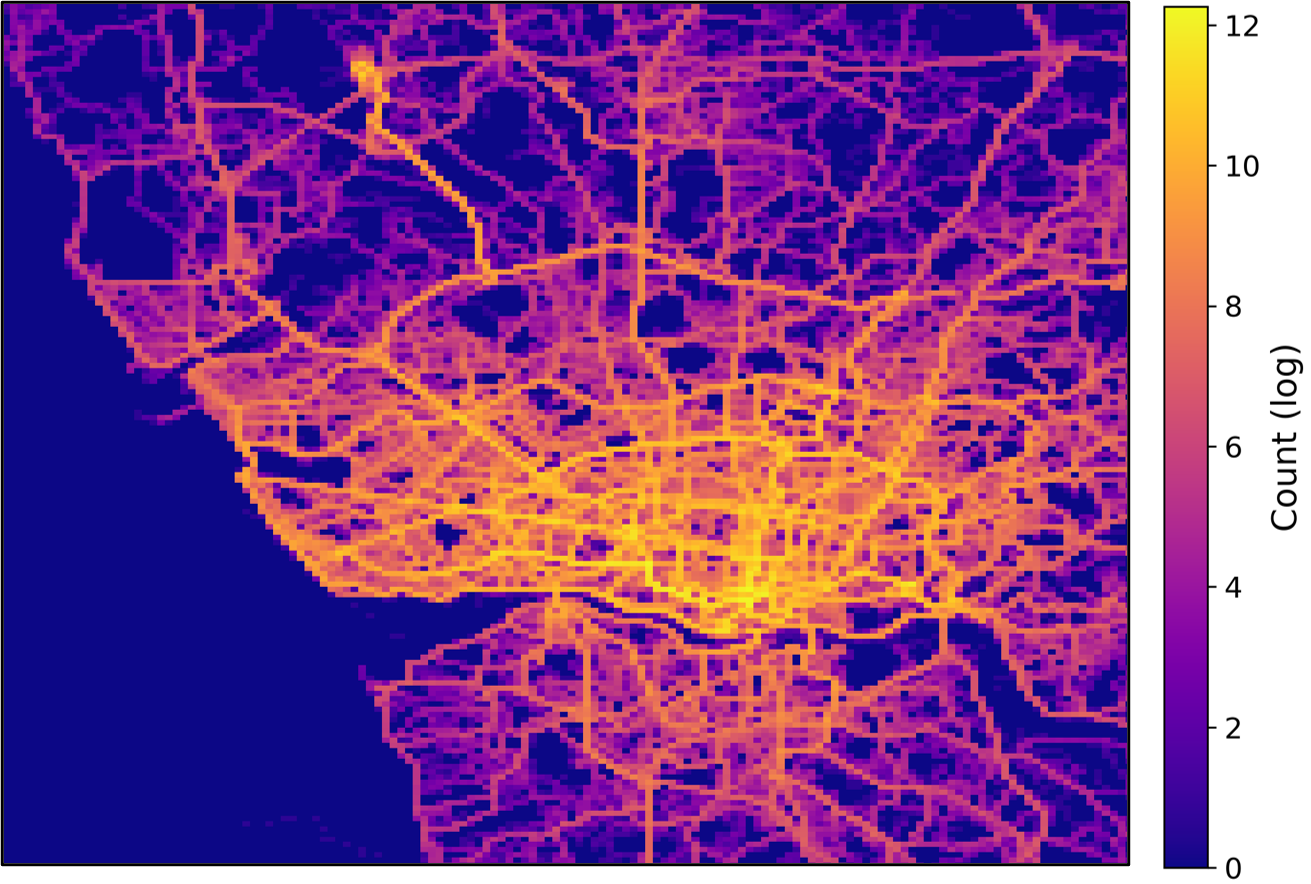}
        \caption{CountGrid $\rho_p$}
    \end{subfigure}%
    \hfill
    \begin{subfigure}[b]{0.245\linewidth}
        \centering
        \includegraphics[width=\linewidth]{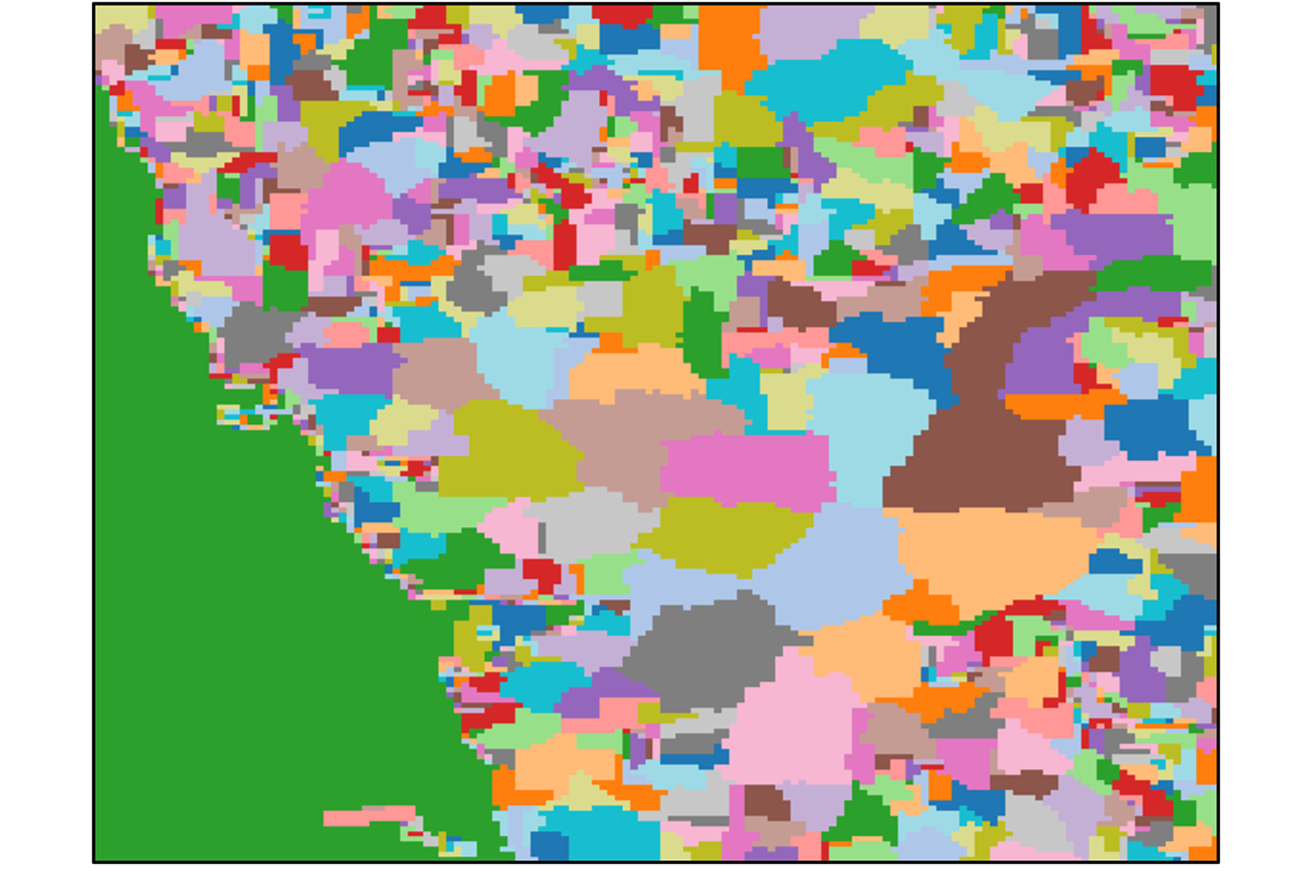}
        \caption{LabelMap}
    \end{subfigure}
    \hfill
    \begin{subfigure}[b]{0.245\linewidth}
        \centering
        \includegraphics[width=\linewidth]{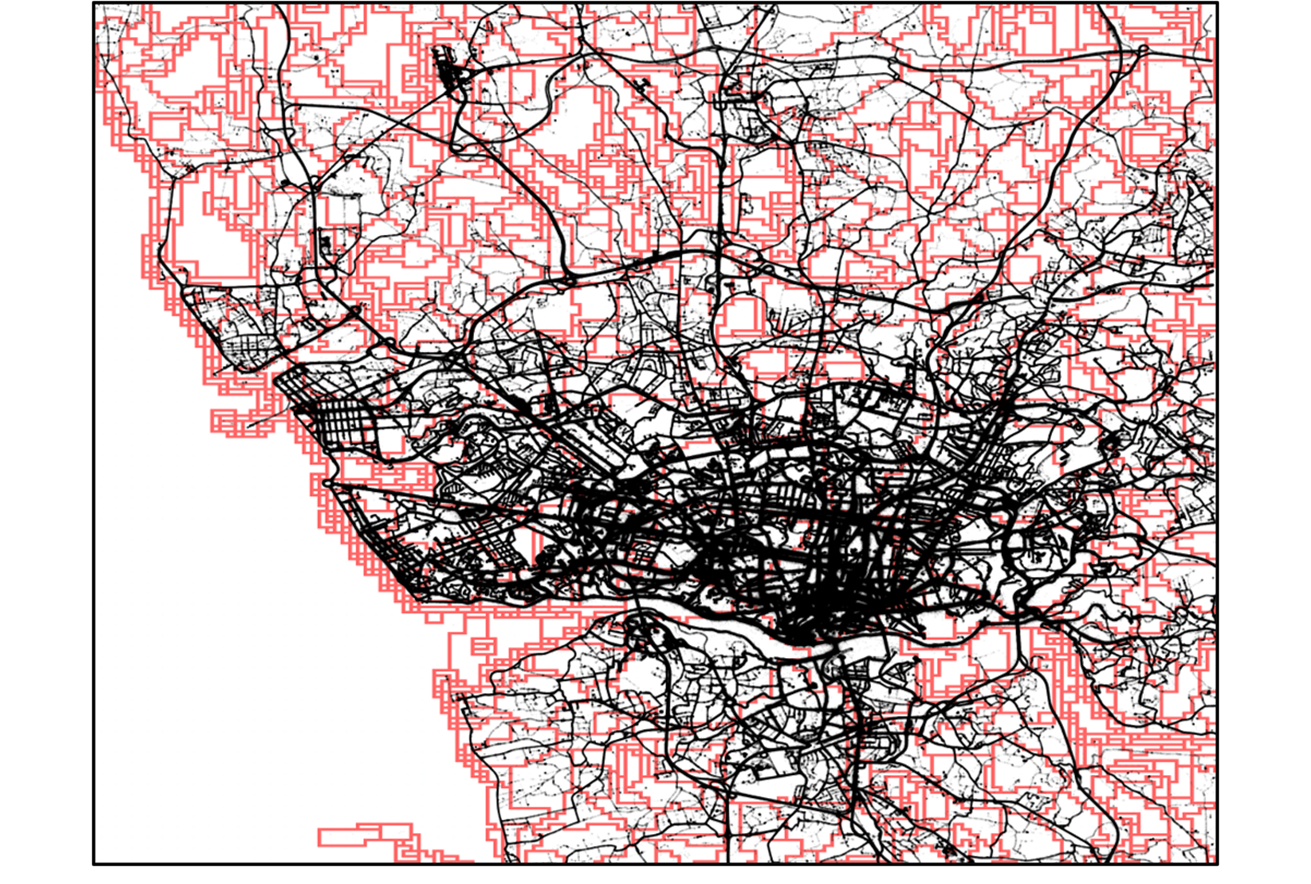}
        \caption{Bucket Boundaries}
    \end{subfigure}

    \caption{Overview of TraHist. (a) The raw trajectory dataset. (b) The corresponding CountGrid $\rho_p$, representing the log-transformed density field. (c) The LabelMap produced by density-adaptive bucket formation, with each color denoting a distinct bucket. (d) The resulting rectilinear bucket boundaries overlaid on the raw data.}
    \Description{Four panels show the progression of TraHist from raw trajectory data to the final partitioning. Panel (a) shows the trajectory dataset, panel (b) its CountGrid density representation, panel (c) the color-coded LabelMap assigning grid cells to buckets, and panel (d) the resulting rectilinear bucket boundaries overlaid on the trajectories.}
    \label{fig:overall_process}
\end{figure}

\begin{figure}[t]
    \centering
    \begin{subfigure}[b]{0.212\linewidth}
        \includegraphics[width=\linewidth]{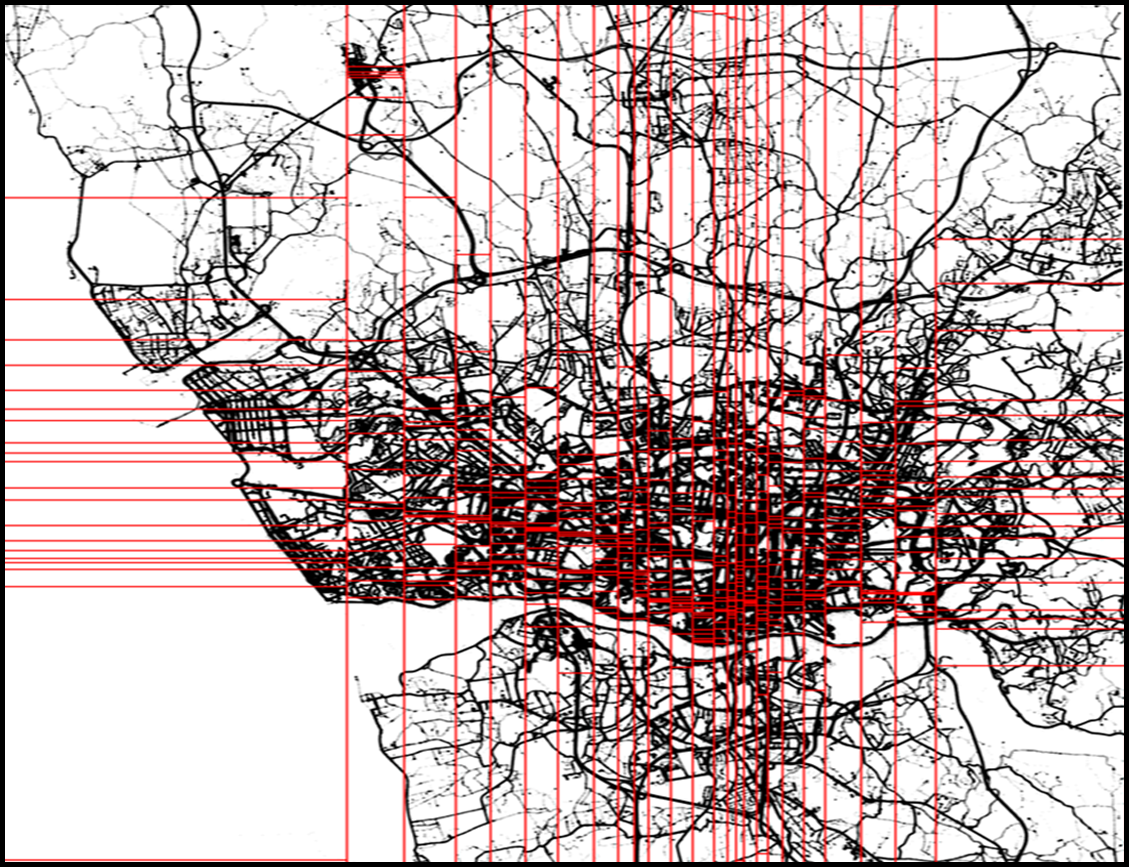}
        \caption{MEH}
        \label{fig:partitioning-meh}
    \end{subfigure}\qquad
    \begin{subfigure}[b]{0.212\linewidth}
        \includegraphics[width=\linewidth]{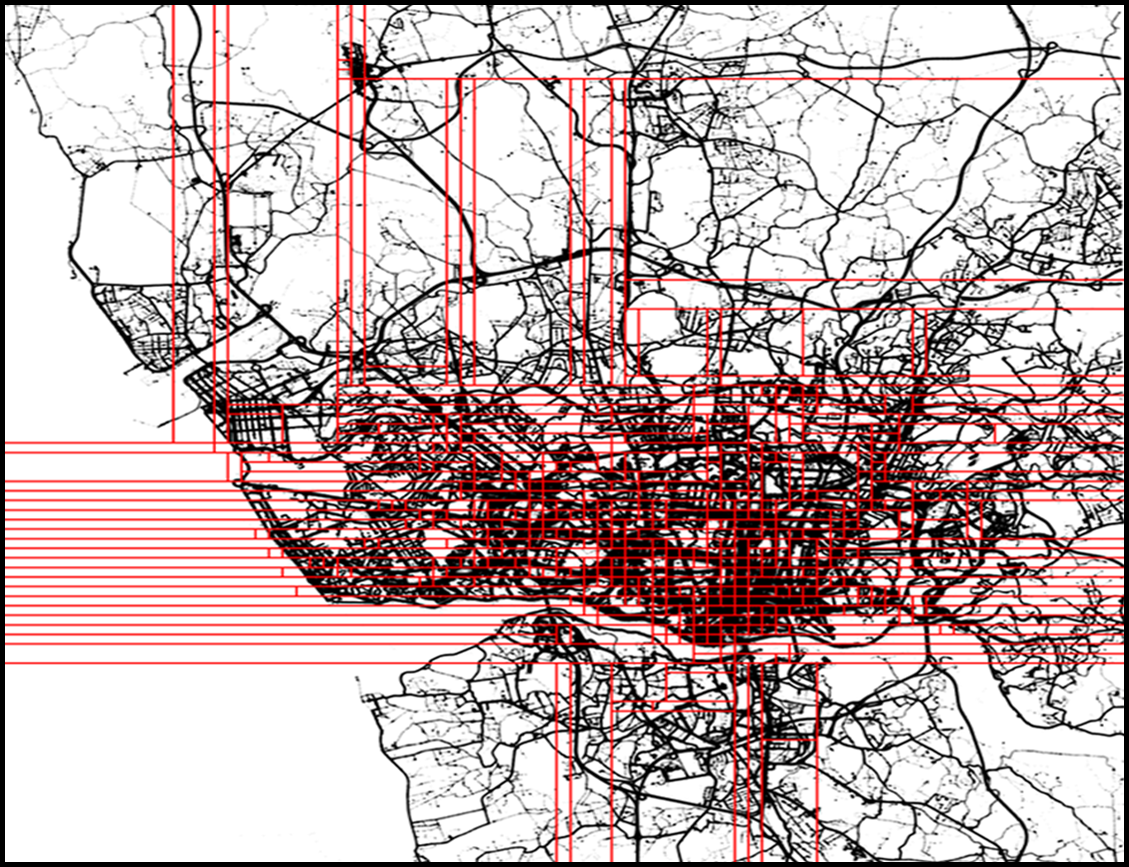}
        \caption{MHist}
        \label{fig:partitioning-mhist}
    \end{subfigure}
    \caption{Axis-aligned partitioning of MEH and MHist on \textit{Porto}, both with 500 buckets as in~\cref{fig:overall_process}.}
    \Description{Two panels show the partition boundaries produced by MEH and MHist on the Porto trajectory dataset. Red lines denote the axis-aligned rectangular bucket boundaries, with both methods using 500 buckets.}
    \label{fig:partitioning}
\end{figure}

\subsection{Density-adaptive Approximation} \label{subsec:implementation}
To make density-aligned partitioning applicable to large-scale trajectory datasets, TraHist adopts a practical approximation replacing direct continuous boundary optimization with an efficient construction on a discretized density field. TraHist approximates density-aligned partitioning through a two-stage process. It first constructs a discretized density representation (CountGrid), and then forms histogram buckets via a density-adaptive aggregation on this grid, providing a tractable surrogate for the ideal objective. The resulting pipeline, from the raw trajectory dataset to the final rectilinear bucket boundaries, is illustrated on a real-world example in~\cref{fig:overall_process}.

\noindentparagraph{\textnormal{\textbf{CountGrid Construction.}}} TraHist partitions the spatial domain into a regular grid $\mathcal{G}$ with cell size $\Delta$, and aggregates points into grid cells to obtain a cell-level representation of the trajectory-induced density field. For each cell $x \in \mathcal{G}$, the CountGrid value $C(x)$ is defined as
\begin{equation}
    \textstyle
    C(x) = \sum_{T \in \mathcal{D}} \sum_{p \in T} \mathbf{1}[p \in x],
\end{equation}
where $\mathcal{D}$ is the trajectory dataset, and $\mathbf{1}[\cdot]$ is the indicator function. This aggregation preserves spatial locality at a fixed resolution while making local density variation explicit, as illustrated in~\cref{fig:overall_process}b. To stabilize local density variation, TraHist applies a log transformation to the raw CountGrid values, yielding a log-transformed CountGrid density field $\rho_p(x)=\log(1+C(x))$, which serves as a discrete surrogate for the continuous density function $\rho(x)$.

\begin{algorithm}[t]
\SetKw{And}{and}
\caption{Density-adaptive Bucket Formation}
\label{alg:density-adaptive_bucket_formation}

\KwInput{Log-transformed CountGrid $\rho_p$; parameters $\alpha$, $\beta$; window size $\kappa$}
\KwOutput{LabelMap $\mathrm{LM}$ and buckets $\{B_k\}$}

Compute local mean map $\mu(x)$ over a $\kappa\times\kappa$ window\;
Initialize cumulative cost $\bar{D}(x)\leftarrow\infty$ and LabelMap $\mathrm{LM}(x)\leftarrow -1$\;
Initialize empty min-heap $H$ and current label $\ell\leftarrow0$\;

Select an unlabeled seed $s$ and initialize
$\ell\leftarrow\ell+1,\ \bar{D}(s)\leftarrow0,\ \mathrm{LM}(s)\leftarrow\ell$\;
$H.\text{push}((0,s))$\;

\While{$H$ is not empty}{
    $(\bar D^{\text{curr}},x)\leftarrow H.\text{pop}()$\;

    \lIf(\tcp*[f]{Better path already found}){$\bar D^{\text{curr}}>\bar D(x)$}{continue}

    Compute threshold $\tau(x)=\beta\cdot\mu(x)$ via \cref{eq:threshold}\;

    \For{each neighbor $n\in\mathcal{N}(x)$}{
        Compute contrast $D(x,n)$ via \cref{eq:local_cost}\;

        \If(\tcp*[f]{\textbf{BOUNDARY}: Start new bucket}){$D(x,n)>\tau(x)$ \And $\mathrm{LM}(n)=-1$}{
            $\ell\leftarrow\ell+1,\ \bar{D}(n)\leftarrow0,\ \mathrm{LM}(n)\leftarrow\ell$\;
            $H.\text{push}((0,n))$\;
        }
        \Else(\tcp*[f]{\textbf{MERGE}: Propagate current label}){
            $\bar D^{\text{temp}}\leftarrow\bar D^{\text{curr}}+D(x,n)$\;
        
            \If{$\bar D^{\text{temp}}<\bar{D}(n)$}{
                $\bar{D}(n)\leftarrow\bar D^{\text{temp}},\
                \mathrm{LM}(n)\leftarrow\mathrm{LM}(x)$\;
                $H.\text{push}((\bar D^{\text{temp}},n))$\;
            }
        }
    }
}

Group cells by $\mathrm{LM}$ to form buckets $\{B_k\}$\;
\Return{$\mathrm{LM},\{B_k\}$}\;
\end{algorithm}

\begin{figure}[t]
    \centering
    \includegraphics[width=\linewidth]{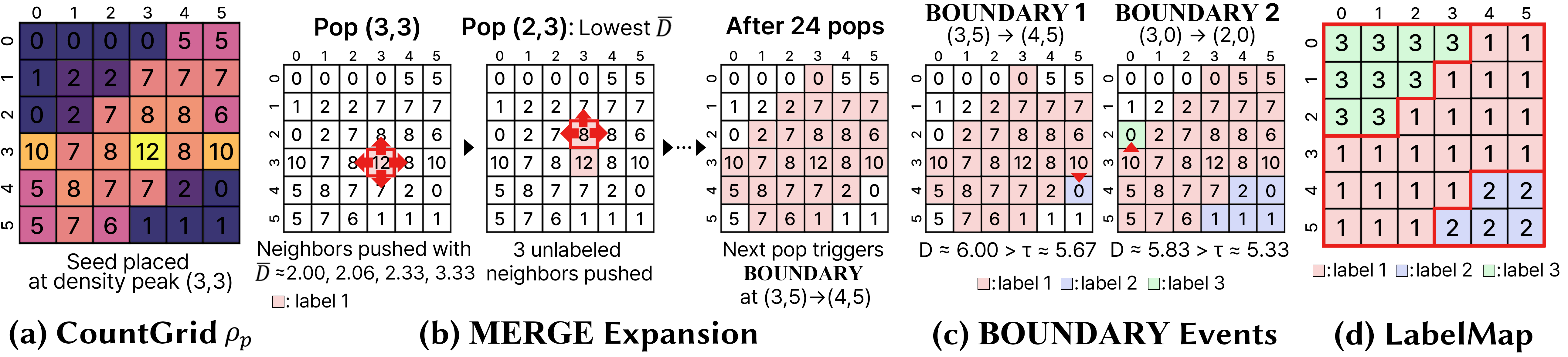}
    \caption{Density-adaptive propagation (Algorithm~\ref{alg:density-adaptive_bucket_formation}) on a $6 \times 6$ CountGrid with $\alpha=0.5$ and $\beta=1.0$.
    (a) A seed is initialized at the density peak.
    (b) MERGE expands to neighboring cells in order of increasing accumulated contrast $\bar{D}$, covering homogeneous high-density regions.
    (c) When the local contrast exceeds the adaptive threshold ($D > \tau$), BOUNDARY initiates a new bucket at an unlabeled neighboring cell.
    (d) The resulting LabelMap forms rectilinear buckets aligned with the underlying density structure.}
    \Description{Four panels illustrate the progression of density-adaptive propagation on a six-by-six CountGrid. The panels show initialization at a density peak, expansion into neighboring cells through MERGE operations, boundary formation at density transitions, and the final LabelMap consisting of distinct rectilinear buckets.}
    \label{fig:trahist-algorithm}
\end{figure}

\noindentparagraph{\textnormal{\textbf{Density-adaptive Bucket Formation.}}}
Given the density field $\rho_p(x)$, TraHist forms histogram buckets by aggregating neighboring grid cells with similar density, while explicitly detecting boundaries induced by sharp local density transitions. This process produces rectilinear partitions that adapt to local density variation while remaining compatible with grid-based estimation, as illustrated by the density-adaptive LabelMap in \cref{fig:overall_process}c and the resulting rectilinear buckets in~\cref{fig:overall_process}d. In contrast, \cref{fig:partitioning} shows conventional axis-aligned histograms, which fail to follow trajectory-induced density structures. To realize these density-adaptive partitions in practice, bucket formation proceeds by expanding from seed cells using a priority queue ordered by cumulative density contrast, causing growth to follow directions of minimal density variation. The complete density-adaptive bucket formation procedure is summarized in~\cref{alg:density-adaptive_bucket_formation}.
\par At each propagation step, when the algorithm considers a neighboring cell $n$ of a currently labeled cell $x$, it first evaluates whether a significant density transition should induce a bucket boundary. \cref{fig:trahist-algorithm} visualizes this process on a $6\times6$ CountGrid, where propagation from a seed at the density peak produces rectilinear buckets through a sequence of \textbf{MERGE} and \textbf{BOUNDARY} decisions. Accordingly, the algorithm makes one of the following decisions:
\begin{itemize}
    \item \textbf{MERGE}: By default, assign $n$ to the same bucket as $x$ and continue propagation;
    \item \textbf{BOUNDARY}: When a significant density transition is detected, initiate a new bucket at $n$.
\end{itemize}
These decisions are governed by a contrast-based cost defined on adjacent grid cells. The local density contrast between neighboring cells $x$ and $n$ is measured by
\begin{equation}\label{eq:local_cost}
    D(x, n) = \alpha|\rho_p(x)-\rho_p(n)| 
             + (1-\alpha)|\mu(x)-\mu(n)|,
\end{equation}
where $\mu(x)$ denotes the local mean of $\rho_p$ in a neighborhood around $x$, and $\alpha \in [0,1]$ controls the balance between sensitivity to sharp local transitions and robustness to noise. The first term captures direct density contrast between adjacent cells, while the second term anchors this contrast to the local density baseline, suppressing fluctuations caused by sparse sampling.

\par To determine whether propagation should continue across a cell boundary, TraHist compares $D(x, n)$ with an adaptive threshold
\begin{equation}\label{eq:threshold}
\tau(x) = \beta \cdot \mu(x),
\end{equation}
where $\beta$ controls the granularity of partitioning. This approach effectively evaluates the relative contrast $D(x,n)/\mu(x)$, providing a scale-consistent criterion that remains stable under strong density variation. In practice, values of $\beta$ near $1.0$ yield stable partitions, while slightly larger values can suppress minor contrasts in noisy or highly imbalanced regions. For an unlabeled neighboring cell $n$, if $D(x,n) > \tau(x)$, \textbf{BOUNDARY} is triggered and $n$ initiates a new bucket; otherwise, $n$ is considered density-compatible with $x$ and \textbf{MERGE} is performed.
\par To enforce coherent bucket shapes, TraHist maintains a cumulative propagation cost $\bar D$, which tracks the minimum accumulated contrast required to reach each cell from its bucket seed. This cost is updated as
\begin{equation} \label{eq:acc_cost}
    \textstyle \bar D(x) = \min_{n \in \mathcal{N}(x)} \left( \bar D(n) + D(x, n) \right),
\end{equation}
where $\mathcal{N}(x)$ denotes the neighbor cells of $x$. Propagation is processed in ascending order of $\bar D$ using the priority queue, causing propagation to follow paths of least aggregate contrast. Consequently, bucket boundaries tend to align with persistent density transitions rather than isolated local fluctuations. By favoring propagation along directions of minimal density contrast and suppressing expansion across sharp transitions, this contrast-driven process provides a discrete approximation to the geometric objective in~\cref{subsec:DAP_principle}. Its role as a surrogate for minimizing intra-bucket variance is further analyzed in~\cref{subsec:theory}.
\par Once propagation converges, the resulting LabelMap $\mathrm{LM}$ induces a partition $\{B_1, ..., B_{\vert B\vert}\}$ of the grid. For each bucket $B_k$, TraHist additionally maintains simple bucket-level statistics, including the average number of trajectory points per cell $S_p^{(k)}$ and that of distinct trajectories per cell $S_T^{(k)}$, which are used during cardinality estimation. These buckets form rectilinear polygons whose geometry reflects local density variation, and they are subsequently used as the histogram synopsis for cardinality estimation.

\subsection{Cardinality Estimation}\label{subsec:estimation}
\par Given the LabelMap $\mathrm{LM}$ and density-adaptive buckets $\{B_1, \dots, B_{|B|}\}$, TraHist estimates the trajectory-level cardinality $|RQ(Q)|$ directly from compact per-bucket statistics. Through grid-aligned sampling, TraHist maps $Q$ to the CountGrid and aggregates the statistics of the selected cells, avoiding segment--cell intersections and geometric overlap computation over irregular bucket polygons. This design retains the benefits of density-coherent partitioning while keeping query time linear in the number of selected cells. Estimation proceeds in two stages: a point-level estimate followed by a trajectory-to-point correction for object-level dependency.

\noindentparagraph{\textnormal{\textbf{Point-level estimation.}}}
Given a query region $Q$, TraHist samples $Q$ at the resolution of the CountGrid. Let $G_Q\subseteq \mathcal{G}$ denote the resulting set of selected grid cells. Each cell $x\in G_Q$ is assigned to a bucket using the LabelMap $\mathrm{LM}$. Let $N_k(Q) = \left| \{ x \in G_Q \mid \mathrm{LM}(x) = k \} \right|$ denote the number of selected grid cells assigned to bucket $B_k$. The point-level cardinality estimate $\hat{N}_p(Q)$ is computed as
\begin{equation} 
    \hat{N}_p(Q) = \sum_{k=1}^{|B|} N_k(Q)S_p^{(k)}.
\end{equation}

\noindentparagraph{\textnormal{\textbf{Trajectory-level estimation.}}}
To estimate the number of distinct trajectories intersecting $Q$, TraHist applies a ratio-based correction to account for object-level dependency among sampled points. For each bucket $B_k$, a trajectory-to-point ratio is defined as
\begin{equation}
    \textstyle \gamma^{(k)} = \frac{S_T^{(k)}}{S_p^{(k)} + \epsilon},
\end{equation}
where $\epsilon$ is a small constant added for numerical stability. A query-specific weighted average ratio is then computed as
\begin{equation}
    \textstyle \bar{\gamma}(Q) = \frac{\sum_{k=1}^{|B|} N_k(Q) \gamma^{(k)}}{\sum_{k=1}^{|B|} N_k(Q)},
\end{equation}
and the final trajectory-level estimate is obtained by
\begin{equation}
\hat{N}_T(Q) = \hat{N}_p(Q) \cdot \bar{\gamma}(Q).
\end{equation}

\par Beyond query-time estimation, we also consider synopsis updates under trajectory insertions and deletions. In~\cref{appendix:update}, we describe exact maintenance of per-bucket statistics under a fixed partition and examine TraHist’s accuracy under a controlled data shift.

%%%%%%%%%%%%%%%%%%%%%%%%%%%%%%%%%%%%%%%%%%%%%%%%%%%%%%%%%%%%%%%%%

\section{Analysis} \label{sec:analysis}
This section analyzes TraHist from both computational and theoretical perspectives. We first examine its time and space complexity, and then discuss theoretical properties that explain how density-adaptive propagation leads to coherent bucket formation.

\subsection{Computational Complexity} \label{subsec:complexity}
Let $N_{\text{pts}}$ denote the number of trajectory points, $N$ the number of grid cells in the CountGrid, and $N_q$ the number of grid cells covered by a query region.

\noindentparagraph{\textnormal{\textbf{Construction cost.}}}
CountGrid construction scans all trajectory points once to compute point and trajectory counts per cell, which takes $O(N_{\text{pts}})$ time and $O(N)$ space. Density-adaptive bucket formation then processes each grid cell while maintaining a priority queue, resulting in $O(N \log N)$ time and $O(N)$ space. Since $N_{\text{pts}}$ is fixed for a given dataset, the overall construction cost is effectively governed by $N$, which is determined by the chosen grid resolution rather than by the number of trajectories or points.

\noindentparagraph{\textnormal{\textbf{Query cost.}}}
For a query covering $N_q$ grid cells, TraHist aggregates per-cell labels and bucket statistics in $O(N_q)$ time, as the final weighted aggregation over buckets is bounded by $\min(|B|, N_q)$. Thus, per-query estimation scales linearly with the number of sampled grid cells, independent of bucket shape complexity.

\noindentparagraph{\textnormal{\textbf{Space usage.}}}
After construction, TraHist stores only the LabelMap and per-bucket statistics, requiring $O(N)$ space determined by the grid resolution. While this is higher than the space cost of traditional histograms that store only per-bucket summaries, it remains significantly smaller than the model footprints of learned estimators.

%%%

\subsection{Theoretical Properties} \label{subsec:theory}

We now examine the structural properties induced by the proposed density-adaptive propagation. We first define the algorithmic cost implicitly minimized by the propagation process and establish local optimality. We then relate the local contrast underlying this cost to density coherence within buckets, quantified by the intra-bucket variance in~\cref{eq:intra-bucket_variance}. We further derive a bound on the difference between the local contrast and the corresponding bucket-mean deviation. Full derivations are provided in~\cref{appendix:proofs}.

\noindentparagraph{\textnormal{\textbf{Algorithmic structure and local optimality of density-adaptive propagation.}}} The propagation-based partitioning process, guided by the cumulative cost $\bar{D}(x)$ in~\cref{eq:acc_cost}, can be interpreted as implicitly minimizing a global algorithmic cost. For a final partition $\Pi=\{B_1, \dots, B_{|B|}\}$ of the domain $\Omega$, we define the accumulated contrast objective
\begin{equation}
E(\Pi) = \sum_{x\in\mathcal{G}}\bar{D}(x),
\end{equation}
which represents the total accumulated contrast required to construct the partition. Since each cell is assigned to the region that reaches it with minimum cumulative cost, reassigning a cell to a neighboring region cannot reduce $E(\Pi)$, establishing local optimality.

\noindentparagraph{\textnormal{\textbf{Connection to the statistical objective.}}}
The ideal discrete statistical objective for histogram-based estimation is the intra-bucket sum of squared errors
\begin{equation}\label{eq:sse}
J(\Pi) = \sum_k \sum_{x\in B_k}\bigl(\rho_p(x)-\bar{\rho}_{p,k}\bigr)^2,
\end{equation}
where $\bar{\rho}_{p,k} = \frac{1}{|B_k|}\sum_{x\in B_k}\rho_p(x)$ is the mean density within bucket $B_k$. This objective serves as a statistical surrogate for cardinality estimation error under a bucket-wise mean model. When a new cell $x$ is added to $B_k$, the increase in $J$ is proportional to $(\rho_p(x)-\bar{\rho}_{p,k})^2$, so minimizing this increment for a given bucket amounts to minimizing $\bigl|\rho_p(x)-\bar{\rho}_{p,k}\bigr|$. However, using this quantity to guide propagation is computationally costly, as evolving bucket means require updating the costs of pending candidate cells. TraHist instead uses the locally computable contrast $D(x, n)$ (\cref{eq:local_cost}) as a tractable surrogate.

\noindentparagraph{\textnormal{\textbf{Bounded deviation from the ideal objective.}}}
For a candidate cell $x$ and an adjacent cell $n\in B_k$, the deviation between $D(x,n)$ and the ideal local objective $\bigl|\rho_p(x)-\bar{\rho}_{p,k}\bigr|$ is provably bounded:
\begin{equation}\label{eq:deviation}
\bigl|D(x,n) - \bigl|\rho_p(x)-\bar{\rho}_{p,k}\bigr|\bigr|
\le (1-\alpha)(\delta_x+\delta_n) + \epsilon_k,
\end{equation}
where $\delta_x = \max_{y\in \mathcal{N}(x)} |\rho_p(y) - \mu(x)|$ captures the maximum deviation from the local mean around cell $x$, and $\epsilon_k = \max_{y\in B_k} |\rho_p(y) - \bar{\rho}_{p,k}|$ measures the maximum deviation from the bucket mean within $B_k$. The bound shows that the discrepancy between the practical contrast cost and the ideal local objective is controlled by two factors: local density smoothness (via $\delta_x$ and $\delta_n$) and intra-bucket homogeneity (via $\epsilon_k$). The first term bounds the discrepancy introduced by local averaging, while the second bounds the error from using the neighboring cell's density as a proxy for the bucket mean. For $0<\alpha\le 1$, $\epsilon_k$ admits a formal bound in terms of the cumulative propagation cost. Together, these bounds support the use of $E(\Pi)$ as a principled surrogate for minimizing $J(\Pi)$, aligning the algorithmic objective with the statistical objective.
%

%%%%%%%%%%%%%%%%%%%%%%%%%%%%%%%%%%%%%%%%%%%%%%%%%%%%%%%%%%%%%%%%%
\section{Experimental Study} \label{sec:experimental_study}
This section presents an extensive experimental evaluation of TraHist, focusing on its \textit{accuracy}, \textit{robustness}, and \textit{efficiency} for the cardinality estimation of trajectory range queries. Our evaluation is designed to address four research questions:
\begin{itemize}
    \item \textbf{RQ1.} Does TraHist provide more accurate estimates under realistic workloads?
    \item \textbf{RQ2.} What structural properties explain its accuracy gains?
    \item \textbf{RQ3.} Does TraHist achieve these benefits with practical construction and query costs?
    \item \textbf{RQ4.} How effective are spatial-object adaptations for trajectory cardinality estimation?
\end{itemize}
All experiments were conducted on a single server running Ubuntu 24.04.3 LTS, equipped with an Intel(R) Core(TM) i9-9900X CPU and 128\,GB of RAM; all estimators were implemented in Python.

\begin{table}[t]
    \centering
    \caption{Statistics of the real-world trajectory datasets.}
    \label{tab:dataset_stats}
    {%
    \renewcommand{\arraystretch}{1}
    \setlength{\aboverulesep}{0pt}
    \setlength{\belowrulesep}{0pt}
    \setlength{\arrayrulewidth}{0.8pt}
    \setlength{\tabcolsep}{4pt}
    \newcommand{\topgap}{\rule{0pt}{2.7ex}}
    \newcommand{\botgap}{\rule[-1.5ex]{0pt}{0pt}}
    \ifdefined\dstra\else\newlength{\dstra}\fi
    \settowidth{\dstra}{0,000,000}
    \newcommand{\dsa}[1]{\makebox[\dstra][r]{#1}}
    \ifdefined\dstrb\else\newlength{\dstrb}\fi
    \settowidth{\dstrb}{00,000,000}
    \newcommand{\dsb}[1]{\makebox[\dstrb][r]{#1}}
    \ifdefined\dstrc\else\newlength{\dstrc}\fi
    \settowidth{\dstrc}{000}
    \newcommand{\dsc}[1]{\makebox[\dstrc][r]{#1}}
    \ifdefined\dstrd\else\newlength{\dstrd}\fi
    \settowidth{\dstrd}{00,000}
    \newcommand{\dsd}[1]{\makebox[\dstrd][r]{#1}}
    \resizebox{0.54\textwidth}{!}{%
    \begin{tabular}{@{}l@{\hspace{6pt}}|@{\hspace{6pt}}cc@{\hspace{14pt}}ccc@{\hspace{6pt}}}
        \toprule[0.8pt]
        \multirow{2}{*}{\textbf{Dataset}}
        & \multirow{2}{*}{\textbf{\# Traj.}}
        & \multirow{2}{*}{\textbf{\# Points}}
        & \multicolumn{3}{c}{\topgap\textbf{Trajectory Length $\lvert T\rvert$}} \\
        \cmidrule(r{6pt}){4-6}
        & & & \topgap\botgap\textbf{Min} & \textbf{Median} & \textbf{Max} \\
        \midrule[0.8pt]
        \topgap\textit{Porto}   & \dsa{1,672,257} & \dsb{80,414,709} & 2 & \dsc{41}  & \dsd{3,836} \\
        \textit{T-Drive}        & \dsa{127,159}   & \dsb{17,502,090} & 2 & \dsc{36}  & \dsd{98,607} \\
        \botgap\textit{GeoLife} & \dsa{17,335}    & \dsb{20,003,131} & 2 & \dsc{487} & \dsd{56,780} \\
        \bottomrule[0.8pt]
    \end{tabular}%
    }
    }
\end{table}

\subsection{Experimental Setup}\label{subsec:experimental-setup}

\noindentparagraph{\textnormal{\textbf{Datasets.}}}
We use three real-world trajectory datasets that are widely adopted in prior studies on trajectory data management: \textit{Porto}~\cite{porto2015ecml,wang2019clustering,yao2022trajgat}, \textit{T-Drive}~\cite{yuan2010tdrive,yuan2011kdd,wang2019clustering}, and \textit{GeoLife}~\cite{geolife1, geolife2, geolife3}. We apply standard preprocessing to all datasets. Specifically, trajectories consisting of fewer than two points were removed. Since \textit{T-Drive} does not explicitly provide trajectory identifiers, individual trajectories were reconstructed by splitting point sequences whenever the temporal gap exceeded 30 minutes. All geographic coordinates were projected into a two-dimensional planar domain using the Universal Transverse Mercator projection, converting them to a uniform, meter-based coordinate system. \Cref{tab:dataset_stats} summarizes the key statistics of the resulting datasets.
\par \textit{Porto} stresses scalability and construction efficiency due to its scale, while \textit{T-Drive} emphasizes trajectory continuity challenges due to its heavy-tailed trajectory length distribution. These datasets form the basis of our primary evaluation. \textit{GeoLife}, in contrast, exhibits distinct sampling characteristics, with dense point distributions from high-frequency and heterogeneous sampling. It also reflects more diverse and irregular human mobility patterns than vehicular trajectories in \textit{Porto} and \textit{T-Drive}. Given these properties, we use \textit{GeoLife} as a complementary dataset to assess generalizability across trajectory characteristics, and present its main results separately in~\cref{appendix:geolife}.

\noindentparagraph{\textnormal{\textbf{Compared Methods.}}}
We evaluate TraHist against five representative histogram-based estimators spanning classical partitioning, learned intra-bucket modeling, and trajectory-specific connection-aware aggregation. We separately evaluate two spatial-object adaptations in~\cref{subsec:spatial-object}.

\begin{itemize}
    \item \textbf{MEH}~\cite{muralikrishna1988meh}: A canonical baseline that partitions the spatial domain into axis-aligned rectangular buckets with an equal number of records.
    \item \textbf{MHist}~\cite{poosala1997mhist}: An iterative partitioning histogram guided by a splitting criterion that recursively divides regions where data values change the most. We adapt it to spatial data using the same CountGrid and grid cell size $\Delta$ as TraHist to ensure identical spatial granularity.
    \item \textbf{LHist}~\cite{liu2021lhist}: A hierarchical histogram that models intra-bucket distributions using local polynomial regression while retaining axis-aligned partitions.
    \item \textbf{DEH}~\cite{deh} and \textbf{CASE+}~\cite{fanaeepour2015case}: Connection-aware, Euler histogram methods for counting moving objects. They improve distinct count estimation through auxiliary connectivity statistics and Euler-style aggregation over a fixed partition, without adapting partition boundaries. DEH estimates distinct connected entries via segment-wise aggregation, whereas CASE+ jointly processes all segments of each trajectory, conditionally adding virtual counts to preserve object-level connectivity. We evaluate CASE+ without its optional differential privacy.
    \item \textbf{Spatial-object adaptations}: We evaluate two estimators using the same per-trajectory minimum bounding rectangle (MBR) representation. CDHistogram-MBR~\cite{cdhistogram} applies a cumulative density histogram to the resulting MBR relation, while DeepDB-MBR~\cite{hilprecht2020deepdb} is a learned non-histogram estimator that models the same relation using a Sum-Product Network (SPN).
\end{itemize}

\begin{figure}[t]
    \centering
    \begin{tabular}{@{}c@{\hspace{3pt}}c@{\hspace{4pt}}c@{\hspace{4pt}}c@{}}
        & \textbf{Low} & \textbf{Medium} & \textbf{High} \\
        \raisebox{-\dimexpr\fboxsep+\fboxrule\relax}{\rotatebox{90}{\makebox[\dimexpr0.126\linewidth+2\fboxsep+2\fboxrule\relax][c]{\textbf{(a)} \textit{Porto}}}}
        & \fbox{\includegraphics[width=0.18\linewidth, height=0.126\linewidth]{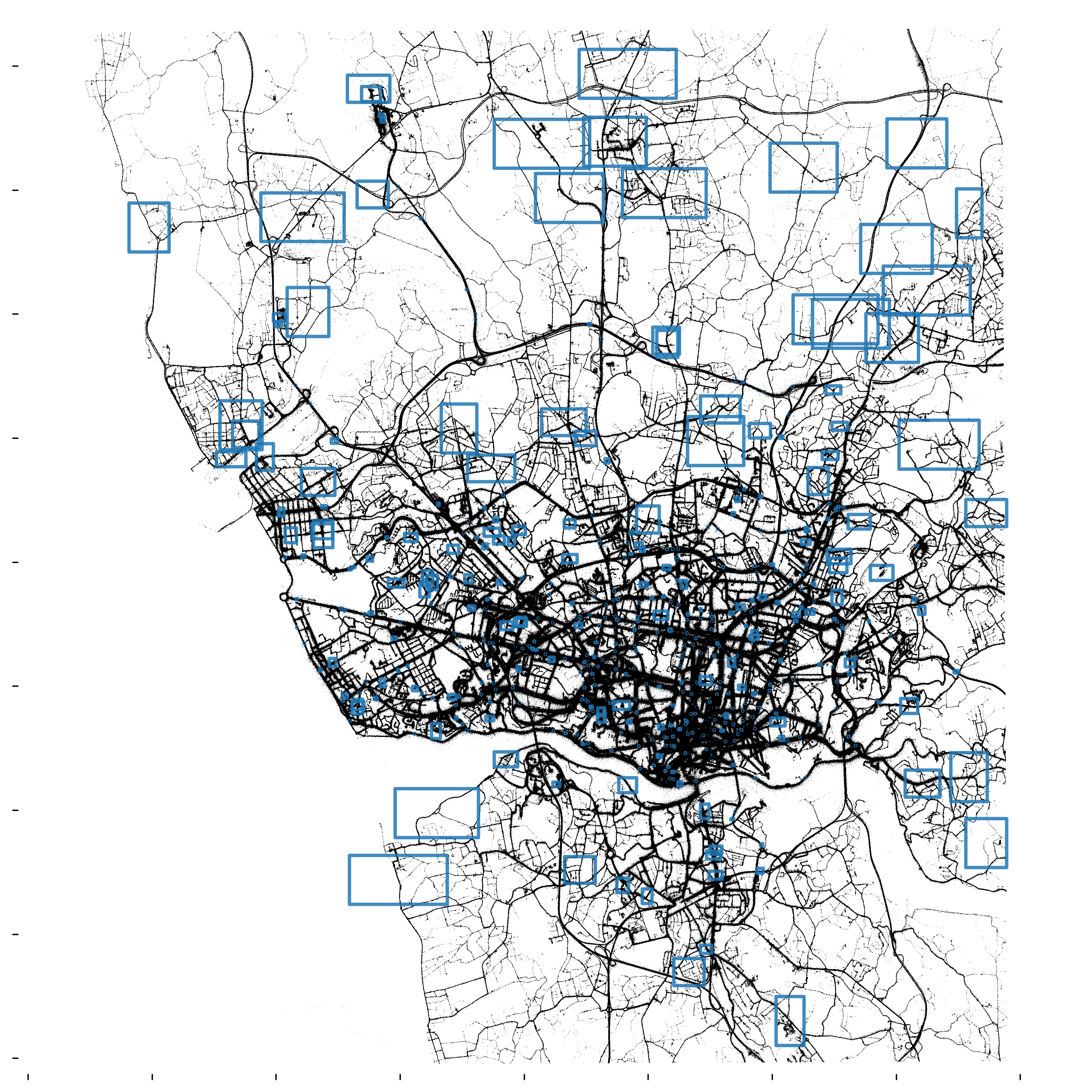}}
        & \fbox{\includegraphics[width=0.18\linewidth, height=0.126\linewidth]{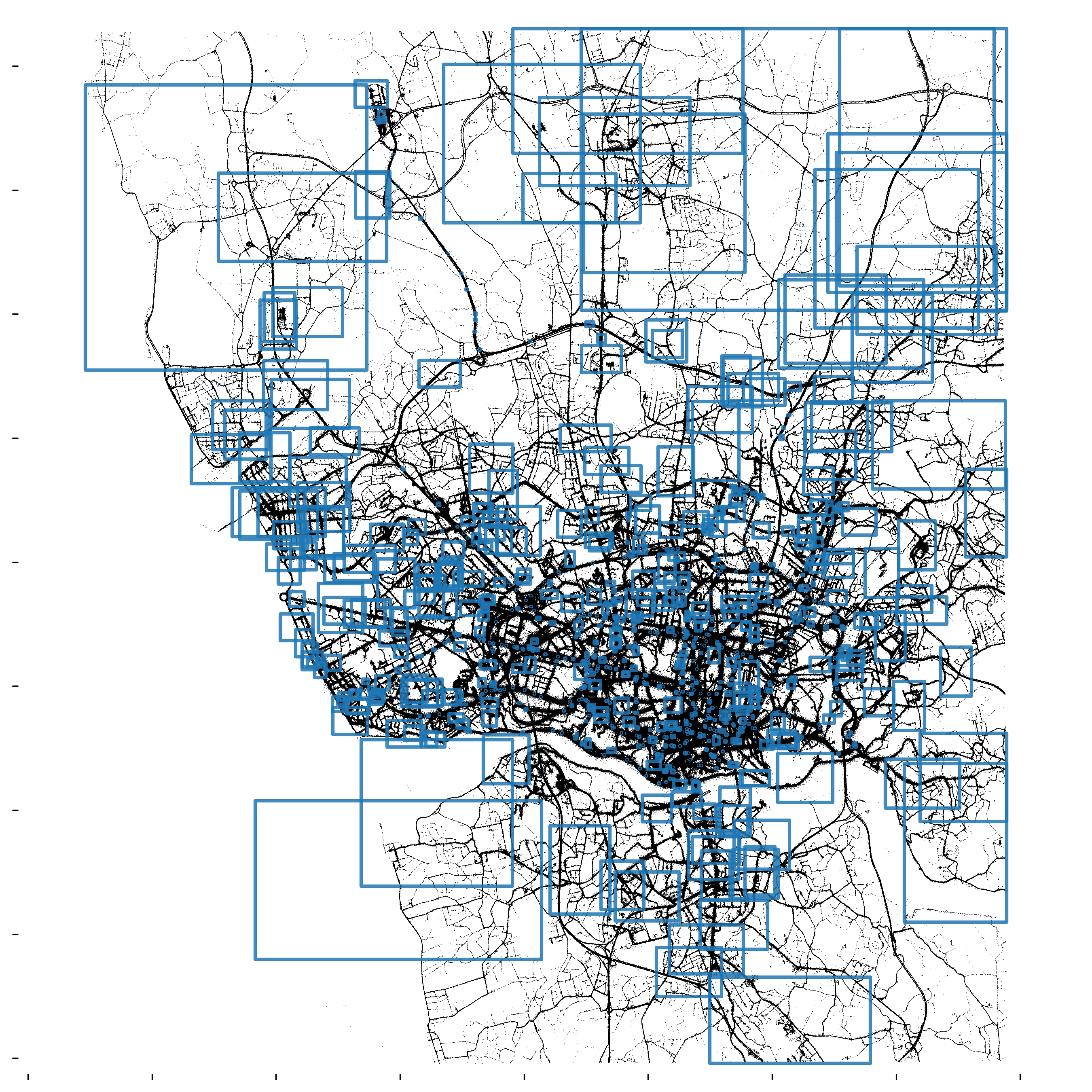}}
        & \fbox{\includegraphics[width=0.18\linewidth, height=0.126\linewidth]{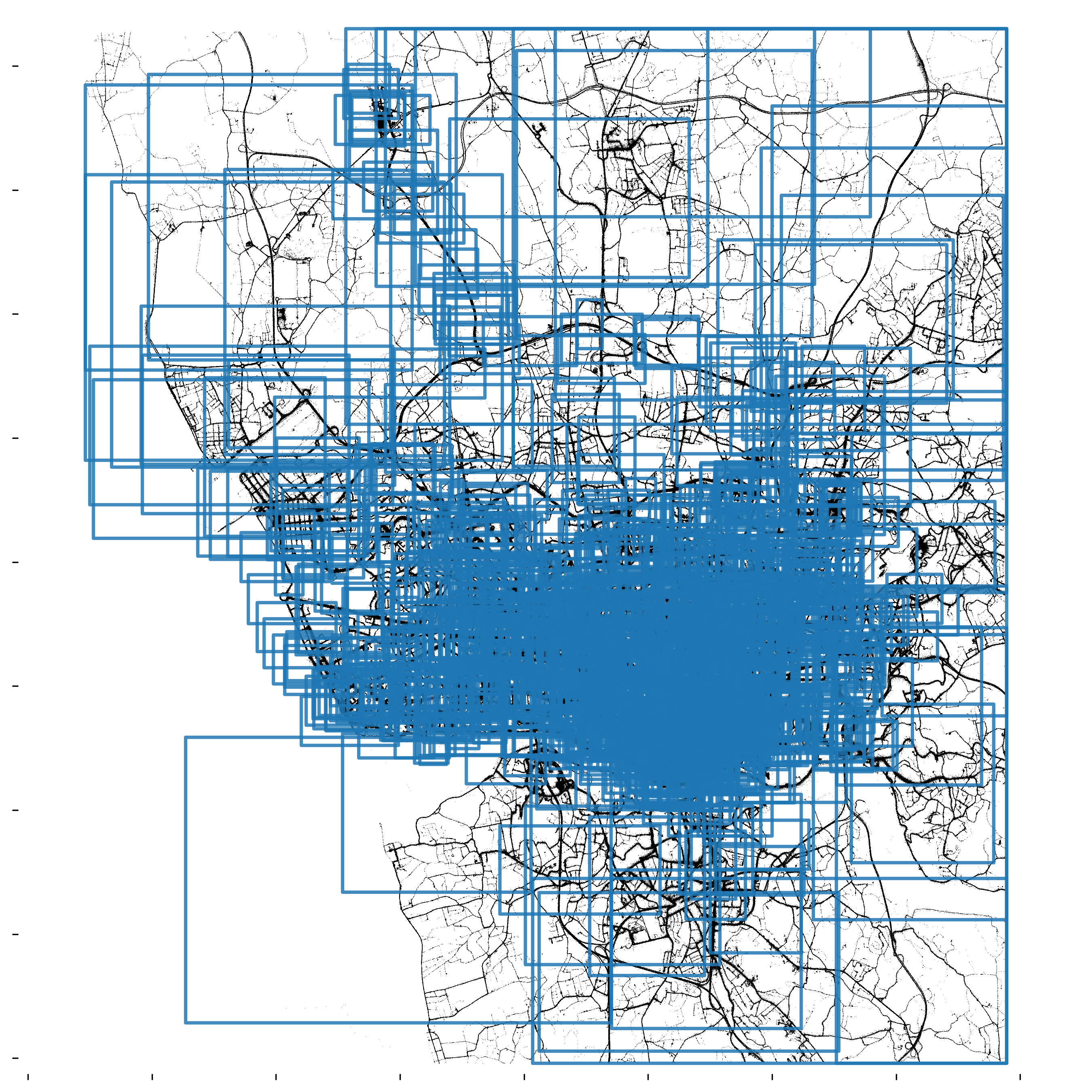}} \\[4pt]
        \raisebox{-\dimexpr\fboxsep+\fboxrule\relax}{\rotatebox{90}{\makebox[\dimexpr0.126\linewidth+2\fboxsep+2\fboxrule\relax][c]{\textbf{(b)} \textit{T-Drive}}}}
        & \fbox{\includegraphics[width=0.18\linewidth, height=0.126\linewidth]{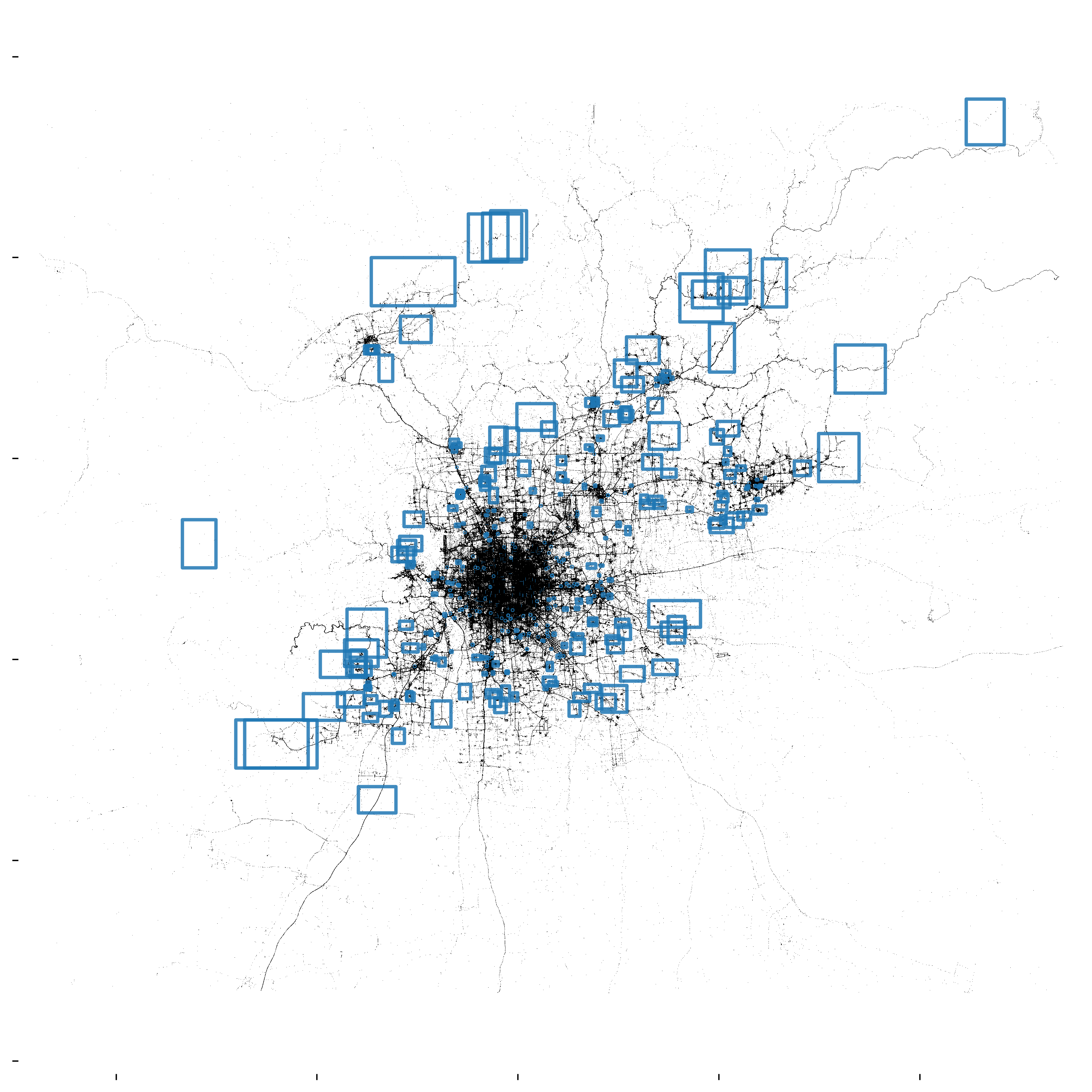}}
        & \fbox{\includegraphics[width=0.18\linewidth, height=0.126\linewidth]{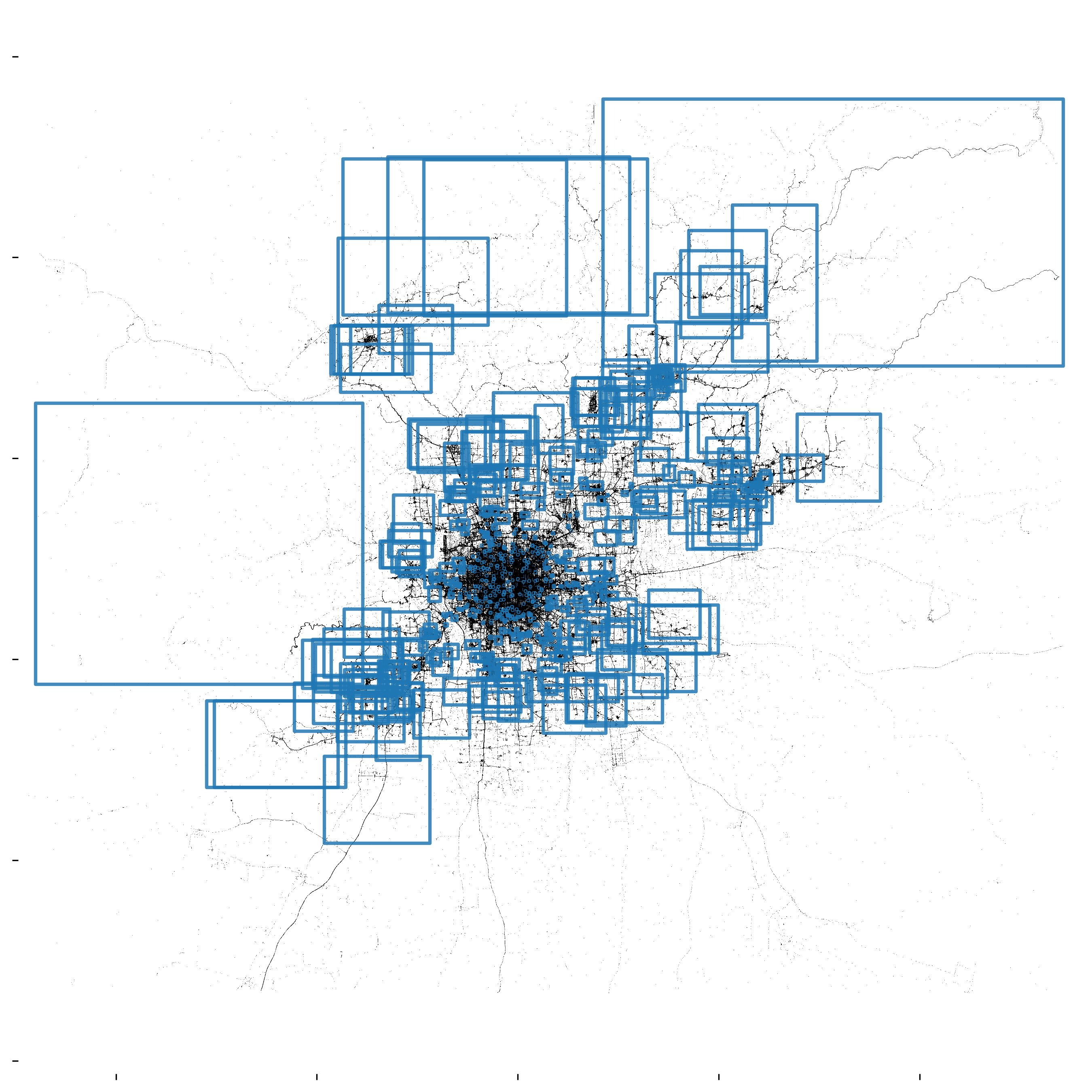}}
        & \fbox{\includegraphics[width=0.18\linewidth, height=0.126\linewidth]{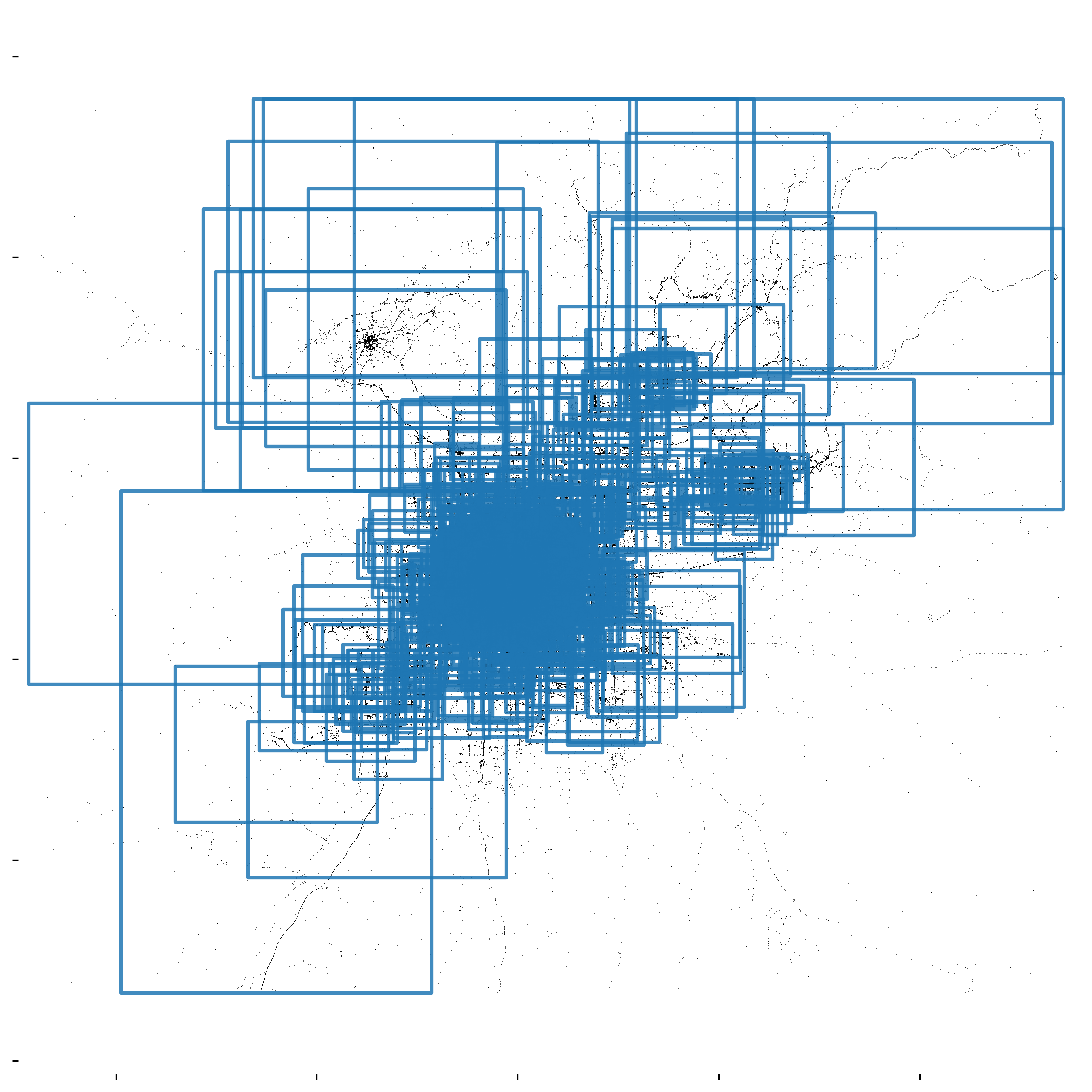}} \\[4pt]
        \raisebox{-\dimexpr\fboxsep+\fboxrule\relax}{\rotatebox{90}{\makebox[\dimexpr0.126\linewidth+2\fboxsep+2\fboxrule\relax][c]{\textbf{(c)} \textit{GeoLife}}}}
        & \fbox{\includegraphics[width=0.18\linewidth, height=0.126\linewidth]{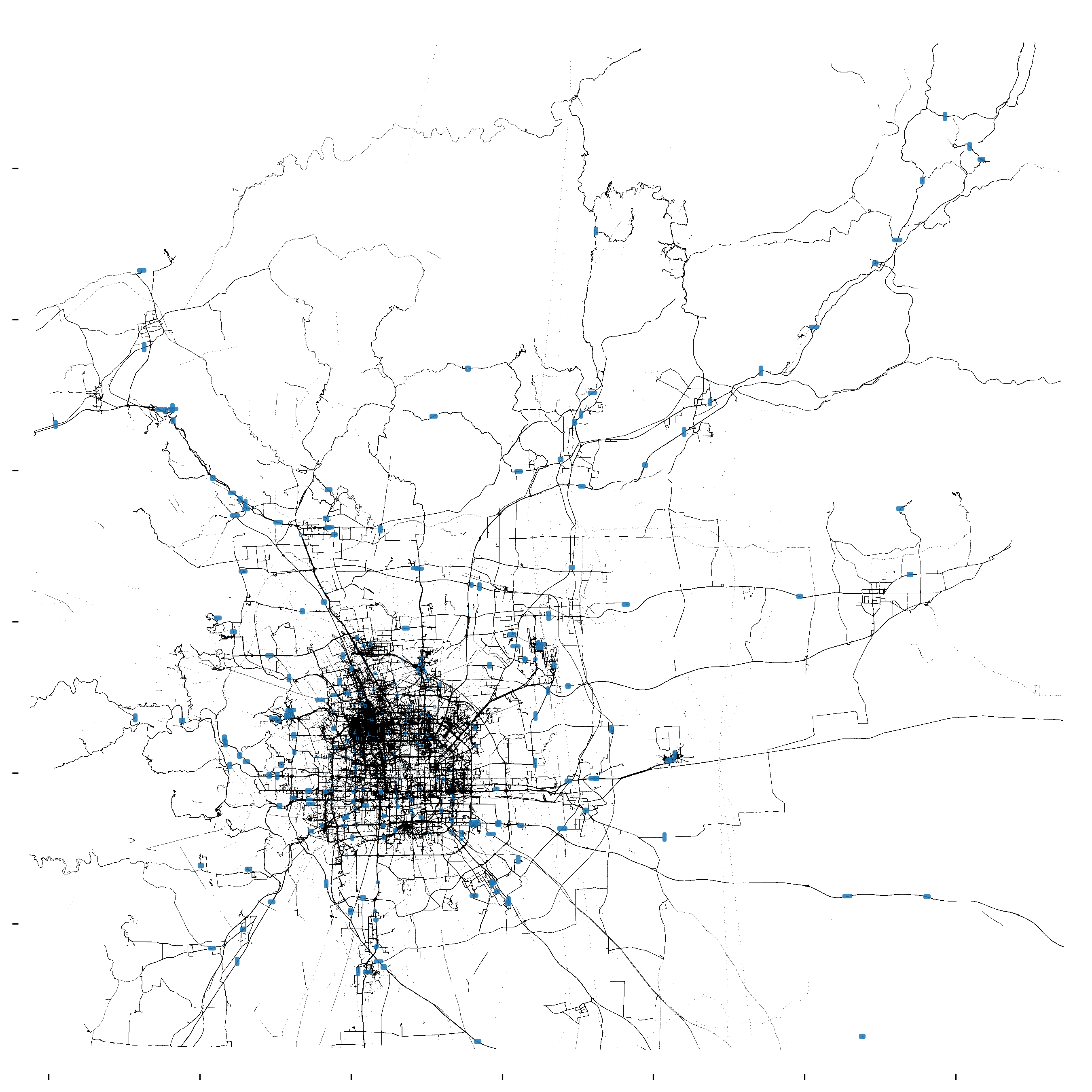}}
        & \fbox{\includegraphics[width=0.18\linewidth, height=0.126\linewidth]{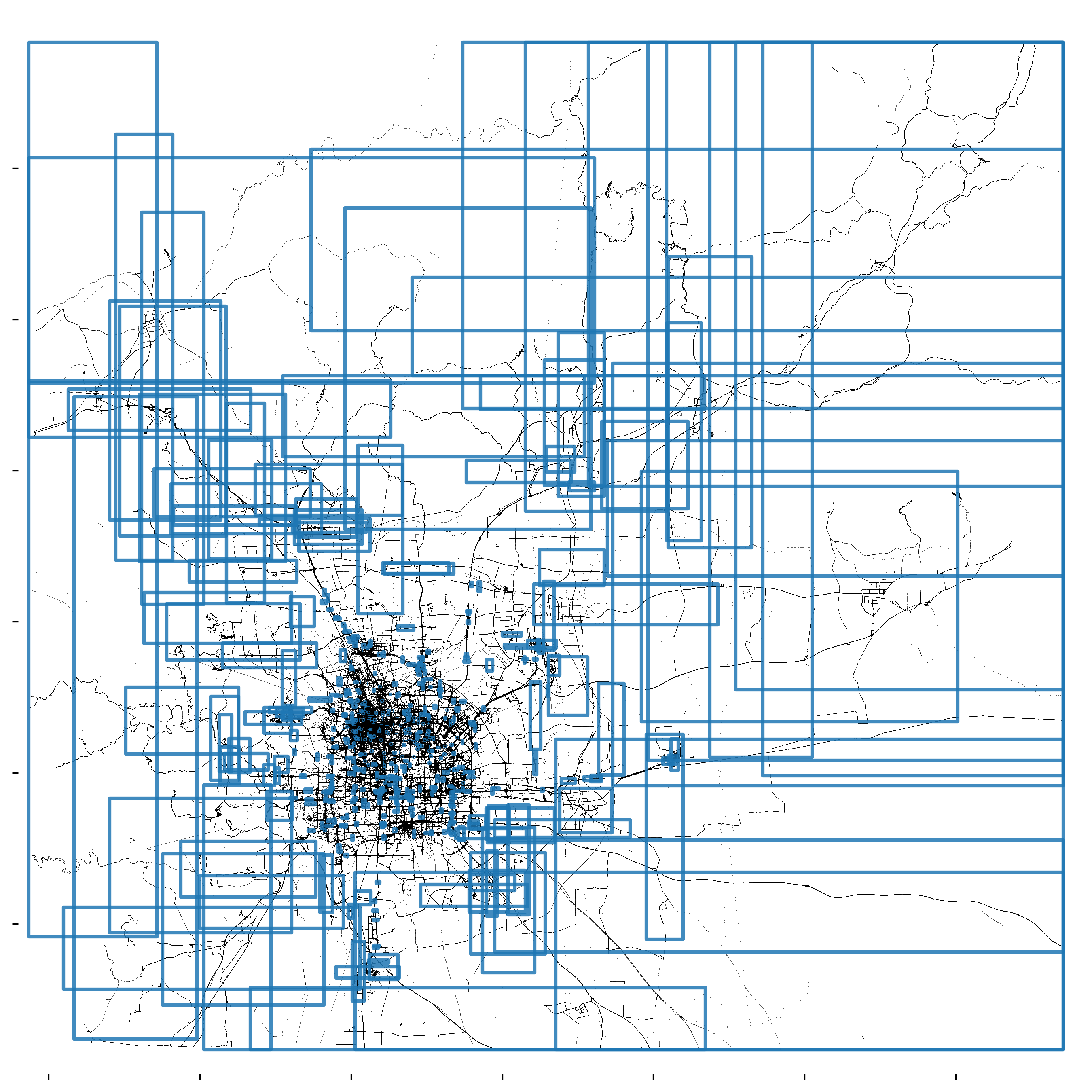}}
        & \fbox{\includegraphics[width=0.18\linewidth, height=0.126\linewidth]{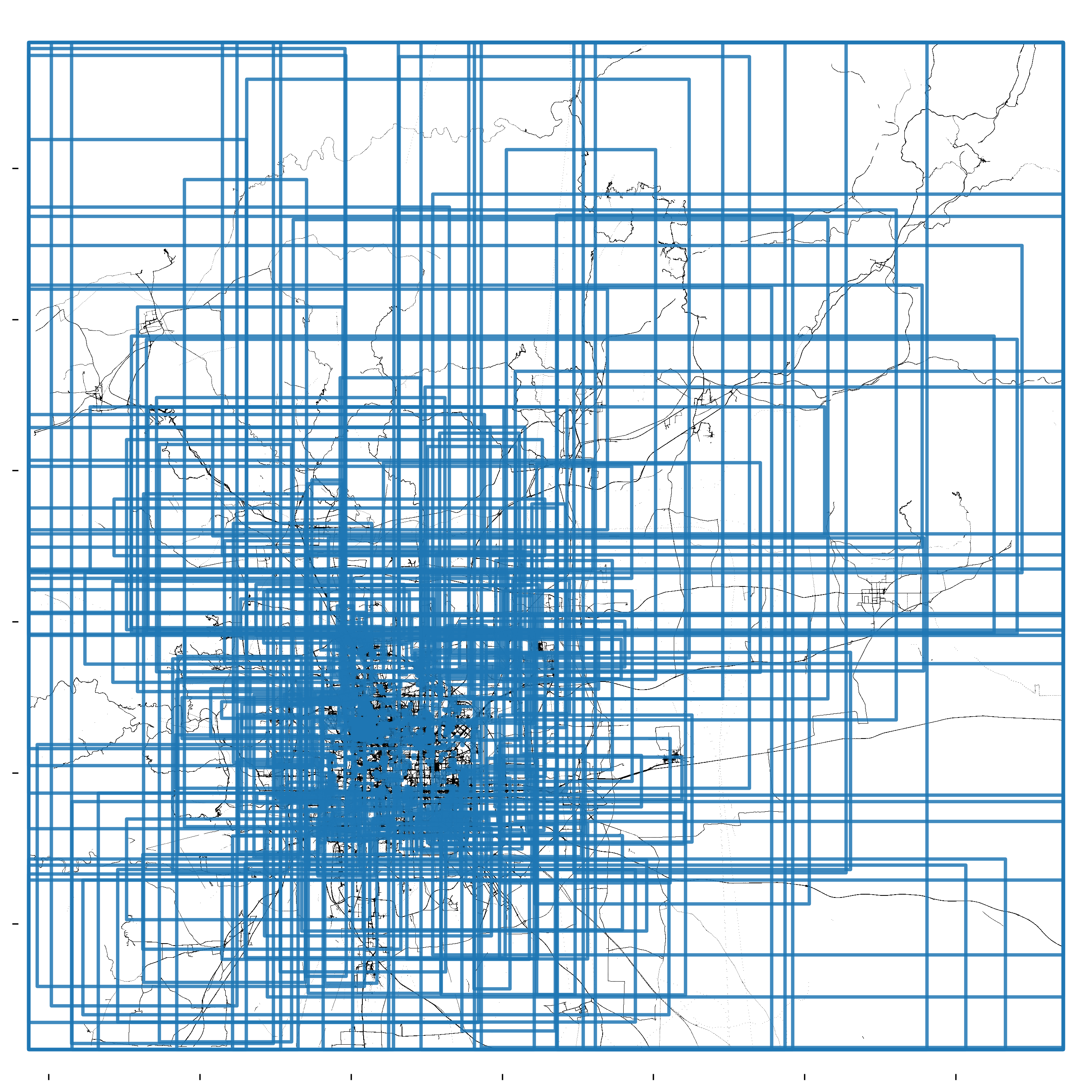}} \\
    \end{tabular}
    \caption{Visualization of trajectory range queries. 1,000 query regions (blue) overlaid on (a)~\textit{Porto}, (b)~\textit{T-Drive}, and (c)~\textit{GeoLife}. Columns show low, medium, and high selectivity regimes based on relative query area.}
    \Description{A three-by-three matrix visualizing trajectory range query workloads. The rows correspond to Porto, T-Drive, and GeoLife, while the columns represent low, medium, and high relative query areas. Each panel overlays blue rectangular query regions on the spatial distribution of trajectory points.}
    \label{fig:query_workloads}
\end{figure}
\noindentparagraph{\textnormal{\textbf{Query Workloads.}}}
We evaluate all estimators using trajectory range queries, which count the number of distinct trajectories intersecting a rectangular query region, as defined in Definition \ref{dfn:trajectory-range-query}. For each dataset, we generate a workload of trajectory range queries designed to reflect realistic, data-driven access patterns commonly observed in spatial query workloads~\cite{pagel1993towards,shekelyan2017digithist}. Specifically, we sample each query center from the empirical distribution of trajectory points, so regions containing more observed samples receive proportionally higher query-center probability. \cref{fig:query_workloads} illustrates the spatial distributions of query regions generated under this workload for each dataset, further stratified by selectivity. Queries are categorized into three selectivity groups---low (0–0.5\%), medium (0.5–2\%), and high (2–100\%)---with each group consisting of 1,000 queries~\cite{yang2019naru, rashedi2025accuracy}. In general, estimation becomes more challenging as query selectivity decreases, as low selectivity queries require capturing fine-grained spatial structure~\cite{yang2019naru} and are prone to large estimation errors that degrade query plan quality~\cite{leis2015good}.

\noindentparagraph{\textnormal{\textbf{Comparison Strategy.}}}
We evaluate the histogram-based methods under two complementary evaluation settings. First, we compare methods under \textit{equivalent memory budgets}. Since the storage cost of TraHist is based on the data-dependent number of emergent buckets $\vert B \vert$, it cannot be fixed a priori. We therefore fix its core parameters ($\alpha=0.5$, $\beta=1.0$) and vary the grid cell size $\Delta$ to obtain a target memory budget. All other methods are configured according to their theoretical cost models to match this budget. Second, we evaluate methods under \textit{a fixed number of buckets}. This setting fixes the number $\vert B \vert$ of buckets across methods, highlighting how different partitioning strategies capture spatial density variation and trajectory continuity. For DEH and CASE+, the controlled bucket count corresponds to the number of cells in their fixed regular grids. We report results at representative budget levels. Detailed analyses of TraHist, covering robustness, CountGrid scalability, and parameter sensitivity, are provided in Appendices~\ref{appendix:robustness} and \ref{appendix:parameter}, while \cref{appendix:larger-budgets} extends the results reported in this section to larger budgets.

\noindentparagraph{\textnormal{\textbf{Evaluation Metrics.}}}
We evaluate estimator performance in terms of estimation accuracy, preservation of trajectory continuity, and computational efficiency.
\par We assess estimation accuracy primarily using the \textit{Q-error}~\cite{moerkotte2009preventing, leis2015good, kipf2019mscn}. For a query $q$, let $y_q$ and $\hat{y}_q$ denote the true and estimated cardinality values, respectively. The Q-error is defined as 
\begin{equation}
    \text{Qerr}(q) = \max\left(\frac{\hat{y}_q}{y_q},\frac{y_q}{\hat{y}_q} \right),
\end{equation}
where the ideal value is 1.0, indicating perfect estimation. Further, larger values reflect greater estimation error. To ensure numerical stability, queries where either $y_q$ or $\hat{y}_q$ equals zero were excluded from Q-error analysis.
\par To quantify how partitioning affects trajectory continuity, we measure the \textit{trajectory fragmentation count} $F(T)$ for each trajectory $T = \langle p_1, \ldots, p_i, \ldots, p_n\rangle$. Let $l_i$ denote the label of the histogram bucket containing each point $p_i$. Then, the fragmentation count is defined as
\begin{equation}\label{eq:trajfrag}
    F(T) = 1+\sum_{i=2}^{n}\textbf{1}[l_i\neq l_{i-1}].
\end{equation} 
Intuitively, $F(T)$ counts the number of contiguous histogram buckets visited by the sequence of trajectory points, incrementing each time it crosses a bucket boundary. Generally, lower average fragmentation indicates better preservation of trajectory continuity under the partition.
\par Finally, we evaluate computational efficiency using wall-clock time measurements. \textit{Estimation time} is defined as the time required to compute a single cardinality estimate $\hat{y}_q$, whereas \textit{construction time} denotes the time required to build the estimator. Throughout the experimental tables, bold and underlined values indicate the best and second-best results, respectively.

% ====================== Accuracy Result ======================

\begin{table}[t]
    \centering
    \caption{Q-error of trajectory range query estimation on \textit{Porto}, comparing estimators under \textit{matched memory budgets}. Errors are summarized by selectivity using percentiles over 1,000 queries per group.}
    \label{tab:qerror-DD-Porto-memory}
    {%
    \renewcommand{\arraystretch}{1}
    \setlength{\aboverulesep}{0pt}
    \setlength{\belowrulesep}{0pt}
    \setlength{\arrayrulewidth}{0.8pt}
    \setlength{\tabcolsep}{3pt}
    \newcommand{\topgap}{\rule{0pt}{2.7ex}}
    \newcommand{\botgap}{\rule[-1.5ex]{0pt}{0pt}}
    \ifdefined\qcolw\else\newlength{\qcolw}\fi
    \settowidth{\qcolw}{0000.0}
    \newcommand{\hd}[1]{\makebox[\qcolw][r]{#1}}
    \ifdefined\qhd\else\newlength{\qhd}\fi
    \settowidth{\qhd}{50th}
    \newcommand{\qlt}{\dimexpr\qcolw-\qhd\relax}
    \resizebox{\textwidth}{!}{%
    \begin{tabular}{@{}c@{\hspace{6pt}}l@{\hspace{6pt}}|@{\hspace{6pt}}rrrr@{\hspace{5pt}}c@{\hspace{5pt}}rrrr@{\hspace{5pt}}c@{\hspace{5pt}}rrrr@{\hspace{6pt}}}
        \toprule[0.8pt]
        \multirow{2}{*}{\begin{tabular}[c]{@{}c@{}}\textbf{Memory}\\\textbf{(KB)}\end{tabular}}
        & \multirow{2}{*}{\textbf{Est.}}
        & \multicolumn{4}{c}{\topgap\textbf{Low (0\%, 0.5\%]}}
        && \multicolumn{4}{c}{\textbf{Medium (0.5\%, 2\%]}}
        && \multicolumn{4}{c}{\textbf{High (2\%, 100\%]}} \\
        \cmidrule(l{\qlt}r{5pt}){3-6} \cmidrule(l{\qlt}r{5pt}){8-11} \cmidrule(l{\qlt}r{6pt}){13-16}
        & & \topgap\botgap\hd{50th} & \hd{95th} & \hd{99th} & \hd{Max}
        && \hd{50th} & \hd{95th} & \hd{99th} & \hd{Max}
        && \hd{50th} & \hd{95th} & \hd{99th} & \hd{Max} \\
        \midrule[0.8pt]
        \multirow{6}{*}{50} & \topgap MEH & 183.6 & 7147.8 & 4.1e4 & 3.5e5
            && 15.2 & 1018.3 & 4484.9 & 9.9e4
            && 4.2 & 7.7 & 8.6 & 15.8 \\
         & MHist & 327.9 & 5504.3 & 1.5e4 & 1.1e5
            && 29.4 & 996.0 & 2893.2 & 2.0e4
            && 3.4 & 5.0 & 8.3 & 19.8 \\
         & LHist & 35.3 & 178.7 & 384.2 & 3.1e4
            && 21.0 & 161.3 & 310.0 & 2.0e4
            && \secondbest{1.5} & 5.4 & 21.5 & 45.4 \\
         & DEH & \secondbest{19.2} & \secondbest{61.6} & \secondbest{87.4} & \secondbest{186.1}
            && \secondbest{5.2} & \secondbest{15.5} & \secondbest{20.0} & \best{38.2}
            && \best{1.4} & \best{2.7} & \best{4.0} & \best{5.7} \\
         & CASE+ & 50.9 & 113.7 & 160.5 & 382.0
            && 11.9 & 26.6 & 34.7 & \secondbest{57.3}
            && 1.6 & \secondbest{3.5} & \secondbest{4.7} & \secondbest{5.9} \\
         & \botgap\textbf{TraHist} & \best{5.0} & \best{11.1} & \best{27.7} & \best{175.6}
            && \best{1.4} & \best{7.9} & \best{16.1} & 62.8
            && 2.6 & 7.9 & 18.6 & 68.3 \\
        \midrule
        \multirow{6}{*}{100} & \topgap MEH & 116.0 & 5901.7 & 2.6e4 & 2.7e5
            && 10.3 & 678.9 & 3745.8 & 5.2e4
            && 4.9 & 8.6 & 9.8 & 17.8 \\
         & MHist & 218.6 & 3295.8 & 9327.0 & 1.3e6
            && 19.8 & 642.2 & 1954.4 & 3.9e5
            && 4.0 & 5.8 & 9.5 & 20.9 \\
         & LHist & 41.8 & 217.5 & 431.5 & 3.4e4
            && 23.3 & 183.6 & 389.9 & 2.2e4
            && 1.7 & 6.4 & 25.8 & 53.6 \\
         & DEH & \secondbest{14.7} & \secondbest{51.5} & \secondbest{83.8} & \secondbest{179.3}
            && \secondbest{4.2} & \secondbest{13.8} & \secondbest{21.6} & \secondbest{38.6}
            && \best{1.4} & \best{2.4} & \best{3.4} & \best{5.2} \\
         & CASE+ & 48.2 & 109.2 & 154.8 & 682.9
            && 11.0 & 25.3 & 36.6 & 60.1
            && \secondbest{1.5} & \secondbest{2.9} & \secondbest{4.4} & \secondbest{5.3} \\
         & \botgap\textbf{TraHist} & \best{4.2} & \best{8.6} & \best{20.2} & \best{45.3}
            && \best{1.4} & \best{7.6} & \best{12.8} & \best{38.0}
            && 3.9 & 9.0 & 13.8 & 28.8 \\
        \midrule
        \multirow{6}{*}{500} & \topgap MEH & 52.1 & 1719.0 & 9772.7 & 5.7e5
            && 5.4 & 219.3 & 1157.2 & 1.0e5
            && 6.1 & 10.2 & 11.6 & 21.6 \\
         & MHist & 95.3 & 1622.3 & 5659.5 & 1.2e5
            && 9.8 & 273.4 & 907.0 & 2.1e4
            && 5.2 & 7.9 & 11.3 & 22.8 \\
         & LHist & 64.1 & 361.3 & 843.2 & 5.1e4
            && 28.1 & 320.1 & 723.2 & 3.3e4
            && 2.2 & 8.4 & 31.9 & 68.6 \\
         & DEH & \secondbest{10.3} & \secondbest{36.0} & \secondbest{50.0} & \secondbest{104.8}
            && \secondbest{2.9} & \secondbest{9.4} & \secondbest{13.5} & \best{23.6}
            && \best{1.2} & \best{1.7} & \best{2.5} & \best{4.0} \\
         & CASE+ & 42.9 & 96.3 & 127.6 & 654.8
            && 7.8 & 22.0 & 27.3 & \secondbest{53.4}
            && \secondbest{1.3} & \secondbest{2.4} & \secondbest{3.4} & \secondbest{5.0} \\
         & \botgap\textbf{TraHist} & \best{1.9} & \best{5.3} & \best{14.7} & \best{29.8}
            && \best{2.6} & \best{7.5} & \best{12.2} & 430.2
            && 5.0 & 8.5 & 10.9 & 13.8 \\
        \bottomrule[0.8pt]
    \end{tabular}%
    }
    }
\end{table}

\begin{figure}[t]
    \centering
    \includegraphics[width=\linewidth]{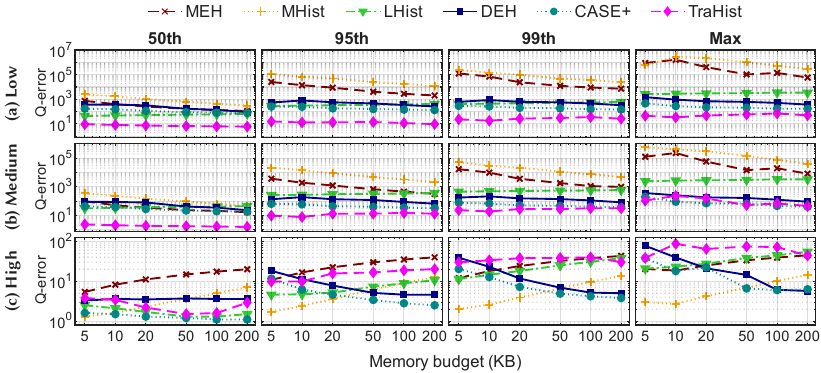}
    \caption{Q-error for trajectory range queries on \textit{T-Drive}. Median and tail Q-error are reported across \textit{memory budgets} for (a) low (0\%, 0.5\%], (b) medium (0.5\%, 2\%], and (c) high (2\%, 100\%] selectivity queries.}
    \Description{Three rows of line plots showing Q-error on a logarithmic scale versus memory budget for six estimators on T-Drive. Rows correspond to low, medium, and high selectivity queries, and columns to the 50th, 95th, and 99th percentile and maximum Q-error.}
    \label{fig:qerror-DD-Tdrive-memory}
\end{figure}

\subsection{Evaluation under Fixed Memory Budgets} \label{subsec:cardinality-estimation-accuracy}
We evaluate the accuracy of trajectory range query cardinality estimation under the same memory budget setting. Results on \textit{Porto} are summarized in~\cref{tab:qerror-DD-Porto-memory}, while corresponding results on \textit{T-Drive} are reported in~\cref{fig:qerror-DD-Tdrive-memory}.

\subsubsection{TraHist achieves the highest accuracy for low and medium selectivity queries (RQ1).}
Under matched memory budgets, TraHist consistently achieves the lowest median Q-error among all estimators for low and medium selectivity queries. On \textit{Porto} (\cref{tab:qerror-DD-Porto-memory}), at 50\,KB under low selectivity, TraHist achieves 37--66$\times$ lower median Q-error than MEH and MHist, and about 7$\times$ lower error than LHist.
Across the reported budgets, TraHist reduces median Q-error relative to DEH, the strongest baseline on \textit{Porto}, by 3.5--5.4$\times$ under low selectivity and 1.1--3.7$\times$ under medium selectivity. On \textit{T-Drive} (\cref{fig:qerror-DD-Tdrive-memory}), it improves over the most competitive baseline by 4.7--11.0$\times$ and 10.7--16.2$\times$ under low and medium selectivity, respectively. The consistency of these gains across datasets with very different scale and trajectory length characteristics highlights the robustness of TraHist under fragmentation-dominated query workloads.

\subsubsection{TraHist reaches low estimation error under tight memory budgets (RQ1).}
As memory budgets increase, estimation error generally decreases. However, methods differ in how quickly they reach practically useful accuracy--low single-digit median Q-error---under tight memory constraints.
\par TraHist consistently attains low absolute Q-error at substantially smaller memory budgets than competing estimators for low and medium selectivity queries.  On \textit{Porto} (\cref{tab:qerror-DD-Porto-memory}), TraHist already achieves median Q-error values in the low single-digit range (approximately 1--3) at 500\,KB. Under low selectivity, no baseline reaches this range at the same budget: all baselines remain above 10, with DEH providing the lowest median Q-error of 10.3. The gap narrows under medium selectivity, where DEH approaches TraHist at 500\,KB (2.9 versus 2.6). However, TraHist reaches the low single-digit range at tighter budgets, achieving 1.4 at both 50 and 100\,KB, compared with 5.2 and 4.2 for DEH, respectively. This shows that TraHist reaches and maintains the same accuracy range with substantially less memory. A similar pattern is observed on \textit{T-Drive} (\cref{fig:qerror-DD-Tdrive-memory}): under medium selectivity, TraHist rapidly reaches median Q-error values in the low single-digit range (approximately 1--3), while under low selectivity it remains stable in the single-digit range (approximately 6--9). These results indicate that TraHist captures the dominant spatial and trajectory-level correlations with a compact synopsis, making it particularly effective when memory is limited.

\subsubsection{Accuracy differences diminish under high selectivity queries (RQ1).} \label{subsubsec:high-selectivity}
High selectivity queries alter the relative ranking of estimators. As query regions grow, fragmentation-induced overcounting becomes less pronounced relative to the true cardinality. On \textit{Porto}, DEH achieves the lowest median Q-error, with CASE+ closely following and LHist also achieving lower error than TraHist across all reported budgets. On \textit{T-Drive}, CASE+ reaches the lowest median Q-error at most budgets. TraHist remains within single-digit median Q-error across the evaluated budgets. This result is consistent with the growing relative importance of connection-aware aggregation and intra-bucket modeling as fragmentation becomes less dominant.
\par However, this disadvantage is considerably smaller than the gains observed under low and medium selectivity. Under high selectivity, TraHist's median Q-error is 1.3--4.2$\times$ that of the most accurate estimator, whereas TraHist improves over the strongest baseline by up to 11.0$\times$ and 16.2$\times$ under low and medium selectivity. These results indicate that TraHist's gains concentrate in queries where fragmentation dominates estimation error.

%%%%%%%%%%%%%%%%%%%%%%%%%%%%%%%%%%%%%%%%%%%%%%%%%%%%%%%%%%%%%%%%%

\subsection{Evaluation under Fixed Bucket Budgets} \label{subsec:Analysis-under-Fixed-Bucket-Budgets}
While TraHist achieves superior accuracy under matched memory budgets, that comparison alone does not reveal the structural sources of these gains. In particular, different estimators incur different per-bucket storage costs, and therefore realize substantially different numbers of buckets under the same memory constraint. To isolate the effect of partitioning structure from memory allocation, we analyze estimation accuracy and trajectory fragmentation under fixed bucket budgets, where all methods are constrained to use the same number of buckets.

\begin{table}[t]
    \centering
    \caption{Q-error of trajectory range query estimation on \textit{T-Drive}, comparing estimators under the \textit{same number of buckets}. Errors are summarized by selectivity using percentiles over 1,000 queries per group.}
    \label{tab:qerror-DD-Tdrive-buckets}
    {%
    \renewcommand{\arraystretch}{1}
    \setlength{\aboverulesep}{0pt}
    \setlength{\belowrulesep}{0pt}
    \setlength{\arrayrulewidth}{0.8pt}
    \setlength{\tabcolsep}{3pt}
    \newcommand{\topgap}{\rule{0pt}{2.7ex}}
    \newcommand{\botgap}{\rule[-1.5ex]{0pt}{0pt}}
    \ifdefined\qcolw\else\newlength{\qcolw}\fi
    \settowidth{\qcolw}{0000.0}
    \newcommand{\hd}[1]{\makebox[\qcolw][r]{#1}}
    \ifdefined\qhd\else\newlength{\qhd}\fi
    \settowidth{\qhd}{50th}
    \newcommand{\qlt}{\dimexpr\qcolw-\qhd\relax}
    \resizebox{\textwidth}{!}{%
    \begin{tabular}{@{}c@{\hspace{6pt}}l@{\hspace{6pt}}|@{\hspace{6pt}}rrrr@{\hspace{5pt}}c@{\hspace{5pt}}rrrr@{\hspace{5pt}}c@{\hspace{5pt}}rrrr@{\hspace{6pt}}}
        \toprule[0.8pt]
        \multirow{2}{*}{\textbf{Buckets}}
        & \multirow{2}{*}{\textbf{Est.}}
        & \multicolumn{4}{c}{\topgap\textbf{Low (0\%, 0.5\%]}}
        && \multicolumn{4}{c}{\textbf{Medium (0.5\%, 2\%]}}
        && \multicolumn{4}{c}{\textbf{High (2\%, 100\%]}} \\
        \cmidrule(l{\qlt}r{5pt}){3-6} \cmidrule(l{\qlt}r{5pt}){8-11} \cmidrule(l{\qlt}r{6pt}){13-16}
        & & \topgap\botgap\hd{50th} & \hd{95th} & \hd{99th} & \hd{Max}
        && \hd{50th} & \hd{95th} & \hd{99th} & \hd{Max}
        && \hd{50th} & \hd{95th} & \hd{99th} & \hd{Max} \\
        \midrule[0.8pt]
        \multirow{6}{*}{50} & \topgap MEH & 1580.3 & 8.7e4 & 4.6e5 & 4.0e6
            && 227.0 & 1.2e4 & 6.7e4 & 8.2e6
            && 3.7 & 7.7 & 25.7 & 60.8 \\
         & MHist & 1924.6 & 8.3e4 & 1.6e5 & 4.3e5
            && 269.1 & 1.6e4 & 3.9e4 & 6.0e5
            && \best{1.5} & \best{2.3} & \best{3.8} & \best{6.4} \\
         & LHist & \secondbest{52.2} & 331.0 & 485.6 & 2919.8
            && \secondbest{38.1} & 303.8 & 518.5 & 2761.3
            && \secondbest{1.6} & \secondbest{5.4} & \secondbest{19.1} & \secondbest{28.0} \\
         & DEH & 328.5 & 480.0 & 608.8 & 801.5
            && 80.2 & 117.7 & 175.3 & 278.7
            && 2.6 & 23.0 & 46.9 & 76.8 \\
         & CASE+ & 183.8 & \secondbest{270.7} & \secondbest{343.4} & \secondbest{452.0}
            && 45.2 & \secondbest{66.4} & \secondbest{98.8} & \secondbest{157.2}
            && 1.7 & 13.0 & 27.8 & 45.1 \\
         & \botgap\textbf{TraHist} & \best{7.3} & \best{13.2} & \best{23.5} & \best{100.1}
            && \best{1.8} & \best{7.7} & \best{23.9} & \best{153.9}
            && 4.1 & 10.0 & 37.9 & 67.3 \\
        \midrule
        \multirow{6}{*}{100} & \topgap MEH & 743.6 & 2.6e4 & 1.3e5 & 8.9e5
            && 99.7 & 3677.2 & 1.7e4 & 1.2e5
            && 5.4 & 10.8 & \secondbest{12.0} & \secondbest{20.3} \\
         & MHist & 1713.5 & 6.8e4 & 1.4e5 & 2.9e6
            && 234.9 & 1.4e4 & 3.0e4 & 4.1e5
            && \secondbest{1.6} & \best{2.4} & \best{2.6} & \best{2.7} \\
         & LHist & \secondbest{57.7} & 374.7 & 549.1 & 3438.5
            && 42.1 & 334.1 & 542.1 & 3251.9
            && \best{1.3} & \secondbest{8.4} & 28.4 & 43.0 \\
         & DEH & 364.1 & 555.5 & 592.4 & 648.7
            && 90.3 & 135.2 & 158.5 & 274.0
            && 2.9 & 16.3 & 31.2 & 57.3 \\
         & CASE+ & 168.8 & \secondbest{257.6} & \secondbest{274.7} & \secondbest{300.8}
            && \secondbest{42.0} & \secondbest{62.9} & \secondbest{73.5} & \best{127.1}
            && \secondbest{1.6} & \secondbest{8.4} & 14.9 & 33.7 \\
         & \botgap\textbf{TraHist} & \best{8.4} & \best{13.3} & \best{18.4} & \best{34.4}
            && \best{2.1} & \best{7.9} & \best{19.2} & \secondbest{233.7}
            && 3.4 & 9.9 & 33.6 & 87.1 \\
        \midrule
        \multirow{6}{*}{500} & \topgap MEH & 239.3 & 8391.9 & 1.7e4 & 6.2e5
            && 32.0 & 1209.4 & 2391.1 & 8.7e4
            && 11.8 & 23.9 & 25.7 & 26.6 \\
         & MHist & 798.5 & 3.5e4 & 7.3e4 & 1.5e6
            && 120.4 & 6508.3 & 1.7e4 & 2.1e5
            && 2.8 & \secondbest{4.7} & \best{5.0} & \best{5.2} \\
         & LHist & \secondbest{96.8} & 574.2 & 975.9 & 4363.3
            && 54.2 & 504.9 & 848.9 & 4125.0
            && 3.0 & 18.6 & 68.4 & 100.1 \\
         & DEH & 325.0 & 738.8 & 829.9 & 895.1
            && 81.0 & 174.7 & 201.1 & 271.2
            && 3.7 & 10.6 & 20.1 & 41.9 \\
         & CASE+ & 101.4 & \secondbest{179.2} & \secondbest{199.7} & \secondbest{215.4}
            && \secondbest{25.1} & \secondbest{42.9} & \secondbest{49.0} & \best{79.6}
            && \best{1.3} & \best{4.3} & \secondbest{7.6} & \secondbest{14.9} \\
         & \botgap\textbf{TraHist} & \best{7.2} & \best{13.9} & \best{25.5} & \best{70.2}
            && \best{1.8} & \best{13.5} & \best{25.7} & \secondbest{156.0}
            && \secondbest{2.0} & 14.8 & 32.4 & 76.8 \\
        \bottomrule[0.8pt]
    \end{tabular}%
    }
    }
\end{table}

\begin{figure}[t]
    \centering
    \includegraphics[width=\linewidth]{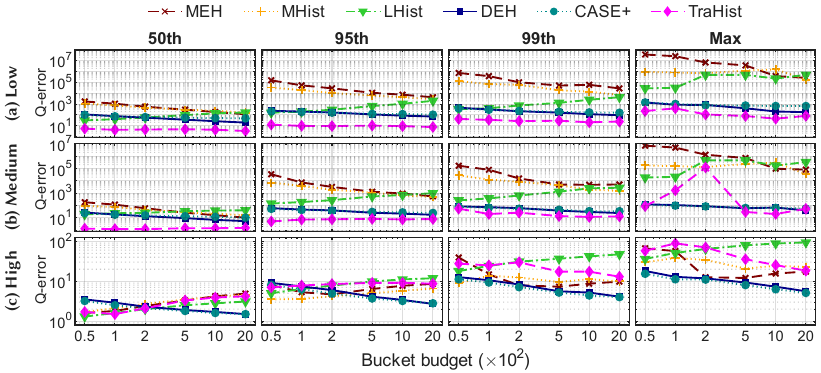}
    \caption{Q-error for trajectory range queries on \textit{Porto}. Median and tail Q-error are reported across \textit{bucket budgets} for (a) low (0\%, 0.5\%], (b) medium (0.5\%, 2\%], and (c) high (2\%, 100\%] selectivity queries.}
    \Description{Three rows of line plots showing Q-error on a logarithmic scale versus bucket budget for six estimators on Porto. Rows correspond to low, medium, and high selectivity queries, and columns to the 50th, 95th, and 99th percentile and maximum Q-error.}
    \label{fig:qerror-DD-Porto-buckets}
\end{figure}

\subsubsection{TraHist maintains high accuracy under fixed bucket budgets (RQ1).}
With the number of buckets fixed across methods, TraHist maintains high estimation accuracy, showing that its gains under matched memory budgets are not solely due to differences in bucket count. This pattern is particularly evident on \textit{T-Drive}, whose long and heterogeneous trajectories amplify boundary-crossing errors when partitioning is misaligned (\cref{tab:dataset_stats}). As shown in \cref{tab:qerror-DD-Tdrive-buckets}, TraHist substantially outperforms all competing estimators under low and medium selectivity queries. At a bucket budget of 50, TraHist reduces the median Q-error by approximately 7.2$\times$ under low selectivity and over 20$\times$ under medium selectivity compared to LHist, while achieving one to two orders of magnitude lower error than classical histogram baselines such as MEH and MHist. Compared with the trajectory-specific baselines, DEH and CASE+, TraHist achieves 25--45$\times$ lower median Q-error at 50 buckets under both low and medium selectivity regimes. The overall accuracy advantage remains substantial across 50--500 buckets. Similar trends are observed on \textit{Porto} (\cref{fig:qerror-DD-Porto-buckets}). DEH and CASE+ generally provide the strongest competing results among the baselines. Compared with the stronger of the two at each budget, TraHist achieves 6.0--18.9$\times$ and 3.5--18.5$\times$ lower median Q-error under low and medium selectivity across 50--2,000 buckets. This comparison indicates that connection-aware aggregation over a fixed grid is effective but does not fully compensate for the limitations of the underlying partition. These results further indicate that TraHist's accuracy gains are closely tied to its partitioning strategy.

\begin{figure}[t]
    \centering
    \includegraphics[width=\linewidth]{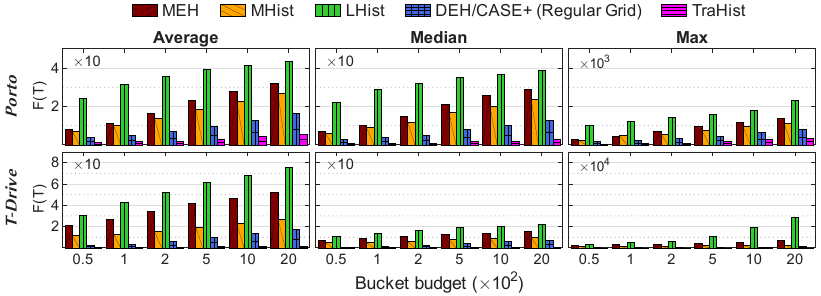}
    \caption{Trajectory fragmentation count $F(T)$ (Eq.~\eqref{eq:trajfrag}) under controlled \textit{bucket budgets}. Average, median, and maximum $F(T)$ are shown for \textit{Porto} (top) and \textit{T-Drive} (bottom) as the bucket budget increases.}
    \Description{Two rows of grouped bar charts showing the trajectory fragmentation count F(T) for five estimators as the bucket budget increases from 50 to 2,000 buckets. The top row shows Porto and the bottom row T-Drive; the columns show the average, median, and maximum F(T).}
    \label{fig:trajfrag}
\end{figure}

\subsubsection{Trajectory continuity is better preserved by TraHist under fixed bucket budgets (RQ2).}
To quantify how partitioning structure affects trajectory continuity, we measure the per-trajectory fragmentation count $F(T)$ (\cref{eq:trajfrag}), which counts the number of contiguous bucket runs into which the partition splits the sampled-point sequence of a trajectory, and is summarized using average, median, and maximum statistics.
\par As shown in~\cref{fig:trajfrag}, TraHist achieves the lowest average and median fragmentation across all bucket budgets. On \textit{Porto}, TraHist yields only 1.3 fragments on average at 50 buckets and remains below 6 fragments even at 2,000 buckets. In contrast, MEH, MHist, and LHist exhibit moderate to high fragmentation, growing from roughly 7--24 fragments at 50 buckets to around 27--44 fragments at 2,000 buckets. DEH and CASE+ share the same regular-grid partition and therefore have identical fragmentation counts, which remain consistently higher than TraHist but lower than those of MEH, MHist and LHist. This contrast is even more pronounced on \textit{T-Drive}. TraHist maintains near-contiguous trajectories with an average fragmentation close to 1 across all budgets, whereas fragmentation under the DEH/CASE+ regular grid increases from 2.9 fragments at 50 buckets to 17.1 at 2,000 buckets, and the other baselines reach 27--76 fragments. These results highlight that fragmentation is not merely reduced but effectively suppressed by our partitioning strategy.
\par These differences reflect the underlying partitioning strategies. MEH, MHist, and LHist optimize point-level distributions and tend to place more boundaries in dense regions, increasing crossings along heavily sampled trajectories. DEH and CASE+ distribute boundaries uniformly through a fixed regular grid, reducing this concentration but remaining trajectory-agnostic. TraHist instead adapts boundaries to trajectory-induced density structure, preserving trajectory continuity while maintaining local density coherence.

%%%%%%%%%%%%%%%%%%%%%%%%%%%%%%%%%%%%%%%%%%%%%%%%%%%%%%%%%%%%%%%%%
\begin{table}[t]
    \centering
    \caption{Query estimation latency (ms) under (a) fixed \textit{memory budgets} (\textit{Porto}: 50--500 KB, \textit{T-Drive}: 5--50 KB) and (b) fixed \textit{bucket budgets}. Average and worst-case (maximum) estimation times are measured across 3,000 trajectory range queries on \textit{Porto} and \textit{T-Drive}, covering all selectivity regimes.}
    \label{tab:qtime}
    {%
    \renewcommand{\arraystretch}{1}
    \setlength{\aboverulesep}{0pt}
    \setlength{\belowrulesep}{0pt}
    \setlength{\arrayrulewidth}{0.8pt}
    \setlength{\tabcolsep}{3pt}
    \newcommand{\topgap}{\rule{0pt}{2.7ex}}
    \newcommand{\botgap}{\rule[-1.5ex]{0pt}{0pt}}
    \ifdefined\qcolw\else\newlength{\qcolw}\fi
    \settowidth{\qcolw}{00.000}
    \newcommand{\hd}[1]{\makebox[\qcolw][c]{#1}}
    \ifdefined\qhd\else\newlength{\qhd}\fi
    \settowidth{\qhd}{Avg}
    \ifdefined\qhm\else\newlength{\qhm}\fi
    \settowidth{\qhm}{Max}
    \newcommand{\qlt}{\dimexpr(\qcolw-\qhd)/2\relax}
    \newcommand{\qrta}{\dimexpr5pt+(\qcolw-\qhm)/2\relax}
    \newcommand{\qrtb}{\dimexpr6pt+(\qcolw-\qhm)/2\relax}
    \ifdefined\tqwlen\else\newlength{\tqwlen}\fi
    \settowidth{\tqwlen}{00.000}
    \newcommand{\tqw}[1]{\makebox[\tqwlen][r]{#1}}
    \ifdefined\tqblen\else\newlength{\tqblen}\fi
    \settowidth{\tqblen}{\textbf{0.000}}
    \newcommand{\tqb}[1]{\makebox[\tqblen][r]{#1}}
    \resizebox{0.95\textwidth}{!}{%
    \begin{tabular}[t]{@{}c@{\hspace{6pt}}l@{\hspace{6pt}}|@{\hspace{6pt}}cc@{\hspace{5pt}}c@{\hspace{5pt}}cc@{\hspace{6pt}}}
        \multicolumn{7}{c}{\textbf{(a) Fixed memory budgets}} \\[2pt]
        \toprule[0.8pt]
        \multirow{2}{*}{\begin{tabular}[c]{@{}c@{}}\textbf{Memory}\\\textbf{(KB)}\end{tabular}}
        & \multirow{2}{*}{\textbf{Est.}}
        & \multicolumn{2}{c}{\topgap\textbf{\textit{Porto}}}
        && \multicolumn{2}{c}{\textbf{\textit{T-Drive}}} \\
        \cmidrule(l{\qlt}r{\qrta}){3-4} \cmidrule(l{\qlt}r{\qrtb}){6-7}
        & & \topgap\botgap\hd{Avg} & \hd{Max} && \hd{Avg} & \hd{Max} \\
        \midrule[0.8pt]
        \multirow{6}{*}{50 / 5} & \topgap MEH & \tqw{\best{0.152}} & \tqw{\best{0.324}} && \tqw{\best{0.053}} & \tqw{\best{0.115}} \\
         & MHist & \tqw{8.732} & \tqw{8.804} && \tqw{1.637} & \tqw{1.651} \\
         & LHist & \tqw{1.211} & \tqw{2.367} && \tqw{0.710} & \tqw{1.077} \\
         & DEH & \tqw{\secondbest{0.535}} & \tqw{\secondbest{0.554}} && \tqw{\secondbest{0.527}} & \tqw{\secondbest{0.531}} \\
         & CASE+ & \tqw{0.548} & \tqw{0.573} && \tqw{0.530} & \tqw{0.533} \\
         & \botgap\textbf{TraHist} & \tqw{0.572} & \tqw{0.588} && \tqw{0.570} & \tqw{0.581} \\
        \midrule
        \multirow{6}{*}{100 / 10} & \topgap MEH & \tqw{\best{0.269}} & \tqw{0.612} && \tqw{\best{0.097}} & \tqw{\best{0.229}} \\
         & MHist & \tqw{17.355} & \tqw{17.386} && \tqw{3.265} & \tqw{3.331} \\
         & LHist & \tqw{1.542} & \tqw{3.190} && \tqw{0.810} & \tqw{1.347} \\
         & DEH & \tqw{\secondbest{0.544}} & \tqw{\best{0.561}} && \tqw{\secondbest{0.527}} & \tqw{\secondbest{0.537}} \\
         & CASE+ & \tqw{0.559} & \tqw{\secondbest{0.586}} && \tqw{0.539} & \tqw{0.543} \\
         & \botgap\textbf{TraHist} & \tqw{0.577} & \tqw{0.592} && \tqw{0.564} & \tqw{0.580} \\
        \midrule
        \multirow{6}{*}{500 / 50} & \topgap MEH & \tqw{1.235} & \tqw{3.048} && \tqw{\best{0.374}} & \tqw{0.985} \\
         & MHist & \tqw{83.195} & \tqw{83.247} && \tqw{16.668} & \tqw{16.706} \\
         & LHist & \tqw{2.912} & \tqw{6.710} && \tqw{1.304} & \tqw{2.610} \\
         & DEH & \tqw{\best{0.546}} & \tqw{\best{0.565}} && \tqw{\secondbest{0.531}} & \tqw{\best{0.548}} \\
         & CASE+ & \tqw{\secondbest{0.561}} & \tqw{\secondbest{0.582}} && \tqw{0.546} & \tqw{\secondbest{0.573}} \\
         & \botgap\textbf{TraHist} & \tqw{0.581} & \tqw{0.610} && \tqw{0.571} & \tqw{0.593} \\
        \bottomrule[0.8pt]
    \end{tabular}%
    \hspace{14pt}%
    \begin{tabular}[t]{@{}c@{\hspace{6pt}}l@{\hspace{6pt}}|@{\hspace{6pt}}cc@{\hspace{5pt}}c@{\hspace{5pt}}cc@{\hspace{6pt}}}
        \multicolumn{7}{c}{\textbf{(b) Fixed bucket budgets}} \\[2pt]
        \toprule[0.8pt]
        \multirow{2}{*}{\textbf{Buckets}}
        & \multirow{2}{*}{\textbf{Est.}}
        & \multicolumn{2}{c}{\topgap\textbf{\textit{Porto}}}
        && \multicolumn{2}{c}{\textbf{\textit{T-Drive}}} \\
        \cmidrule(l{\qlt}r{\qrta}){3-4} \cmidrule(l{\qlt}r{\qrtb}){6-7}
        & & \topgap\botgap\hd{Avg} & \hd{Max} && \hd{Avg} & \hd{Max} \\
        \midrule[0.8pt]
        \multirow{6}{*}{50} & \topgap MEH & \tqb{\best{0.023}} & \tqw{\best{0.033}} && \tqb{\best{0.036}} & \tqw{\best{0.072}} \\
         & MHist & \tqb{\secondbest{0.417}} & \tqw{\secondbest{0.420}} && \tqb{\secondbest{0.427}} & \tqw{\secondbest{0.435}} \\
         & LHist & \tqb{0.958} & \tqw{1.707} && \tqb{0.988} & \tqw{1.805} \\
         & DEH & \tqb{0.533} & \tqw{0.534} && \tqb{0.525} & \tqw{0.527} \\
         & CASE+ & \tqb{0.540} & \tqw{0.542} && \tqb{0.535} & \tqw{0.537} \\
         & \botgap\textbf{TraHist} & \tqb{0.566} & \tqw{0.583} && \tqb{0.569} & \tqw{0.585} \\
        \midrule
        \multirow{6}{*}{100} & \topgap MEH & \tqb{\best{0.032}} & \tqw{\best{0.051}} && \tqb{\best{0.053}} & \tqw{\best{0.115}} \\
         & MHist & \tqb{0.820} & \tqw{0.822} && \tqb{0.815} & \tqw{0.816} \\
         & LHist & \tqb{1.453} & \tqw{2.971} && \tqb{1.538} & \tqw{3.220} \\
         & DEH & \tqb{\secondbest{0.526}} & \tqw{\secondbest{0.530}} && \tqb{\secondbest{0.523}} & \tqw{\secondbest{0.527}} \\
         & CASE+ & \tqb{0.537} & \tqw{0.540} && \tqb{0.540} & \tqw{0.541} \\
         & \botgap\textbf{TraHist} & \tqb{0.567} & \tqw{0.585} && \tqb{0.564} & \tqw{0.580} \\
        \midrule
        \multirow{6}{*}{500} & \topgap MEH & \tqb{\best{0.087}} & \tqw{\best{0.172}} && \tqb{\best{0.191}} & \tqw{\best{0.482}} \\
         & MHist & \tqb{4.065} & \tqw{4.083} && \tqb{4.103} & \tqw{4.146} \\
         & LHist & \tqb{5.167} & \tqw{12.572} && \tqb{5.577} & \tqw{13.685} \\
         & DEH & \tqb{\secondbest{0.534}} & \tqw{\secondbest{0.552}} && \tqb{\secondbest{0.533}} & \tqw{\secondbest{0.550}} \\
         & CASE+ & \tqb{0.549} & \tqw{0.576} && \tqb{0.547} & \tqw{0.576} \\
         & \botgap\textbf{TraHist} & \tqb{0.571} & \tqw{0.590} && \tqb{0.567} & \tqw{0.584} \\
        \bottomrule[0.8pt]
    \end{tabular}%
    }
    }
\end{table}

\begin{figure}[t]
    \centering
    \includegraphics[width=\linewidth]{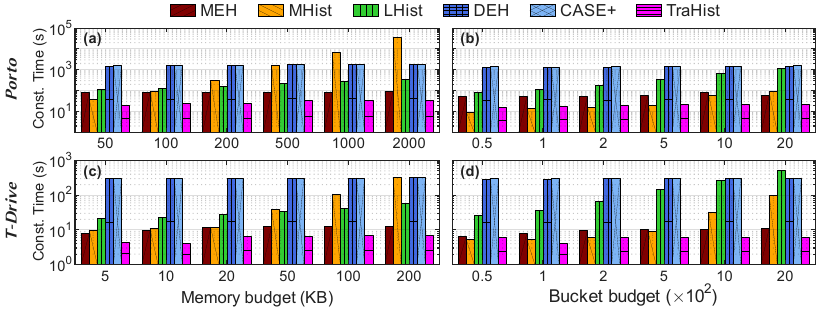}
    \caption{Estimator construction time in seconds on \textit{Porto} (top row) and \textit{T-Drive} (bottom row), under varying \textit{memory budgets} (left column) and \textit{bucket budgets} (right column).}
    \Description{Four grouped bar charts on a logarithmic scale showing the construction time in seconds of six estimators. The top row shows Porto and the bottom row T-Drive; the left column varies the memory budget and the right column varies the bucket budget. The panels are labeled (a) to (d).}
    \label{fig:consttime}
\end{figure}

\subsection{Evaluation of Computational Efficiency} \label{subsec:Computational-Efficiency}
Beyond accuracy, we evaluate computational efficiency to assess the practical viability of each estimator for large-scale query optimization. We consider two complementary evaluation settings: fixed memory budgets, which reflect realistic deployment scenarios, and fixed bucket budgets, which isolate the effect of partition granularity by controlling the number of buckets.

\subsubsection{TraHist enables fast query estimation with low and stable latency (RQ3).}
\cref{tab:qtime} reports the average and worst-case query estimation time measured over 3,000 trajectory range queries on \textit{Porto} and \textit{T-Drive} under both fixed memory and bucket budgets. Query-time performance varies substantially across estimators. Under fixed memory budgets (\cref{tab:qtime}a), DEH, CASE+, and TraHist consistently achieve sub-millisecond average latency that remains stable across all budgets. All three use lightweight grid-based query processing over precomputed statistics, while TraHist performs additional bucket mapping and trajectory-to-point correction, resulting in slightly higher latency than DEH and CASE+. MEH is the fastest estimator at small budgets but slows as the budget grows, whereas MHist incurs significantly higher latency that grows rapidly with the memory budget, exceeding 80\,ms on \textit{Porto} at 500\,KB. LHist also exhibits higher latency than TraHist despite using fewer buckets under the same memory budget, due to the additional per-bucket computation required by its expressive bucket models.
\par Under fixed bucket budgets (\cref{tab:qtime}b), DEH, CASE+, and TraHist maintain stable sub-millisecond average latency as the bucket budget increases, while MEH remains the fastest estimator despite increasing latency. This stability also extends to worst-case query time: while MHist and LHist exhibit noticeable tail-latency growth as the bucket budget increases, DEH, CASE+, and TraHist maintain consistently low and tightly bounded latency across all evaluated settings.
\par These results suggest that TraHist provides predictable and stable query-time behavior---a property particularly desirable for query optimizers, where robustness to worst-case estimation latency is critical in large-scale trajectory workloads.

\subsubsection{TraHist shows efficient construction with stable cost growth across bucket budgets (RQ3).}
\cref{fig:consttime} reports construction time on \textit{Porto} in the top row and \textit{T-Drive} in the bottom row, with memory budgets on the left and bucket budgets on the right. Under fixed memory budgets (Figures \ref{fig:consttime}a and \ref{fig:consttime}c), TraHist consistently achieves the lowest construction time across all evaluated memory budgets on both \textit{Porto} and \textit{T-Drive}. DEH and CASE+ require approximately 46--75$\times$ longer construction time than TraHist across the evaluated budgets, reflecting the cost of constructing the additional connectivity statistics used by their connection-aware aggregation over the regular grid. MHist also exhibits a dramatic increase in construction cost as memory grows, reaching several orders of magnitude higher cost at large budgets. This behavior is consistent with the smaller per-bucket memory footprint of MHist, which allows it to realize a substantially larger number of buckets under the same memory budget.
\par Under fixed bucket budgets (Figures~\ref{fig:consttime}b and~\ref{fig:consttime}d), TraHist again exhibits consistently low construction overhead as bucket count increases. On \textit{Porto}, construction time grows only modestly from approximately 15\,s at 50 buckets to about 23\,s at 2,000 buckets, while on \textit{T-Drive} it remains nearly constant at around 6--7\,s. The competing estimators exhibit different cost characteristics: DEH and CASE+ maintain substantially higher construction costs across all bucket budgets, whereas MHist and LHist become increasingly expensive as the number of buckets grows. MEH remains comparatively lightweight, but generally requires more construction time than TraHist.

% ====================== Spatial-Object Adaptation ======================

\subsection{Comparison with Spatial-Object Estimators}\label{subsec:spatial-object}
Generic spatial-object estimators can be adapted to trajectory range CE by replacing each trajectory $T$ with a spatial surrogate $S(T)$ and estimating $|\{T\in\mathcal{D}\mid S(T)\cap Q \neq \emptyset \}|$. We consider two representative choices for $S(T)$: the per-trajectory minimum bounding rectangle (MBR) and convex hull. Since the resulting surrogate cardinality may differ from the target cardinality $|RQ(Q)|$ defined over the original trajectory geometry $L(T)$, we first evaluate surrogate fidelity using exact counting. We then examine whether accurate MBR cardinality estimation translates into accurate end-to-end trajectory estimates of $|RQ(Q)|$.

\begin{table}[t]
    \centering
    \caption{Fidelity of enclosing spatial-object surrogates. Median and 95th-percentile Q-error are measured against the true trajectory range cardinality $\lvert RQ(Q)\rvert$. Exact intersection counting isolates surrogate-representation error.}
    \label{tab:exact-surrogate}
    {%
    \renewcommand{\arraystretch}{1}
    \setlength{\aboverulesep}{0pt}
    \setlength{\belowrulesep}{0pt}
    \setlength{\arrayrulewidth}{0.8pt}
    \setlength{\tabcolsep}{3pt}
    \newcommand{\topgap}{\rule{0pt}{2.7ex}}
    \newcommand{\botgap}{\rule[-1.5ex]{0pt}{0pt}}
    \ifdefined\qcolw\else\newlength{\qcolw}\fi
    \settowidth{\qcolw}{000.00}
    \newcommand{\hd}[1]{\makebox[\qcolw][r]{#1}}
    \ifdefined\qhd\else\newlength{\qhd}\fi
    \settowidth{\qhd}{50th}
    \newcommand{\qlt}{\dimexpr\qcolw-\qhd\relax}
    \resizebox{0.68\textwidth}{!}{%
    \begin{tabular}{@{}c@{\hspace{6pt}}l@{\hspace{6pt}}|@{\hspace{6pt}}rr@{\hspace{5pt}}c@{\hspace{5pt}}rr@{\hspace{5pt}}c@{\hspace{5pt}}rr@{\hspace{6pt}}}
        \toprule[0.8pt]
        \multirow{2}{*}{\textbf{Dataset}}
        & \multirow{2}{*}{\textbf{Surrogate}}
        & \multicolumn{2}{c}{\topgap\textbf{Low}}
        && \multicolumn{2}{c}{\textbf{Medium}}
        && \multicolumn{2}{c}{\textbf{High}} \\
        \cmidrule(l{\qlt}r{5pt}){3-4} \cmidrule(l{\qlt}r{5pt}){6-7} \cmidrule(l{\qlt}r{6pt}){9-10}
        & & \topgap\botgap\hd{50th} & \hd{95th} && \hd{50th} & \hd{95th} && \hd{50th} & \hd{95th} \\
        \midrule[0.8pt]
        \multirow{2}{*}{\textbf{\textit{Porto}}}
        & \topgap Exact-MBR & 42.53 & 89.03 && 10.92 & 22.05 && 1.77 & 3.55 \\
        & \botgap Exact-ConvexHull & 20.99 & 43.44 && 5.47 & 11.30 && 1.25 & 2.13 \\
        \midrule
        \multirow{2}{*}{\textbf{\textit{T-Drive}}}
        & \topgap Exact-MBR & 61.68 & 114.72 && 14.89 & 27.98 && 1.07 & 2.28 \\
        & \botgap Exact-ConvexHull & 39.85 & 88.93 && 9.70 & 21.86 && 1.03 & 1.31 \\
        \bottomrule[0.8pt]
    \end{tabular}%
    }
    }
\end{table}

\begin{table}[t]
    \centering
    \caption{Comparison of spatial-object adaptations and TraHist. Each entry reports median Q-error against the exact MBR cardinality and $\lvert RQ(Q)\rvert$, shown as \textit{MBR}\,/\,$\lvert RQ(Q)\rvert$. DeepDB-MBR$^{\dagger}$ denotes the configuration with the lowest low selectivity MBR Q-error.}
    \label{tab:spatial-object-estimators}
    {%
    \renewcommand{\arraystretch}{1}
    \setlength{\aboverulesep}{0pt}
    \setlength{\belowrulesep}{0pt}
    \setlength{\arrayrulewidth}{0.8pt}
    \setlength{\tabcolsep}{3pt}
    \newcommand{\topgap}{\rule{0pt}{2.7ex}}
    \newcommand{\botgap}{\rule[-1.5ex]{0pt}{0pt}}
    \newcommand{\qhead}{\textbf{Median Q-error (}\textit{MBR}\,/\,$\lvert RQ(Q)\rvert$\textbf{)}}
    \resizebox{\textwidth}{!}{%
    \begin{tabular}{@{}l@{\hspace{6pt}}|@{\hspace{6pt}}cccc@{\hspace{5pt}}c@{\hspace{5pt}}cccc@{\hspace{6pt}}}
        \toprule[0.8pt]
        \multirow{3}{*}{\textbf{Method}}
        & \multicolumn{4}{c}{\topgap\textbf{\textit{Porto}}}
        && \multicolumn{4}{c}{\textbf{\textit{T-Drive}}} \\
        \cmidrule(r{5pt}){2-5} \cmidrule{7-10}
        & \multirow{2}{*}{\textbf{Memory}} & \multicolumn{3}{c}{\topgap\qhead}
        && \multirow{2}{*}{\textbf{Memory}} & \multicolumn{3}{c}{\qhead} \\
        \cmidrule(r{5pt}){3-5} \cmidrule{8-10}
        & & \topgap\botgap\textbf{Low} & \textbf{Medium} & \textbf{High}
        && & \textbf{Low} & \textbf{Medium} & \textbf{High} \\
        \midrule[0.8pt]
        \topgap\botgap\textit{Exact-MBR (reference)}
        & \textemdash{} & 1.00\,/\,42.53 & 1.00\,/\,10.92 & 1.00\,/\,1.77
        && \textemdash{} & 1.00\,/\,61.68 & 1.00\,/\,14.89 & 1.00\,/\,1.07 \\
        \midrule
        \topgap CDHistogram-MBR
        & 50\,KB & 1.02\,/\,42.16 & 1.03\,/\,10.67 & 1.02\,/\,\best{1.76}
        && 50\,KB & 1.05\,/\,59.63 & 1.05\,/\,14.56 & 1.03\,/\,\best{1.04} \\
        DeepDB-MBR$^{\dagger}$
        & 103\,MB & \best{1.00}\,/\,42.41 & \best{1.00}\,/\,10.90 & \best{1.00}\,/\,1.77
        && 8\,MB & \best{1.01}\,/\,61.52 & \best{1.01}\,/\,14.77 & \best{1.00}\,/\,1.07 \\
        \botgap TraHist
        & 50\,KB & \textemdash{}\,/\,\best{5.00} & \textemdash{}\,/\,\best{1.40} & \textemdash{}\,/\,2.60
        && 50\,KB & \textemdash{}\,/\,\best{6.70} & \textemdash{}\,/\,\best{1.70} & \textemdash{}\,/\,1.50 \\
        \bottomrule[0.8pt]
    \end{tabular}%
    }
    }
\end{table}

\subsubsection{Surrogate fidelity limits spatial-object adaptations under selective queries (RQ4).}
\Cref{tab:exact-surrogate} first isolates surrogate-representation error using exact intersection counts over per-trajectory MBRs and convex hulls. Exact-MBR yields low selectivity median Q-errors of 42.53 on \textit{Porto} and 61.68 on \textit{T-Drive}. Exact-ConvexHull reduces these errors by 1.5--2.0$\times$, but they remain substantial at 20.99 and 39.85, since even a tighter enclosure includes untraversed regions. Under high selectivity, both surrogates become substantially more accurate, with median Q-errors ranging from 1.03 to 1.77.
\par \Cref{tab:spatial-object-estimators} next examines whether accurate MBR cardinality estimation removes this representation error. CDHistogram-MBR is evaluated under the same 50\,KB memory budget as TraHist. For DeepDB-MBR, we vary the RDC threshold $\rho\in\{0.3,0.5\}$ and dataset-appropriate slice sizes. Since DeepDB-MBR does not produce histogram buckets and its SPNs require substantially more memory, we report the configuration with the lowest low-selectivity median Q-error against Exact-MBR at its corresponding model size (103\,MB on \textit{Porto} and 8\,MB on \textit{T-Drive}). Both methods estimate the MBR cardinality nearly exactly, with native median Q-errors of 1.00--1.05, yet their end-to-end errors remain close to the Exact-MBR reference. This identifies the trajectory-to-MBR representation, rather than estimator error, as the dominant error source under low and medium selectivity.
\par Compared with the evaluated MBR-based estimators, TraHist achieves 7.6--9.2$\times$ lower median Q-error under low and medium selectivity, while its median Q-error is 1.4--1.5$\times$ higher under high selectivity. This contrast indicates that TraHist's accuracy advantage is concentrated in selective queries, where preserving local trajectory support is critical.

%%%%%%%%%%%%%%%%%%%%%%%%%%%%%%%%%%%%%%%%%%%%%%%%%%%%%%%%%%%%%%%%%

\section{Related Work}\label{sec:related-work}

\noindentparagraph{\textnormal{\textbf{Multi-dimensional Histograms.}}}
Histograms are fundamental synopsis structures for selectivity estimation. Traditional data-driven histograms~\cite{muralikrishna1988meh, ioannidis1995voptimal, poosala1996maxdiff, poosala1997mhist, jagadish1998voptimal2} partition the data domain into axis-aligned rectangular buckets. Workload-aware extensions such as STGrid~\cite{aboulnaga1999stgrid} and STHoles~\cite{bruno2001stholes} refine histogram buckets using query feedback to adapt to frequently accessed regions, focusing on workload patterns rather than intrinsic data characteristics. Our work is complementary to these workload-aware extensions, focusing on data-driven synopses independent of access patterns. More recently, LHist~\cite{liu2021lhist} incorporates machine learning models to approximate intra-bucket distributions while preserving the same rectangular partitioning framework. Despite these advances, existing methods largely retain axis-aligned bucket constraints, which limit their ability to capture complex or directional data distributions. Our work addresses this core limitation by fundamentally improving the partitioning strategy itself.

\noindentparagraph{\textnormal{\textbf{Histograms for Moving Objects.}}}
Several studies~\cite{choi2002selectivity, hadjieleftheriou2003performance, deh, fanaeepour2015case} have extended histograms to moving objects. Spatiotemporal histograms~\cite{choi2002selectivity, hadjieleftheriou2003performance} incorporate velocity information into their bucket representations to estimate the selectivity of queries involving the future positions of moving objects, whereas TraHist focuses on trajectory-aware spatial partitioning for range cardinality estimation. DEH~\cite{deh} and CASE~\cite{fanaeepour2015case} address the distinct-counting problem~\cite{braz2007approximate, distinctcountproblem2, orlando2007spatio}---the overcounting of objects across partition boundaries---by maintaining explicit connection metadata between adjacent partitions to post-hoc correct for overcounting during aggregation, in distributed sensing and privacy-aware settings, respectively. However, these methods operate on a given partitioning and do not optimize the partition structure itself. TraHist instead structurally reduces fragmentation by aligning boundaries with trajectory-induced density patterns.

\noindentparagraph{\textnormal{\textbf{Learned Cardinality Estimation for Relational Data.}}}
Learned estimators~\cite{kipf2019mscn, yang2019naru, hilprecht2020deepdb} use trained neural or probabilistic models to predict cardinalities from query predicates or to approximate the underlying data distribution. These approaches either model complex correlations or learn directly from query workloads and often achieve high accuracy on relational workloads. However, their original formulations estimate cardinalities for predicates over relational tuples rather than trajectory polylines. In contrast, TraHist is designed for trajectory range CE without training.

\noindentparagraph{\textnormal{\textbf{Spatial Cardinality Estimation.}}}
Spatial CE has been studied for point range queries~\cite{ji2024safe}, rectangular objects and spatial joins~\cite{cdhistogram, an2001selectivity}, and polygon-intersection queries~\cite{ji2025polycard}. These methods natively estimate cardinalities over different spatial units and therefore do not directly target the trajectory range cardinality considered in this work. Point-based methods estimate the number of points contained in $Q$, whereas applying object-based methods to trajectories requires replacing each polyline $L(T)$ with a spatial-object surrogate $S(T)$, such as an MBR or an enclosing polygon. The resulting surrogate cardinality can differ from $|RQ(Q)|$ when $S(T)$ intersects $Q$ in a region not traversed by $L(T)$. We evaluate this distinction in~\cref{subsec:spatial-object} using exact MBR and convex hull surrogates, CDHistogram-MBR~\cite{cdhistogram}, and DeepDB-MBR~\cite{hilprecht2020deepdb}. TraHist instead targets trajectory range CE without collapsing each trajectory into a single enclosing object.

%%%%%%%%%%%%%%%%%%%%%%%%%%%%%%%%%%%%%%%%%%%%%%%%%%%%%%%%%%%%%%%%%

\section{Conclusion}\label{sec:conclusion}
In this work, we presented TraHist, a histogram-based cardinality estimator for trajectory data with density-adaptive partitioning. By approximating density-aligned partitioning via grid-based propagation, TraHist constructs rectilinear buckets that adapt to anisotropic density patterns, mitigating the overestimation caused by trajectory fragmentation. Extensive experiments on large-scale real-world datasets demonstrate Q-error reductions of over 20$\times$ against the strongest competing estimators, with fast, stable query estimation and competitive construction cost. These properties make TraHist a practical solution for query optimization in large-scale spatial and trajectory database systems. Future work includes adaptive multi-resolution indexing and extending density-aligned partitioning to spatiotemporal cardinality estimation while accounting for heterogeneous sampling and temporal predicate semantics.

%%%%%%%%%%%%%%%%%%%%%%%%%%%%%%%%%%%%%%%%%%%%%%%%%%%%%%%%%%%%%%%%%

\begin{acks}
    This work was partly supported by the Institute of Information \& Communications Technology Planning \& Evaluation (IITP) grants funded by the Korea government (MSIT) (IITP-2026-RS-2020-II201819, ICT Creative Consilience Program; RS-2026-25522620), the National Research Foundation of Korea (NRF) grants funded by the Korea government (MSIT) (RS-2026-25616569 and RS-2026-25494942), and the NRF grant funded by the Ministry of Education (RS-2021-NR060143, Basic Science Research Program).
\end{acks}


\bibliographystyle{ACM-Reference-Format}
\bibliography{reference}

\clearpage

% \begin{center}
\noindent{\Large\sffamily\bfseries Appendix}
% \end{center}

\appendix
\crefalias{section}{appendix}
\crefalias{subsection}{appendix}

\setcounter{table}{0}
\renewcommand{\thetable}{\Alph{section}\arabic{table}} 

\setcounter{figure}{0}
\renewcommand{\thefigure}{\Alph{section}\arabic{figure}}

%%%%%%%%%%%%%%%%%%%%%%%%%%%%%%%%%%%%%%%%%%%%%%%%%%%%%%%%%%%%%%%%%
\setcounter{table}{0}
\setcounter{figure}{0}

\section{Topological Predicates Beyond \textsc{Intersects}}
\label{app:topological-predicates}
\par TraHist targets the \textsc{Intersects} predicate in Definition \ref{dfn:trajectory-range-query}, counting distinct trajectories whose polyline $L(T)$ shares at least one point with the query region $Q$. More specific spatial range queries may ask whether $L(T)$ is \textsc{Within} $Q$, meaning that the complete polyline lies in $Q$ but not entirely on its boundary; whether it \textsc{Crosses} $Q$, passing through its interior and extending outside $Q$; or whether it \textsc{Touches} $Q$, meeting the boundary without entering the interior. Each predicate selects a different subset of trajectories intersecting $Q$, and therefore defines a distinct cardinality function rather than a rescaling of the intersection count. These relationships are defined by the interiors, boundaries, and exteriors of the two geometries under the standard topological model~\cite{ogc_sfa_2011}.
\par TraHist's synopsis consists of the LabelMap and two summaries for each bucket: the average point count per cell $S_p^{(k)}$ and the average distinct trajectory count per cell $S_T^{(k)}$. These local summaries do not capture trajectory identifiers across cells or buckets, complete trajectory extent, segment connectivity or orientation, or contact with a query boundary. Determining whether an individual trajectory satisfies the other topological predicates may require its points, endpoints, and connecting segments. \cite{bornholdt2025trajectory} formulate such predicates using recorded samples, endpoints, and, when necessary, the straight-line segments between consecutive samples. TraHist instead estimates cardinalities from a compact synopsis, so extending it to these predicates would require aggregate summaries of the corresponding trajectory and segment information.
\par For \textsc{Within}, local counts inside and outside $Q$ do not reveal whether they belong to the same trajectories. For an axis-aligned rectangular $Q$, the coordinate extrema $(x_{\min}, y_{\min}, x_{\max}, y_{\max})$ of each trajectory determine exactly whether $L(T)\subseteq Q$. Estimating this condition without storing one extent tuple per trajectory would require a separate aggregate synopsis over these four correlated values. Additional geometric information would be needed to exclude trajectories that lie entirely on the boundary of $Q$. For \textsc{Crosses}, local counts inside and outside $Q$ likewise do not indicate whether the same trajectory has support on both sides of the query boundary. Candidate summaries could record distinct trajectory transitions between neighboring cells or buckets, grouped by transition direction or segment orientation, and aggregate them along the query boundary. Since one trajectory may generate several transition events, the current trajectory-to-point ratio $\gamma^{(k)}$ cannot be applied directly; a separate correction from transition events to distinct trajectories would be required. For \textsc{Touches}, a cell intersected by the query boundary may contain both a segment that enters the interior of $Q$ and one that only contacts its boundary. Distinguishing these cases requires finer summaries of segment position and orientation that the current $S_p^{(k)}$ and $S_T^{(k)}$ do not have.
\par Collectively, these requirements indicate that the existing partition can organize some additional local statistics, but does not by itself provide all information required by the other predicates. Nevertheless, any trajectory satisfying \textsc{Within}, \textsc{Crosses}, or \textsc{Touches} also satisfies \textsc{Intersects}. The result set of each predicate is therefore contained in $RQ(Q)$. A query plan may first form a candidate set of trajectories intersecting $Q$ and then evaluate the additional topological condition on these candidates. In such a plan, TraHist estimates the intermediate candidate cardinality, providing an input for costing candidate generation and the subsequent geometric evaluation. Estimating the reduction from the intersecting candidates to the final result would still require the additional statistics described above. Thus, the method and claims of this paper remain limited to \textsc{Intersects}, while its estimates can also support query processing for more specific topological predicates.

%%%%%%%%%%%%%%%%%%%%%%%%%%%%%%%%%%%%%%%%%%%%%%%%%%%%%%%%%%%%%%%%%
\setcounter{table}{0}
\setcounter{figure}{0}

\section{Updating TraHist}\label{appendix:update}

\par This section examines maintenance of TraHist's per-bucket statistics under trajectory updates and its accuracy under data shifts. We consider three strategies: \emph{No refresh} reuses both the LabelMap and per-bucket statistics; \emph{Statistics only} fixes the initial LabelMap while updating the statistics for the current trajectories; and \emph{Full rebuild} reconstructs the CountGrid, LabelMap, and statistics from the current trajectories. \Cref{appendix:maintenance} describes exact insertion and deletion under a fixed LabelMap, while \cref{appendix:data-shift} compares these strategies under a controlled data shift.

\subsection{Exact Maintenance under a Fixed LabelMap}\label{appendix:maintenance}
Under \emph{Statistics only}, TraHist fixes the spatial domain, CountGrid cell size $\Delta$, and LabelMap while updating the per-bucket statistics. For each bucket $B_k$, let $A_k$ be its CountGrid cell count, $P_k$ the sum of their point counts, and $U_k$ the sum of their distinct trajectory counts. The means are recovered exactly as $S_p^{(k)}=P_k/A_k$ and $S_T^{(k)}=U_k/A_k$, so the maintained synopsis consists of the LabelMap and $\{A_k,P_k,U_k\}$. Relative to the original synopsis, this replaces $\{S_p^{(k)},S_T^{(k)}\}$ with $P_k$ and $U_k$ and adds only $A_k$. The CountGrid serves as temporary workspace for full rebuilds and is not stored.
\par For a whole trajectory $T$, let $p_k(T)$ be the number of its sampled points mapped to $B_k$, and let $u_k(T)$ be the number of distinct CountGrid cells in $B_k$ occupied by these points. Inserting $T$ adds $p_k(T)$ to $P_k$ and $u_k(T)$ to $U_k$, while deleting $T$ subtracts the same quantities. The occupied cell count $u_k(T)$ is required because $S_T^{(k)}$ averages cell-level distinct trajectory counts. These updates maintain $S_p^{(k)}$ and $S_T^{(k)}$ exactly under the fixed LabelMap. Exact deletion requires either the previous trajectory record or an auxiliary per-trajectory record of $p_k(T)$ and $u_k(T)$ for its visited buckets; the latter is maintained outside the compact query synopsis. A modification is processed as deletion followed by insertion, while updates outside the fixed domain require grid extension or reconstruction.

\begin{table}[t]
    \centering
    \caption{Maintenance under a fixed LabelMap. \emph{State} includes the LabelMap and $\{A_k,P_k,U_k\}$. \emph{Upd.\ pts} reports inserted and deleted points per checkpoint; \emph{Speedup} is rebuild time divided by maintenance time. Values span five checkpoints.}
    \label{tab:maint}
    {%
    \renewcommand{\arraystretch}{1}
    \setlength{\aboverulesep}{0pt}
    \setlength{\belowrulesep}{0pt}
    \setlength{\arrayrulewidth}{0.8pt}
    \setlength{\tabcolsep}{5pt}
    \newcommand{\topgap}{\rule{0pt}{2.7ex}}
    \newcommand{\botgap}{\rule[-1.5ex]{0pt}{0pt}}
    \resizebox{0.73\textwidth}{!}{%
    \begin{tabular}{@{}l@{\hspace{6pt}}|@{\hspace{6pt}}ccccc@{\hspace{6pt}}}
        \toprule[0.8pt]
        \topgap\botgap\textbf{Dataset} & \textbf{State (KB)} & \textbf{Upd.\ pts (M)} & \textbf{Maint.\ (s)} & \textbf{Rebuild (s)} & \textbf{Speedup} \\
        \midrule[0.8pt]
        \topgap\textbf{\textit{Porto}}   & 53.8--54.2 & 15.2--16.6 & 10.1--10.8 & 23.4--26.6 & 2.3--2.5$\times$ \\
        \textbf{\textit{T-Drive}}        & 64.4--64.7 & 3.1--3.8   & 1.2--1.4   & 3.2--3.4   & 2.5--2.8$\times$ \\
        \botgap\textbf{\textit{GeoLife}} & 57.4--59.7 & 3.2--4.7   & 1.0--1.4   & 2.7--3.2   & 1.9--3.0$\times$ \\
        \bottomrule[0.8pt]
    \end{tabular}%
    }
    }
\end{table}
\par \Cref{tab:maint} reports the synopsis state and processing cost of incremental maintenance and compares it with rebuilding TraHist at the same checkpoint. We use the CountGrid cell sizes that produce approximately 50~KB TraHist synopses on the full datasets, fix their spatial domains, and provide the previous trajectory record for each deletion. Each checkpoint replaces approximately one fifth of the stored trajectories while keeping their total number nearly constant, involving 3.1--16.6 million inserted and deleted points. The process time includes computing $p_k(T)$ and $u_k(T)$ from these records and applying the per-bucket updates, but excludes record retrieval and access to the optional per-trajectory structure. Adding $A_k$ increases the original 49.1--55.0~KB synopsis by 6.2--17.9\%, yielding the 53.8--64.7~KB states reported in the table. Incremental maintenance takes 1.0--10.8 seconds and is 1.9--3.0$\times$ faster than full rebuild because it reuses the fixed LabelMap and avoids CountGrid construction and bucket formation.

\begin{figure}[t]
    \centering
    \includegraphics[width=\linewidth]{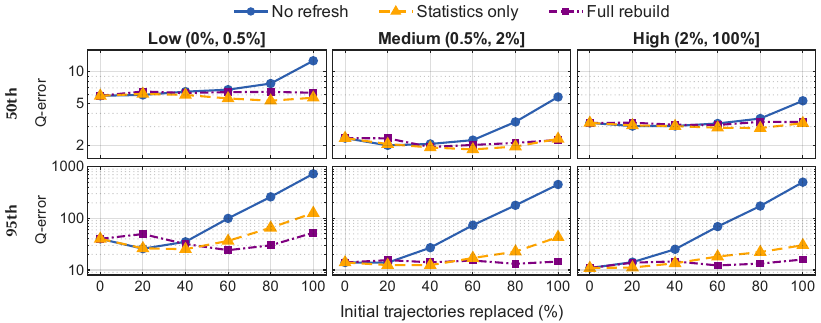}
    \caption{Estimation error under a controlled data shift on \textit{Porto}. Median (top) and 95th percentile (bottom) trajectory-level Q-error are shown for low, medium, and high selectivity as 0--100\% of the initial trajectories are replaced. Lines compare \emph{No refresh}, \emph{Statistics only}, and \emph{Full rebuild}.}
    \Description{Two rows of line plots showing trajectory-level Q-error on a logarithmic scale as the fraction of replaced initial trajectories increases from 0 to 100 percent on Porto. The top row shows the median and the bottom row the 95th percentile; the columns show low, medium, and high selectivity. Three lines compare no refresh, statistics-only refresh, and full rebuild.}
    \label{fig:datashift}
\end{figure}

\begin{table}[t]
    \centering
    \caption{Trajectory-level Q-error before and after complete trajectory replacement. \emph{Initial} is shared across modes; other columns show checkpoint~5. Full rebuild may produce a different bucket count.}
    \label{tab:datashift}
    {%
    \renewcommand{\arraystretch}{1}
    \setlength{\aboverulesep}{0pt}
    \setlength{\belowrulesep}{0pt}
    \setlength{\arrayrulewidth}{0.8pt}
    \setlength{\tabcolsep}{3pt}
    \newcommand{\topgap}{\rule{0pt}{2.7ex}}
    \newcommand{\botgap}{\rule[-1.5ex]{0pt}{0pt}}
    \ifdefined\qcolw\else\newlength{\qcolw}\fi
    \settowidth{\qcolw}{Rebuild}
    \newcommand{\hd}[1]{\makebox[\qcolw][c]{#1}}
    \ifdefined\qva\else\newlength{\qva}\fi
    \settowidth{\qva}{00.0}
    \newcommand{\va}[1]{\makebox[\qva][r]{#1}}
    \ifdefined\qvb\else\newlength{\qvb}\fi
    \settowidth{\qvb}{\textbf{0.0}}
    \newcommand{\vb}[1]{\makebox[\qvb][r]{#1}}
    \ifdefined\qvc\else\newlength{\qvc}\fi
    \settowidth{\qvc}{000}
    \newcommand{\vc}[1]{\makebox[\qvc][r]{#1}}
    \ifdefined\qvd\else\newlength{\qvd}\fi
    \settowidth{\qvd}{\textbf{00}}
    \newcommand{\vd}[1]{\makebox[\qvd][r]{#1}}
    \resizebox{0.80\textwidth}{!}{%
    \begin{tabular}{@{}c@{\hspace{6pt}}l@{\hspace{6pt}}|@{\hspace{6pt}}c@{\hspace{8pt}}ccc@{\hspace{7pt}}c@{\hspace{7pt}}c@{\hspace{8pt}}ccc@{\hspace{6pt}}}
        \toprule[0.8pt]
        \multirow{2}{*}{\textbf{Dataset}}
        & \multirow{2}{*}{\textbf{Sel.}}
        & \multicolumn{4}{c}{\topgap\textbf{50th Q-error}}
        && \multicolumn{4}{c}{\textbf{95th Q-error}} \\
        \cmidrule(r{7pt}){3-6} \cmidrule(r{6pt}){8-11}
        & & \topgap\botgap\hd{Initial} & \hd{No ref.} & \hd{Maint.} & \hd{Rebuild}
        && \hd{Initial} & \hd{No ref.} & \hd{Maint.} & \hd{Rebuild} \\
        \midrule[0.8pt]
        \multirow{3}{*}{\textbf{\textit{Porto}}} & \topgap Low & \va{5.9} & \va{12.6} & \vb{\best{5.6}} & \va{\secondbest{6.3}}
            && \vc{40} & \vc{727} & \vc{\secondbest{127}} & \vd{\best{53}} \\
         & Med & \va{2.3} & \va{5.7} & \vb{\secondbest{2.3}} & \va{\best{2.2}}
            && \vc{14} & \vc{452} & \vc{\secondbest{43}} & \vd{\best{15}} \\
         & \botgap High & \va{3.3} & \va{5.3} & \vb{\best{3.2}} & \va{\secondbest{3.3}}
            && \vc{11} & \vc{501} & \vc{\secondbest{30}} & \vd{\best{16}} \\
        \midrule
        \multirow{3}{*}{\textbf{\textit{T-Drive}}} & \topgap Low & \va{9.4} & \va{10.5} & \vb{\best{8.3}} & \va{\secondbest{10.2}}
            && \vc{42} & \vc{92} & \vc{\secondbest{83}} & \vd{\best{43}} \\
         & Med & \va{2.8} & \va{3.0} & \vb{\best{2.4}} & \va{\secondbest{2.9}}
            && \vc{12} & \vc{77} & \vc{\secondbest{75}} & \vd{\best{14}} \\
         & \botgap High & \va{1.7} & \va{\secondbest{1.5}} & \vb{\best{1.4}} & \va{1.7}
            && \vc{14} & \vc{\secondbest{129}} & \vc{199} & \vd{\best{22}} \\
        \midrule
        \multirow{3}{*}{\textbf{\textit{GeoLife}}} & \topgap Low & \va{10.5} & \va{12.5} & \vb{\secondbest{7.4}} & \va{\best{6.3}}
            && \vc{113} & \vc{150} & \vc{\secondbest{96}} & \vd{\best{90}} \\
         & Med & \va{2.8} & \va{4.9} & \vb{\secondbest{2.4}} & \va{\best{2.2}}
            && \vc{30} & \vc{79} & \vc{\secondbest{24}} & \vd{\best{19}} \\
         & \botgap High & \va{2.6} & \va{\best{3.3}} & \vb{\secondbest{3.4}} & \va{\secondbest{3.4}}
            && \vc{36} & \vc{236} & \vc{\secondbest{105}} & \vd{\best{47}} \\
        \bottomrule[0.8pt]
    \end{tabular}%
    }
    }
\end{table}
\subsection{Accuracy under Data Shift}\label{appendix:data-shift}
To evaluate TraHist under a controlled data shift, we use the update protocol and approximately 50~KB configurations from \cref{appendix:maintenance}. We apply PCA to trajectory centroids solely to order trajectories by their projection onto the first principal component, divide them into approximately equal groups, and replace these groups sequentially. This gradually shifts the distribution along its dominant spatial direction. Starting from checkpoint~0, five updates increase trajectory replacement from 20\% to 100\%.
\par \emph{No refresh} and \emph{Statistics only} share the fixed LabelMap and differ only in whether the per-bucket statistics are updated, so their comparison isolates the effect of stale statistics under the same partition. \emph{Full rebuild} instead reconstructs the partition from the current trajectories under the same spatial domain and construction parameters. Since the resulting bucket count is data-dependent, the rebuilt synopsis can differ in size; this comparison therefore measures the practical benefit of rebuilding rather than a partition-only effect under an identical synopsis size.
\par \Cref{fig:datashift} shows how these differences develop as the shift accumulates on \textit{Porto}. Before any trajectories are replaced, the three strategies coincide, and with limited replacement their differences remain small and show no consistent ordering. As more trajectories are replaced, however, leaving the synopsis unchanged substantially increases estimation error. At complete replacement, the median Q-errors under \emph{Statistics only} are 5.6, 2.3, and 3.2 for low, medium, and high selectivity, respectively, close to the 6.3, 2.2, and 3.3 obtained under \emph{Full rebuild}; \emph{No refresh} yields 12.6, 5.7, and 5.3. The difference is larger in the tail: the corresponding 95th percentile Q-errors are 53, 15, and 16 under \emph{Full rebuild}, 127, 43, and 30 under \emph{Statistics only}, and 727, 452, and 501 under \emph{No refresh}.

\par \Cref{tab:datashift} shows that this complete-replacement pattern extends across the three datasets. Across the nine dataset and selectivity combinations, the median Q-error under \emph{Statistics only} remains close to that under \emph{Full rebuild}, with their ratio ranging from 0.81 to 1.19. \emph{Statistics only} also lowers the 95th percentile Q-error relative to \emph{No refresh} in eight of the nine combinations, while \emph{Full rebuild} achieves the lowest 95th percentile error in all nine. Thus, updating only the per-bucket statistics generally provides median accuracy comparable to \emph{Full rebuild} at lower processing cost, whereas rebuilding TraHist becomes preferable after substantial trajectory replacement when tail accuracy is important.

%%%%%%%%%%%%%%%%%%%%%%%%%%%%%%%%%%%%%%%%%%%%%%%%%%%%%%%%%%%%%%%%%
\setcounter{table}{0}
\setcounter{figure}{0}

\section{Derivations for Theoretical Properties}
\label{appendix:proofs}

%%%
This appendix presents the derivations underlying the theoretical results in~\cref{subsec:theory}.

\noindentparagraph{\textnormal{\textbf{Local optimality.}}}
For the final partition $\Pi$, consider reassigning a cell to a neighboring region while keeping the seed locations and local contrast costs unchanged. We compare the minimum cumulative costs of reaching the cell from its current region and from the neighboring region. During propagation, TraHist updates a cell's cumulative cost whenever a path with a lower cost is found and reinserts the cell into the priority queue. Thus, at termination, $\bar D(n)\le\bar D(x)+D(x,n)$ holds for every pair of adjacent cells. Each cell’s stored cost is the accumulated contrast along the propagation path that assigns its label. Applying this inequality along propagation paths shows that no alternative path has lower cost than the stored value. Consequently, reassigning a cell to a neighboring region cannot reduce $E(\Pi)$, establishing local optimality.

\noindentparagraph{\textnormal{\textbf{Marginal cost of cell assignment.}}}
Let
\begin{equation}\label{eq:app-bucket-mean}
\bar{\rho}_{p,k} =
\frac{1}{|B_k|}\sum_{x\in B_k}\rho_p(x)
\end{equation}
denote the mean density within bucket $B_k$. When a new cell $x$ is added to $B_k$, the increase in the intra-bucket SSE objective $J(\Pi)$ (\cref{eq:sse}) is
\begin{equation}
\Delta J_k = \frac{|B_k|}{|B_k|+1} \bigl(\rho_p(x) - \bar{\rho}_{p,k}\bigr)^2.
\end{equation}
Thus, minimizing $\Delta J_k$ amounts to minimizing $\bigl|\rho_p(x) - \bar{\rho}_{p,k}\bigr|$. TraHist replaces this with the locally computable contrast $D(x,n)$ (\cref{eq:local_cost}), which avoids recomputing bucket means at each step.

\noindentparagraph{\textnormal{\textbf{Proof of the deviation bound (\cref{eq:deviation}).}}} For a candidate cell $x$ and an adjacent cell $n\in B_k$, expanding $D(x,n)$ (\cref{eq:local_cost}) and applying the triangle inequality with respect to the bucket mean $\bar{\rho}_{p,k}$ yields
\begin{equation*}
\begin{aligned}
&\bigl|D(x,n)
      -\bigl|\rho_p(x)-\bar{\rho}_{p,k}\bigr|\bigr|\\
&\quad\le (1-\alpha)
\bigl(|\mu(x)-\rho_p(n)|+|\mu(n)-\rho_p(x)|\bigr)
+|\rho_p(n)-\bar{\rho}_{p,k}|\\
&\quad\le (1-\alpha)(\delta_x+\delta_n)+\epsilon_k.
\end{aligned}
\end{equation*}
The last inequality follows from $n\in\mathcal N(x)$, $x\in\mathcal N(n)$, and the definitions of $\delta_x$, $\delta_n$, and $\epsilon_k$ in~\cref{subsec:theory}. The first term arises from local smoothing, while the second term reflects intra-bucket non-uniformity.

\noindentparagraph{\textnormal{\textbf{Bounding $\epsilon_k$ via cumulative propagation cost.}}}
For $0 < \alpha \le 1$, let $s_k$ be the seed of $B_k$ and define $\bar D_{\max}(B_k)=\max_{y\in B_k}\bar D(y)$. For any cell $y \in B_k$, each edge $(u,v)$ along its propagation path from $s_k$ contributes at least $\alpha|\rho_p(u)-\rho_p(v)|$ to $\bar D(y)$. Repeated application of the triangle inequality yields
\begin{equation}\label{eq:app-seed-deviation}
|\rho_p(y)-\rho_p(s_k)|
\le \frac{1}{\alpha}\bar D(y)
\le \frac{1}{\alpha}\bar D_{\max}(B_k).
\end{equation}
Since $\bar{\rho}_{p,k}$ lies between the extrema of $\rho_p$ within $B_k$, it follows that
\begin{equation}\label{eq:app-internal-deviation-bound}
\epsilon_k
\le \frac{2}{\alpha}\bar D_{\max}(B_k).
\end{equation}
Together with~\cref{eq:deviation}, this bound provides a theoretical basis for using $D(x,n)$ as a surrogate for the ideal local objective, with approximation error controlled by local density variation and cumulative propagation cost.

%%%%%%%%%%%%%%%%%%%%%%%%%%%%%%%%%%%%%%%%%%%%%%%%%%%%%%%%%%%%%%%%%
\setcounter{table}{0}
\setcounter{figure}{0}

\section{Estimation Accuracy on the \textit{GeoLife} Dataset}\label{appendix:geolife}

\begin{figure}[t]
    \centering
    \includegraphics[width=\linewidth]{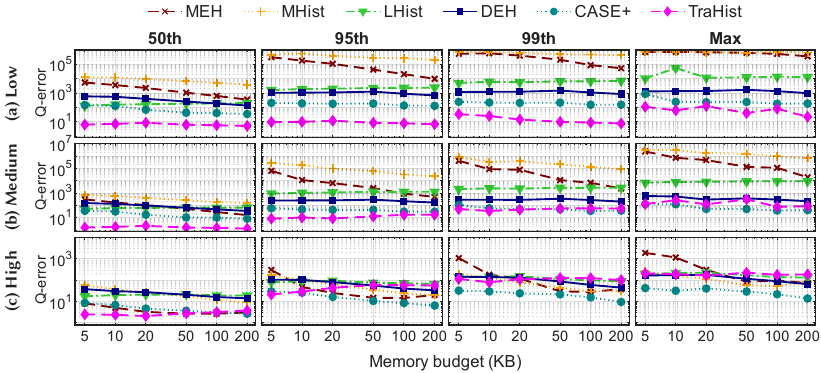}
    \caption{Q-error for trajectory range queries on \textit{GeoLife}. Median and tail Q-error are reported across \textit{memory budgets} for (a) low (0\%, 0.5\%], (b) medium (0.5\%, 2\%], and (c) high (2\%, 100\%] selectivity queries.}
    \Description{Three rows of line plots showing Q-error on a logarithmic scale versus memory budget for six estimators on GeoLife. Rows correspond to low, medium, and high selectivity queries, and columns to the 50th, 95th, and 99th percentile and maximum Q-error.}
    \label{fig:qerror-DD-GeoLife-memory}
\end{figure}

\begin{figure}[t]
    \centering
    \includegraphics[width=\linewidth]{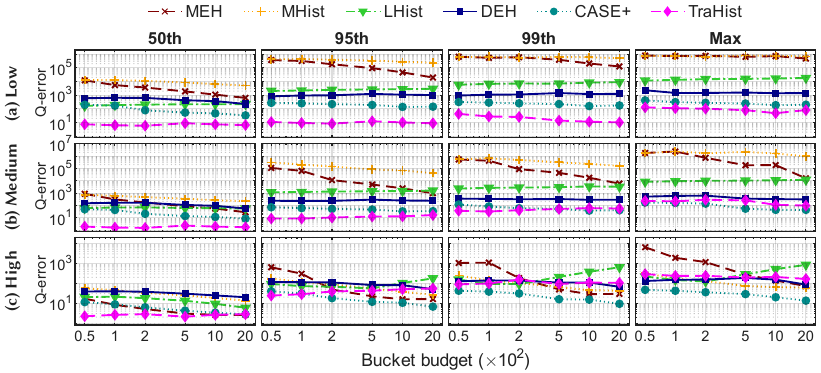}
    \caption{Q-error for trajectory range queries on \textit{GeoLife}. Median and tail Q-error are reported across \textit{bucket budgets} for (a) low (0\%, 0.5\%], (b) medium (0.5\%, 2\%], and (c) high (2\%, 100\%] selectivity queries.}
    \Description{Three rows of line plots showing Q-error on a logarithmic scale versus bucket budget for six estimators on GeoLife. Rows correspond to low, medium, and high selectivity queries, and columns to the 50th, 95th, and 99th percentile and maximum Q-error.}
    \label{fig:qerror-DD-GeoLife-buckets}
\end{figure}

To assess the consistency of TraHist accuracy beyond traffic trajectory datasets, we evaluate on the \textit{GeoLife} dataset, which captures diverse human mobility patterns---including walking, cycling, bus, and driving---from 182 users in the Beijing metropolitan area~\cite{geolife1, geolife_userguide}. Unlike \textit{Porto} and \textit{T-Drive}, which primarily capture structured vehicular movements, \textit{GeoLife} exhibits greater heterogeneity in trajectory length and sampling frequency. This contrast provides a complementary test of whether TraHist's accuracy trends persist beyond road-constrained vehicular trajectories.
\par Figures \ref{fig:qerror-DD-GeoLife-memory} and~\ref{fig:qerror-DD-GeoLife-buckets} report the Q-error results under fixed memory and fixed bucket budgets, respectively. Consistent with the main experiments, TraHist achieves substantially lower estimation error than competing methods under low and medium selectivity queries across all evaluated settings. In the high selectivity regime, the accuracy gap between estimators narrows, consistent with the trends observed on \textit{Porto} and \textit{T-Drive} (\cref{subsubsec:high-selectivity}). TraHist continues to achieve single-digit median Q-error across all configurations on \textit{GeoLife}, ranging from 2.1 to 3.8, and remains within 1.4$\times$ of the lowest median at every budget. Its median accuracy advantage over LHist is more pronounced than on the taxi datasets: TraHist yields 2$\times$--9$\times$ lower median Q-error than LHist across all memory and bucket budgets on \textit{GeoLife}, whereas on \textit{Porto} and \textit{T-Drive} LHist occasionally attains lower median error at large query scales. The tail ranking differs, however: CASE+ achieves the lowest 99th percentile and maximum Q-error at every evaluated budget and the lowest 95th percentile Q-error at most budgets. Thus, TraHist remains competitive in median accuracy, while CASE+ provides stronger tail control under high selectivity.
\par We attribute these patterns to two factors. First, the heterogeneous trajectory distribution in \textit{GeoLife} likely limits the effectiveness of LHist's intra-bucket regression models in capturing local density variation, preserving TraHist's relative advantage. Second, CASE+'s stronger tail control under high selectivity is consistent with the reduced role of fragmentation as query regions span more buckets. Overall, these results confirm that the accuracy benefits of TraHist persist across datasets with fundamentally different trajectory characteristics, including both road-constrained vehicle trajectories and heterogeneous human mobility data.

%%%%%%%%%%%%%%%%%%%%%%%%%%%%%%%%%%%%%%%%%%%%%%%%%%%%%%%%%%%%%%%%%

\setcounter{table}{0}
\setcounter{figure}{0}

\section{Robustness and Scalability Analysis} \label{appendix:robustness}
To evaluate the robustness and scalability of TraHist, we examine trajectory subsampling and query region size across varying CountGrid cell sizes $\Delta$, and assess the accuracy and resource costs as $\Delta$ decreases. We also examine whether a sparse storage representation can mitigate the storage growth associated with smaller CountGrid cells without changing the underlying partition. Unless otherwise specified, we use $\alpha=0.5$, $\beta=1.0$, and $\kappa=3$.

%%%%%%%%%%%%%%%%%%%%%%%%D1 Subsampling%%%%%%%%%%%%%%%%%%%%%%%%%%%%%%%%%%%%%%%%%

\begin{figure}[t]
    \centering
    \includegraphics[width=\linewidth]{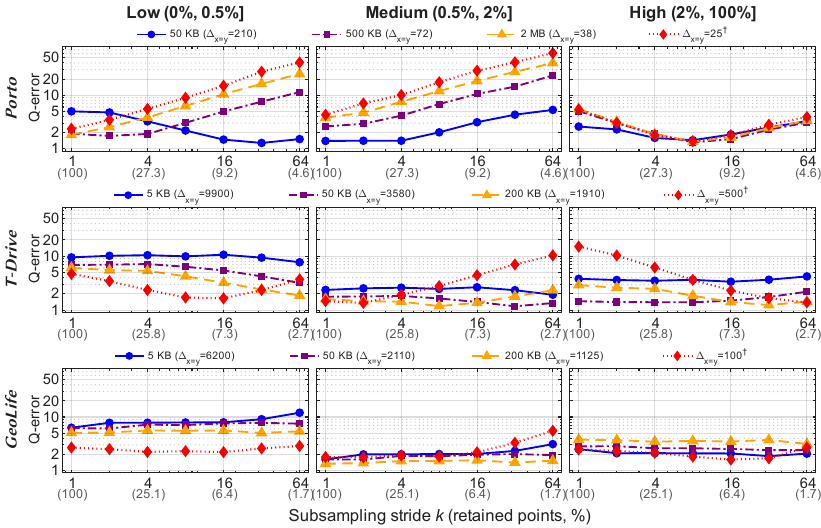}
    \caption{Median Q-error of TraHist under trajectory subsampling. Parentheses show the percentage of retained points. The CountGrid cell size is $\Delta=(\Delta_x,\Delta_y)$; in the legend, $\Delta_{x=y}$ indicates $\Delta_x=\Delta_y$, with values in meters. $\dagger$ marks the additional smaller cell size.}
    \Description{Three rows of log-log line plots, one row each for Porto, T-Drive, and GeoLife, showing the median Q-error of TraHist as the subsampling stride increases from 1 to 64. Columns correspond to low, medium, and high selectivity queries, and each panel compares four CountGrid cell sizes.}
    \label{fig:subsampling}
\end{figure}

\subsection{Robustness to Trajectory Subsampling} \label{appendix:subsampling}
To evaluate TraHist under increasingly sparse samples, we retain every $k$-th point of each trajectory for $k\in\{1,2,4,8,16,32,64\}$, while preserving both endpoints. For each $k$ and $\Delta$, we rebuild TraHist using the setup in~\cref{sec:experimental_study}, while keeping the spatial domain, query workload, and ground truth cardinalities computed from the non-subsampled trajectories fixed. For each dataset, \cref{fig:subsampling} reports three grid cell sizes corresponding to the memory budgets used in our main and extended evaluations, together with one additional smaller cell size (marked $\dagger$) without a corresponding target memory budget.
\par TraHist remains robust to trajectory subsampling across the evaluated cell sizes. At $k=64$, only 4.62\%, 2.74\%, and 1.69\% of the original points remain on \textit{Porto}, \textit{T-Drive}, and \textit{GeoLife}, respectively. Even under this aggressive subsampling, median Q-error remains low across most evaluated memory budgets on all three datasets, and high selectivity estimates are affected even less, remaining stable across subsampling strides and cell sizes. Together, these results indicate that the aggregate CountGrid retains sufficient spatial density information for accurate estimation even when only a small fraction of trajectory samples is retained.
\par The accuracy retained under subsampling remains competitive even against estimators built from the full trajectories. On \textit{T-Drive} and \textit{GeoLife}, TraHist at $k=64$ remains the most accurate estimator across all evaluated memory budgets under low and medium selectivity. At 50\,KB, its median Q-error is 18.1$\times$ and 16.6$\times$ lower than LHist and CASE+ under low and medium selectivity, respectively, on \textit{T-Drive}, and 5.7$\times$ and 5.5$\times$ lower than CASE+ on \textit{GeoLife}. On \textit{Porto} at 50\,KB, TraHist achieves 12.7$\times$ lower median Q-error than DEH under low selectivity and comparable accuracy under medium selectivity (5.3 versus 5.2). Notably, these comparisons favor the competing estimators, which use all trajectory points, whereas TraHist uses only a small fraction of them.
\par Subsampling degrades accuracy when $\Delta$ becomes sufficiently small that each CountGrid cell retains too few points to represent its local density, an effect most apparent on \textit{Porto} under low and medium selectivity. At $\Delta=(72, 72)$, increasing $k$ from 1 to 64 raises the median Q-error from 1.90 to 11.41 under low selectivity and from 2.60 to 23.40 under medium selectivity, and the additional smaller cell size $\Delta = (25, 25)$ further increases the corresponding values to 40.91 and 61.30. Across the evaluated range, \textit{T-Drive} and \textit{GeoLife} show substantially weaker resolution dependence. This contrast shows that the effect of subsampling varies with CountGrid resolution rather than uniformly with the retained sample count.
\par Overall, TraHist remains robust to substantial subsampling across a broad range of cell sizes, with degradation occurring mainly when fine-grid cells contain too few observations. Its accuracy therefore depends primarily on CountGrid resolution relative to the retained sampling density rather than on sample count alone.

%%%%%%%%%%%%%%%%%%%%%%%%D2 Query Size%%%%%%%%%%%%%%%%%%%%%%%%%%%%%%%%%%%%%%%%%

\begin{figure}[t]
    \centering
    \includegraphics[width=\linewidth]{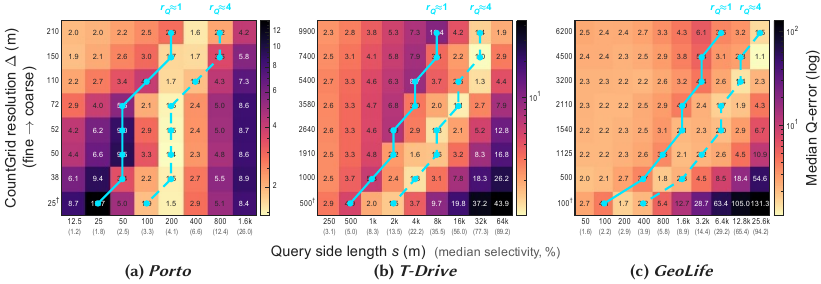}
    \caption{Median Q-error under varying query side length $s$ and CountGrid cell size $\Delta_{x=y}$. The cyan lines connect the evaluated query sizes closest to $r_Q=1$ and $r_Q=4$, where $r_Q=s/\Delta_{x=y}$. Parenthesized values report median selectivity (\%). $\dagger$ marks the additional smallest cell size.}
    \Description{Three heatmaps for Porto, T-Drive, and GeoLife showing the median Q-error of TraHist for each combination of CountGrid cell size (rows) and query side length (columns), with the value printed in each cell. Cyan guide lines connect the query sizes closest to a query-to-grid ratio of 1 and 4.}
    \label{fig:qsize}
\end{figure}

\subsection{Sensitivity to Query Region Size} \label{appendix:query-size}
Building on~\cref{appendix:subsampling}, we next examine how query region size relative to the CountGrid cell size $\Delta$ affects estimation accuracy. Using 1,000 fixed query centers following~\cref{sec:experimental_study}, we evaluate square queries with side length $s$ across the same cell sizes as in~\cref{fig:subsampling}. For each dataset, we double $s$ from half the smallest evaluated cell width until it reaches at least four times the largest cell width. For square CountGrid cells, we define $r_Q=s/\Delta_{x=y}$, where $\Delta_{x=y}=\Delta_x=\Delta_y$; $r_Q$ measures the query side length in units of cell width. We use $r_Q\approx1$ and $r_Q\approx4$ as representative points for queries with side lengths comparable to one and four cell widths, respectively.
\par \Cref{fig:qsize} shows a consistent dependence on query size relative to the CountGrid cell width. The query size nearest to $r_Q=4$ yields lower median Q-error than that nearest to $r_Q=1$ at all evaluated cell sizes on \textit{T-Drive} and \textit{GeoLife}, and at seven of eight cell sizes on \textit{Porto}. At larger $r_Q$, median Q-error increases again for most evaluated cell sizes, particularly at finer resolutions where a query covers many cells; such large queries also fall in the high selectivity range discussed in~\cref{subsec:cardinality-estimation-accuracy}.

\begin{table}[t]
    \centering
    \caption{Query-size sensitivity at the coarsest matched-memory and additional finest ($\dagger$) CountGrid resolutions, near $r_Q=1$ and $r_Q=4$. Sel., Med., and Log denote median selectivity (\%), median Q-error, and median $\log_2(\widehat{y}/y)$; negative and positive Log indicate under- and overestimation.}
    \label{tab:query-size}
    {%
    \renewcommand{\arraystretch}{1}
    \setlength{\aboverulesep}{0pt}
    \setlength{\belowrulesep}{0pt}
    \setlength{\arrayrulewidth}{0.8pt}
    \setlength{\tabcolsep}{3pt}
    \newcommand{\topgap}{\rule{0pt}{2.7ex}}
    \newcommand{\botgap}{\rule[-1.5ex]{0pt}{0pt}}
    \ifdefined\qcolw\else\newlength{\qcolw}\fi
    \settowidth{\qcolw}{$-0.00$}
    \newcommand{\hd}[1]{\makebox[\qcolw][c]{#1}}
    \ifdefined\qhd\else\newlength{\qhd}\fi
    \settowidth{\qhd}{Sel.}
    \ifdefined\qhl\else\newlength{\qhl}\fi
    \settowidth{\qhl}{Log}
    \newcommand{\qlt}{\dimexpr(\qcolw-\qhd)/2\relax}
    \newcommand{\qrta}{\dimexpr5pt+(\qcolw-\qhl)/2\relax}
    \newcommand{\qrtb}{\dimexpr6pt+(\qcolw-\qhl)/2\relax}
    \ifdefined\qva\else\newlength{\qva}\fi
    \settowidth{\qva}{00.00}
    \newcommand{\va}[1]{\makebox[\qva][r]{#1}}
    \ifdefined\qvb\else\newlength{\qvb}\fi
    \settowidth{\qvb}{0.00}
    \newcommand{\vb}[1]{\makebox[\qvb][r]{#1}}
    \ifdefined\qvl\else\newlength{\qvl}\fi
    \settowidth{\qvl}{$-0.00$}
    \newcommand{\vl}[1]{\makebox[\qvl][r]{#1}}
    \ifdefined\qdw\else\newlength{\qdw}\fi
    \settowidth{\qdw}{0000}
    \newcommand{\dv}[1]{\makebox[\qdw][r]{#1}}
    \resizebox{0.55\textwidth}{!}{%
    \begin{tabular}{@{}c@{\hspace{6pt}}c@{\hspace{6pt}}|@{\hspace{6pt}}ccc@{\hspace{5pt}}c@{\hspace{5pt}}ccc@{\hspace{6pt}}}
        \toprule[0.8pt]
        \multirow{2}{*}{\textbf{Dataset}}
        & \multirow{2}{*}{\begin{tabular}[c]{@{}c@{}}$\boldsymbol{\Delta_{x=y}}$\\\textbf{(m)}\end{tabular}}
        & \multicolumn{3}{c}{\topgap$\boldsymbol{r_Q\approx1}$}
        && \multicolumn{3}{c}{$\boldsymbol{r_Q\approx4}$} \\
        \cmidrule(l{\qlt}r{\qrta}){3-5} \cmidrule(l{\qlt}r{\qrtb}){7-9}
        & & \topgap\botgap\hd{Sel.} & \hd{Med.} & \hd{Log}
        && \hd{Sel.} & \hd{Med.} & \hd{Log} \\
        \midrule[0.8pt]
        \multirow{2}{*}{\textbf{\textit{Porto}}}
        & \topgap\dv{210} & \va{4.05} & \va{2.87} & \vl{$-1.37$} && \va{12.36} & \vb{2.18} & \vl{$+1.13$} \\
        & \botgap\dv{25}\rlap{$^{\dagger}$} & \va{1.81} & \va{13.66} & \vl{$-3.77$} && \va{3.31} & \vb{1.92} & \vl{$-0.81$} \\
        \midrule
        \multirow{2}{*}{\textbf{\textit{T-Drive}}}
        & \topgap\dv{9900} & \va{35.46} & \va{10.42} & \vl{$-3.35$} && \va{77.29} & \vb{1.41} & \vl{$-0.47$} \\
        & \botgap\dv{500}\rlap{$^{\dagger}$} & \va{5.04} & \va{4.92} & \vl{$-2.29$} && \va{13.45} & \vb{1.51} & \vl{$+0.53$} \\
        \midrule
        \multirow{2}{*}{\textbf{\textit{GeoLife}}}
        & \topgap\dv{6200} & \va{29.15} & \va{3.62} & \vl{$-1.52$} && \va{94.25} & \vb{1.51} & \vl{$-0.56$} \\
        & \botgap\dv{100}\rlap{$^{\dagger}$} & \va{2.19} & \va{4.14} & \vl{$-2.06$} && \va{3.91} & \vb{2.16} & \vl{$+1.00$} \\
        \bottomrule[0.8pt]
    \end{tabular}%
    }
    }
\end{table}

\par To quantify this pattern, \cref{tab:query-size} compares the evaluated query sizes nearest to $r_Q=1$ and $r_Q=4$ at the coarsest and finest resolutions ($\dagger$) for each dataset. At the finest resolutions, median Q-error decreases from 13.66 to 1.92 on \textit{Porto}, from 4.92 to 1.51 on \textit{T-Drive}, and from 4.14 to 2.16 on \textit{GeoLife}; the same improvement occurs at the coarsest resolutions. The median log ratio $\log_2(\widehat{y}_q/y_q)$ is negative near $r_Q=1$ at both reported resolutions on all three datasets ($-1.37$ to $-3.77$), but moves closer to zero near $r_Q=4$ ($-0.81$ to $+1.13$), showing that the tendency toward underestimation also decreases as queries span several cell widths.
\par Together with~\cref{appendix:subsampling}, these results show that estimation accuracy depends on $\Delta$ relative to both the retained sampling density and the target query sizes. Smaller cells allow a query to span more cells, but can leave insufficient observations within each cell. We next examine the accuracy and resource costs as $\Delta$ decreases.

%%%%%%%%%%%%%%%%%%%%%%%%D3 Delta %%%%%%%%%%%%%%%%%%%%%%%%%%%%%%%%%%%%%%%%%

\begin{figure}[t]
    \centering
    \ifdefined\qdw\else\newlength{\qdw}\fi
    \ifdefined\qdh\else\newlength{\qdh}\fi
    \ifdefined\qcbw\else\newlength{\qcbw}\fi
    \setlength{\qdw}{0.22\linewidth}
    \settoheight{\qdh}{\includegraphics[width=\qdw]{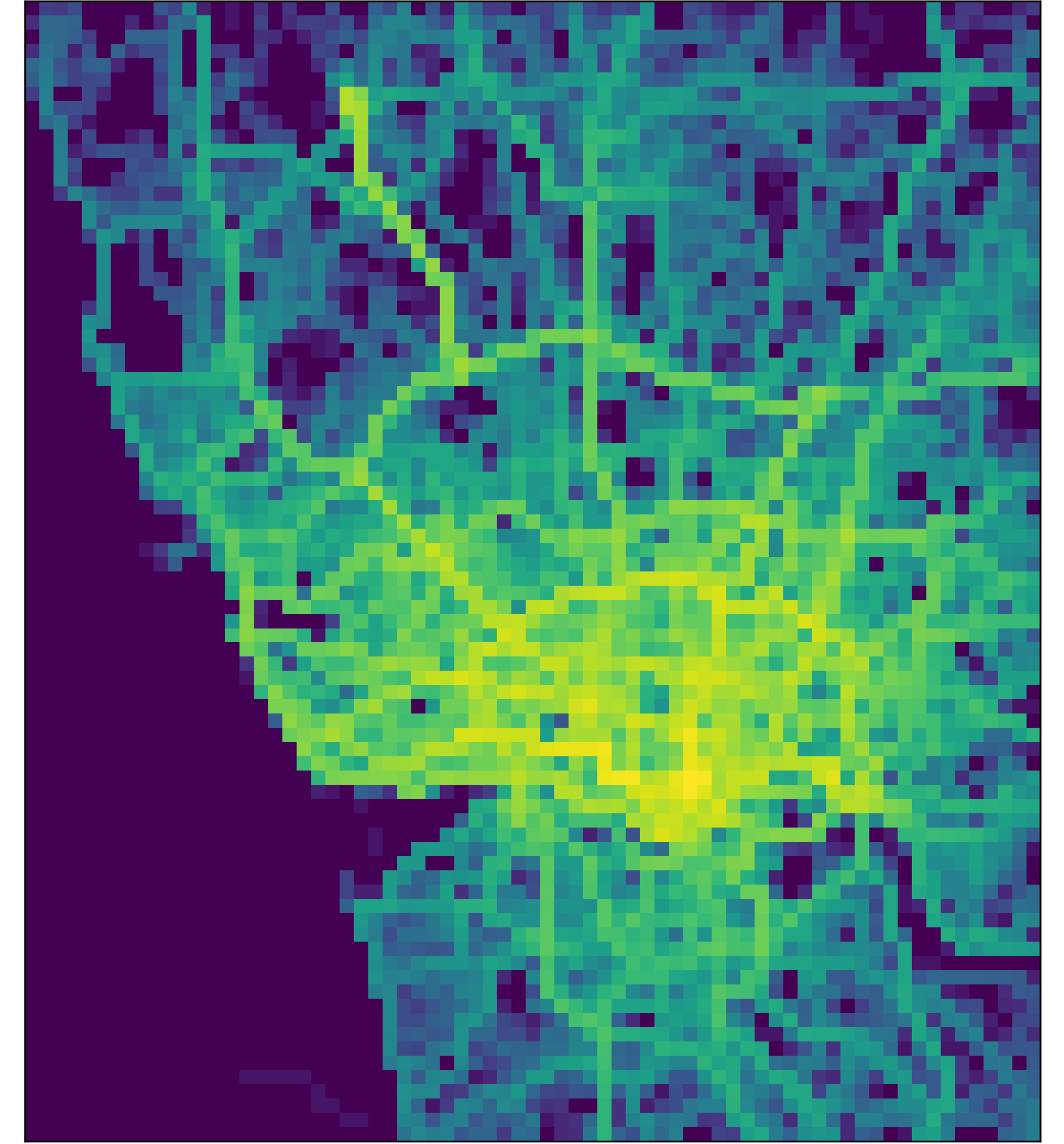}}
    \settowidth{\qcbw}{\includegraphics[height=\qdh]{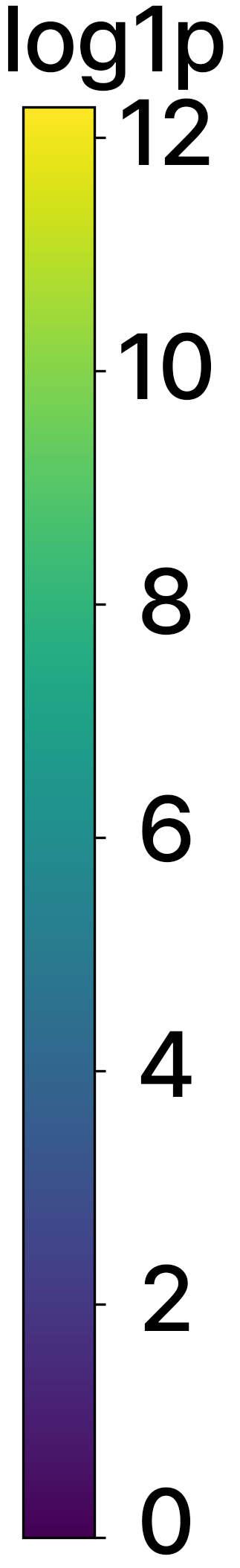}}
    \begin{subfigure}[b]{\qdw}
        \centering
        \includegraphics[width=\linewidth]{figures/appendix/delta210.png}
        \caption{$\Delta{=}(210,210)$}
    \end{subfigure}\hspace{4pt}%
    \begin{subfigure}[b]{\qdw}
        \centering
        \includegraphics[width=\linewidth]{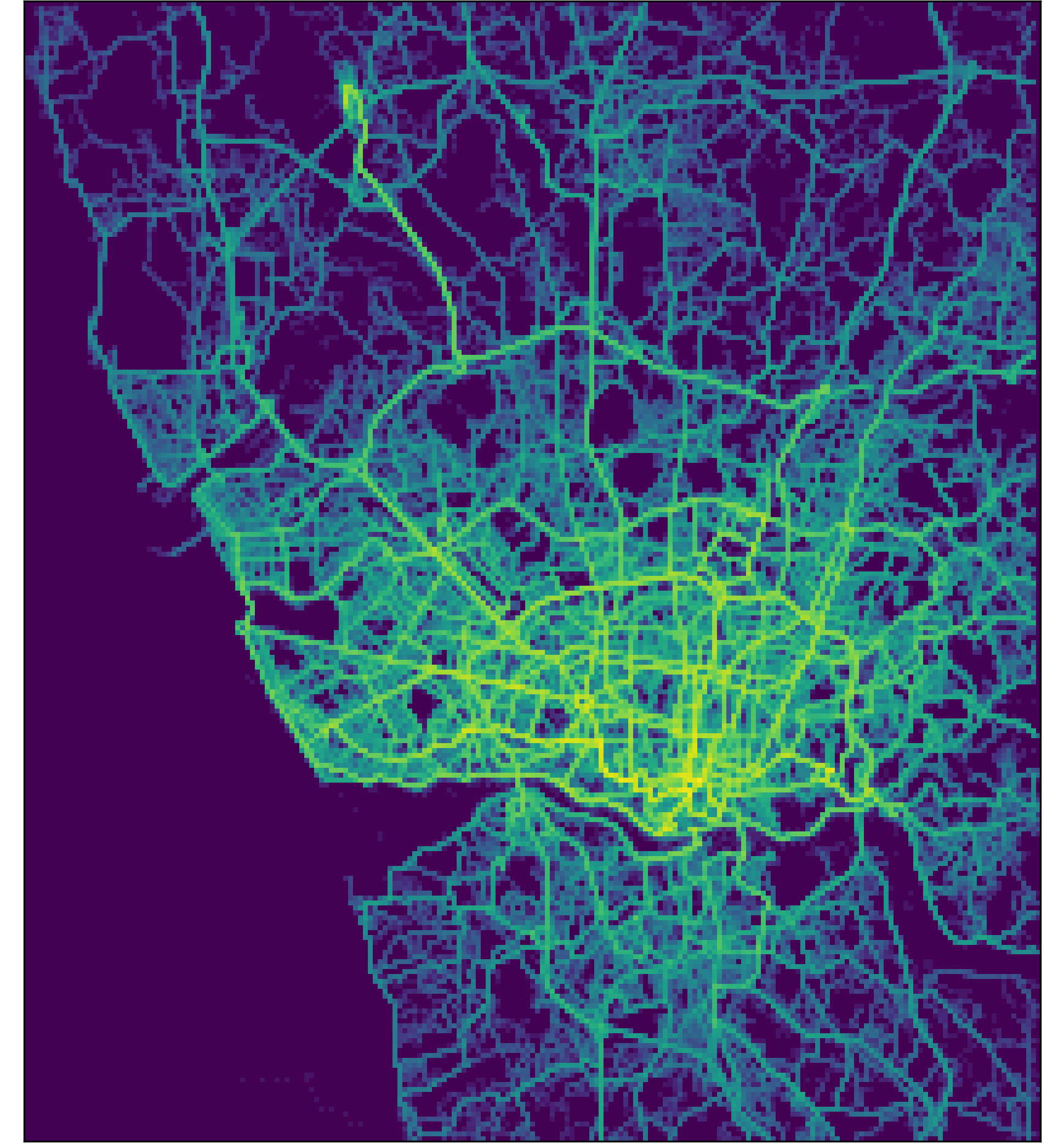}
        \caption{$\Delta{=}(72,72)$}
    \end{subfigure}\hspace{4pt}%
    \begin{subfigure}[b]{\qdw}
        \centering
        \includegraphics[width=\linewidth]{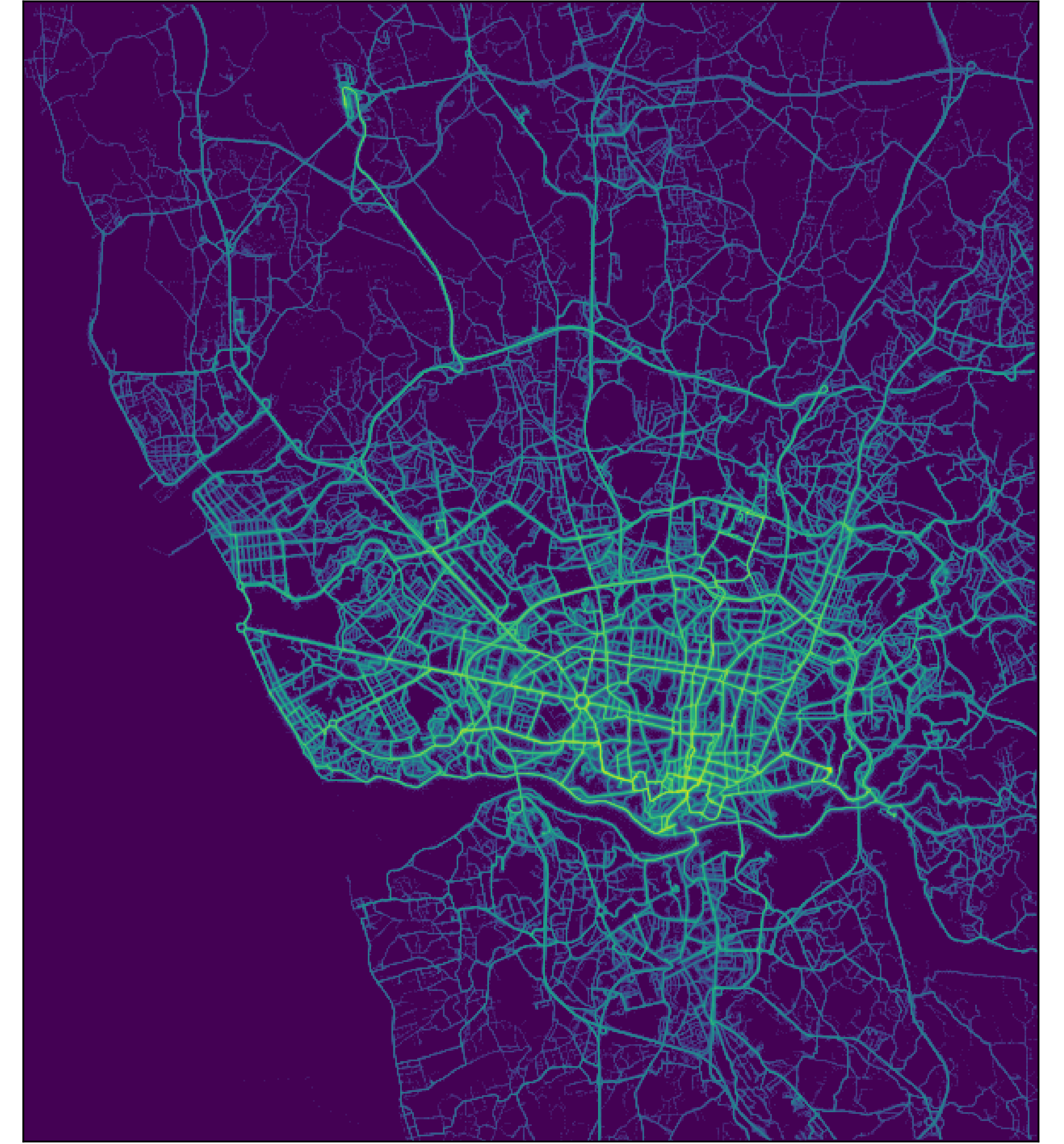}
        \caption{$\Delta{=}(25,25)$}
    \end{subfigure}\hspace{3pt}%
    \begin{subfigure}[b]{\qcbw}
        \centering
        \includegraphics[height=\qdh]{figures/appendix/delta-colorbar.png}
        \caption*{\phantom{$\Delta$}}
    \end{subfigure}
    \caption{Visualization of CountGrid density field on \textit{Porto} under varying cell sizes $\Delta = (\Delta_x, \Delta_y)$, given in meters. (a)~$\Delta=(210, 210)$, (b)~$\Delta=(72, 72)$, (c)~$\Delta=(25, 25)$. Colors represent log-transformed point counts $\rho_p$.}
    \Description{Three heatmaps of the CountGrid density field on Porto at cell sizes of 210, 72, and 25 meters, with a shared color bar of log-transformed point counts. Finer cells resolve individual roads, while coarser cells blur them into broad regions.}
    \label{fig:delta-comparison}
\end{figure}

\subsection{Grid Cell Size $\Delta$ and Synopsis Scalability}\label{appendix:dense-grid}
We now examine how decreasing the CountGrid cell size $\Delta$ affects estimation accuracy and resource cost. To assess whether a sparse storage representation can reduce storage cost without changing the underlying partition, we use a lossless tiled encoding that divides the CountGrid and LabelMap into fixed-size square blocks of adjacent cells, called \textit{tiles}, and stores sparse or uniform tiles compactly. As in~\cref{appendix:subsampling,appendix:query-size}, we evaluate cell sizes spanning the same range, using three configurations selected for the target memory budgets and an additional finest setting ($\dagger$). TraHist memory includes the LabelMap and per-bucket statistics, while CountGrid storage is reported separately.

\par As~\cref{fig:delta-comparison} illustrates, reducing $\Delta$ exposes progressively finer spatial structure in the CountGrid density field. The associated scalability cost is quantified in~\cref{tab:dense-scalability}a, which reports the resulting growth in grid cells, emergent buckets, and memory. On \textit{GeoLife}, the number of cells grows from 506 to 1.83~million, the emergent bucket count from 66 to over one million, and TraHist memory from 5.0\,KB to 30.2\,MB; \textit{Porto} and \textit{T-Drive} follow the same trend. Construction time grows much more slowly: on \textit{GeoLife}, it increases from 4.6 to 39.2s despite a roughly 3,600$\times$ increase in cells, and remains within tens of seconds on all three datasets. Thus, memory is the primary scalability cost of decreasing $\Delta$.

\begin{table}[t]
    \centering
    \caption{CountGrid scalability and lossless $64\times64$ tiling across three target memory configurations and an additional finest setting ($\dagger$). (a) reports grid cells ($N$), buckets ($|B|$), TraHist memory, and construction time. (b) reports median Q-error, CountGrid occupancy, storage reductions after tiling, and query latency overhead.}
    \label{tab:dense-scalability}
    \label{tab:dense-scalability-a}
    \label{tab:dense-scalability-b}
    {%
    \renewcommand{\arraystretch}{1}
    \setlength{\aboverulesep}{0pt}
    \setlength{\belowrulesep}{0pt}
    \setlength{\arrayrulewidth}{0.8pt}
    \setlength{\tabcolsep}{3pt}
    \newcommand{\topgap}{\rule{0pt}{2.7ex}}
    \newcommand{\botgap}{\rule[-1.5ex]{0pt}{0pt}}
    \newcommand{\twoline}[2]{\begin{tabular}[c]{@{}c@{}}#1\\#2\end{tabular}}
    \ifdefined\qnzw\else\newlength{\qnzw}\fi
    \settowidth{\qnzw}{00.0}
    \newcommand{\nz}[1]{\makebox[\qnzw][r]{#1}}
    \resizebox{0.96\textwidth}{!}{%
    \begin{tabular}{@{}c@{\hspace{6pt}}c@{\hspace{6pt}}|@{\hspace{6pt}}rrrc@{\hspace{9pt}}|@{\hspace{6pt}}rrr@{\hspace{4pt}}c@{\hspace{4pt}}c@{\hspace{4pt}}c@{\hspace{4pt}}rrr@{\hspace{6pt}}}
        \toprule[0.8pt]
        \multirow{3}{*}{\textbf{Dataset}}
        & \multirow{3}{*}{\twoline{$\boldsymbol{\Delta_{x=y}}$}{\textbf{(m)}}}
        & \multicolumn{4}{c|@{\hspace{6pt}}}{\topgap\textbf{(a) Scalability and cost}}
        & \multicolumn{9}{c}{\textbf{(b) Accuracy and tiling}} \\
        \cmidrule(r{9pt}){3-6} \cmidrule(l{0pt}){7-15}
        & & \multicolumn{1}{@{}c}{\multirow{2}{*}{$\boldsymbol{N}$}} & \multicolumn{1}{c}{\multirow{2}{*}{$\boldsymbol{|B|}$}}
        & \multirow{2}{*}{\twoline{\textbf{Mem.}}{\textbf{(KB)}}} & \multirow{2}{*}{\twoline{\textbf{Const.}}{\textbf{(s)}}}
        & \multicolumn{3}{c}{\topgap\textbf{Median Q-error}}
        && \multirow{2}{*}{\twoline{\textbf{Nonzero}}{\textbf{(\%)}}}
        && \multicolumn{3}{c}{\textbf{After tiling}} \\
        \cmidrule(r{4pt}){7-9} \cmidrule{13-15}
        & & & & & & \botgap Low & Med. & High && && \twoline{$C(x)$}{(\%)} & \twoline{Mem.}{(\%)} & \twoline{Lat.}{($\times$)} \\
        \midrule[0.8pt]
        \multirow{4}{*}{\textbf{\textit{Porto}}} & \topgap 210 & 5,680 & 360 & 50.0 & \nz{20.5}
            & 4.98 & \best{1.39} & \best{2.59} && \nz{75.1} && $-0.2$ & $-0.1$ & 2.4 \\
         & 72 & 48,024 & 7,914 & 498.8 & \nz{26.0}
            & 1.90 & 2.60 & 5.01 && \nz{56.2} && 16.1 & $-0.1$ & 2.5 \\
         & 38 & 172,480 & 42,645 & 2,014.0 & \nz{27.3}
            & \best{1.82} & 3.79 & 5.50 && \nz{43.2} && 28.7 & 4.7 & 2.7 \\
         & \botgap25$^{\dagger}$ & 397,460 & 118,778 & 4,961.0 & \nz{34.4}
            & 2.35 & 4.29 & 5.40 && \nz{35.4} && 40.1 & 6.4 & 2.9 \\
        \midrule
        \multirow{4}{*}{\textbf{\textit{T-Drive}}} & \topgap 9900 & 621 & 30 & 5.3 & \nz{3.9}
            & 9.45 & 2.36 & 3.78 && \nz{91.9} && $-0.5$ & $-0.4$ & 2.3 \\
         & 3580 & 4,536 & 983 & 50.8 & \nz{4.1}
            & 6.84 & 1.76 & \best{1.46} && \nz{66.5} && 2.5 & $-0.1$ & 2.3 \\
         & 1910 & 15,795 & 4,904 & 200.0 & \nz{4.4}
            & 6.05 & 1.60 & 2.89 && \nz{45.1} && 13.6 & $-0.1$ & 2.4 \\
         & \botgap500$^{\dagger}$ & 229,620 & 108,382 & 3,487.0 & \nz{10.6}
            & \best{4.68} & \best{1.46} & 14.98 && \nz{16.0} && 72.9 & 0.0 & 2.8 \\
        \midrule
        \multirow{4}{*}{\textbf{\textit{GeoLife}}} & \topgap 6200 & 506 & 66 & 5.0 & \nz{4.6}
            & 6.31 & 1.62 & 2.51 && \nz{73.3} && $-0.6$ & $-0.5$ & 2.3 \\
         & 2110 & 4,160 & 1,109 & 49.8 & \nz{4.7}
            & 6.17 & 1.61 & 2.81 && \nz{50.4} && 1.1 & $-0.1$ & 2.4 \\
         & 1125 & 14,518 & 5,381 & 197.5 & \nz{4.9}
            & 5.07 & \best{1.34} & 3.78 && \nz{36.5} && 34.6 & 0.0 & 2.4 \\
         & \botgap100$^{\dagger}$ & 1,826,210 & 1,067,362 & 30,945.0 & \nz{39.2}
            & \best{2.68} & 1.78 & \best{2.48} && \nz{6.4} && 88.2 & 0.2 & 3.0 \\
        \bottomrule[0.8pt]
    \end{tabular}%
    }
    }
\end{table}

\par The median Q-error columns of~\cref{tab:dense-scalability}b show that decreasing $\Delta$ does not yield consistent accuracy gains. Finer spatial localization benefits low selectivity queries, but at larger query scales, finer cells increase the number of cells and buckets covered by a query, allowing partial-overlap and estimation errors to accumulate. This trade-off varies across datasets: on \textit{Porto}, decreasing $\Delta$ generally improves low-selectivity accuracy while increasing medium- and high-selectivity error; on \textit{T-Drive}, the finest setting lowers the low-selectivity median Q-error to 4.68 but raises the high-selectivity error to 14.98; and \textit{GeoLife} shows a more mixed response across selectivity groups. These responses are consistent with the query-size effects in~\cref{appendix:query-size} and explain why no single evaluated cell size minimizes median Q-error across all selectivity groups. We therefore select $\Delta$ primarily to meet the target memory budget while avoiding cells too fine relative to the available sampling density and expected query sizes, as indicated by~\cref{appendix:subsampling,appendix:query-size}.

\par We now assess whether the sparse representation reduces storage in practice. Sparse CountGrid tiles store only the positions and values of nonzero cells, while uniform LabelMap tiles store a single label; other tiles remain stored cell by cell. We evaluate tile side lengths of 32, 64, and 128 cells and report 64, the prespecified intermediate size; all three yield the same qualitative conclusions. As $\Delta$ decreases, the fraction of nonzero CountGrid cells falls---from 73.3\% to 6.4\% on \textit{GeoLife}, for example---and CountGrid storage reductions grow from essentially zero at the coarsest cell sizes to 40.1--88.2\% at the finest settings. In contrast, TraHist memory changes by at most 6.4\%, because most LabelMap tiles span multiple buckets and the per-bucket statistics are unaffected by tiling. Since query estimation uses the LabelMap and per-bucket statistics rather than the CountGrid, compressing the CountGrid does not materially reduce TraHist memory; the tiled LabelMap representation, however, increases average query latency by 2.3--3.0$\times$. Thus, lossless tiling reduces CountGrid storage at fine cell sizes, but improving TraHist memory scalability requires reducing the number of logical cells rather than only changing their storage representation.

%%%%%%%%%%%%%%%%%%%%%%%%%%%%%%%%%%%%%%%%%%%%%%%%%%%%%%%%%%%%%%%%%
\setcounter{table}{0}
\setcounter{figure}{0}

\section{Parameter Sensitivity Analysis} \label{appendix:parameter}

\begin{figure}[t]
    \centering
    % same panel width as the CountGrid figure; each panel gets a thin frame
    \ifdefined\qpw\else\newlength{\qpw}\fi
    \setlength{\qpw}{0.22\linewidth}
    {%
    \setlength{\fboxsep}{0pt}
    \begin{subfigure}[b]{\qpw}
        \centering
        \fbox{\includegraphics[width=\dimexpr\linewidth-2\fboxrule\relax]{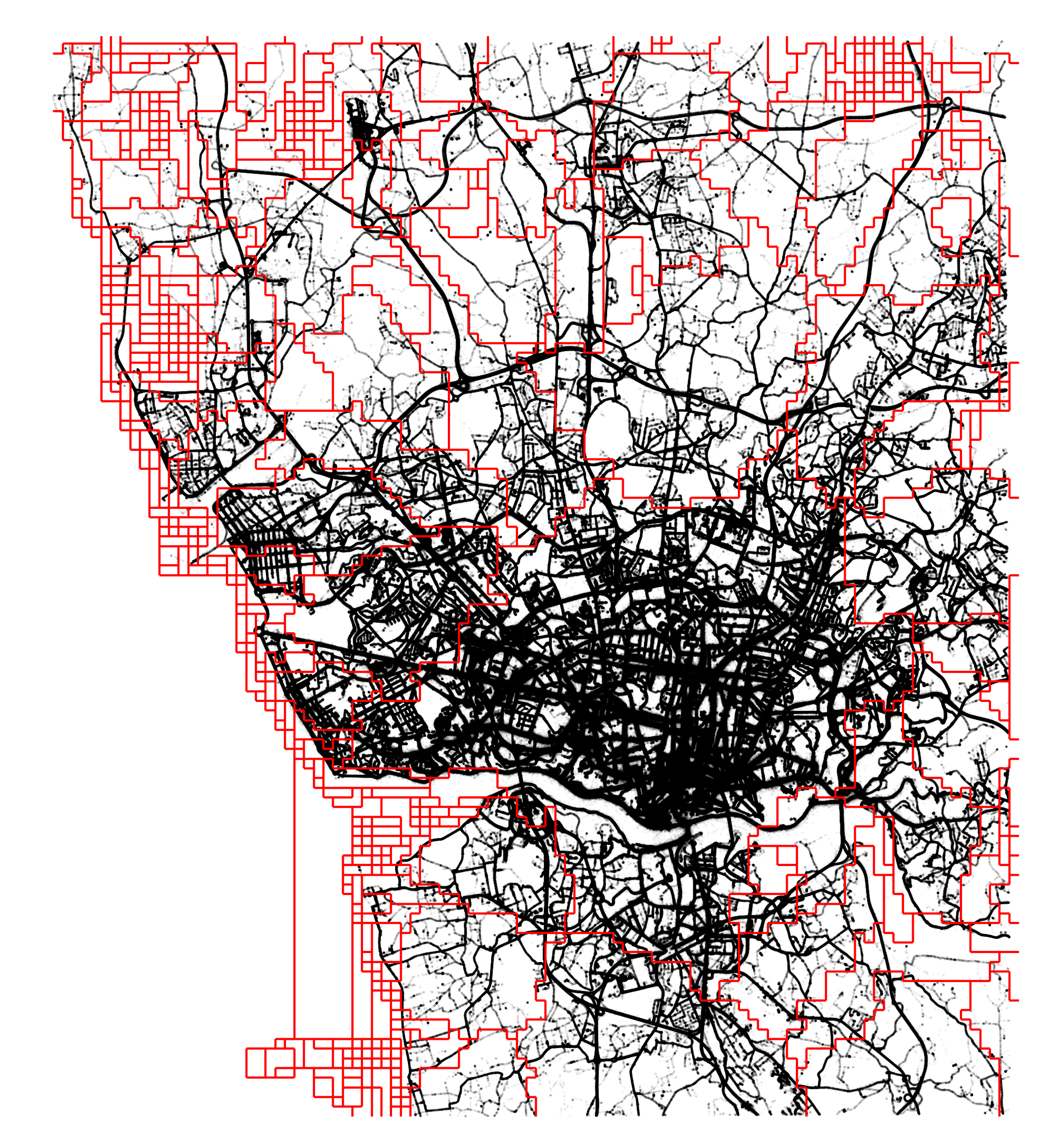}}
        \caption{$\alpha=0.0$}
    \end{subfigure}\hspace{4pt}%
    \begin{subfigure}[b]{\qpw}
        \centering
        \fbox{\includegraphics[width=\dimexpr\linewidth-2\fboxrule\relax]{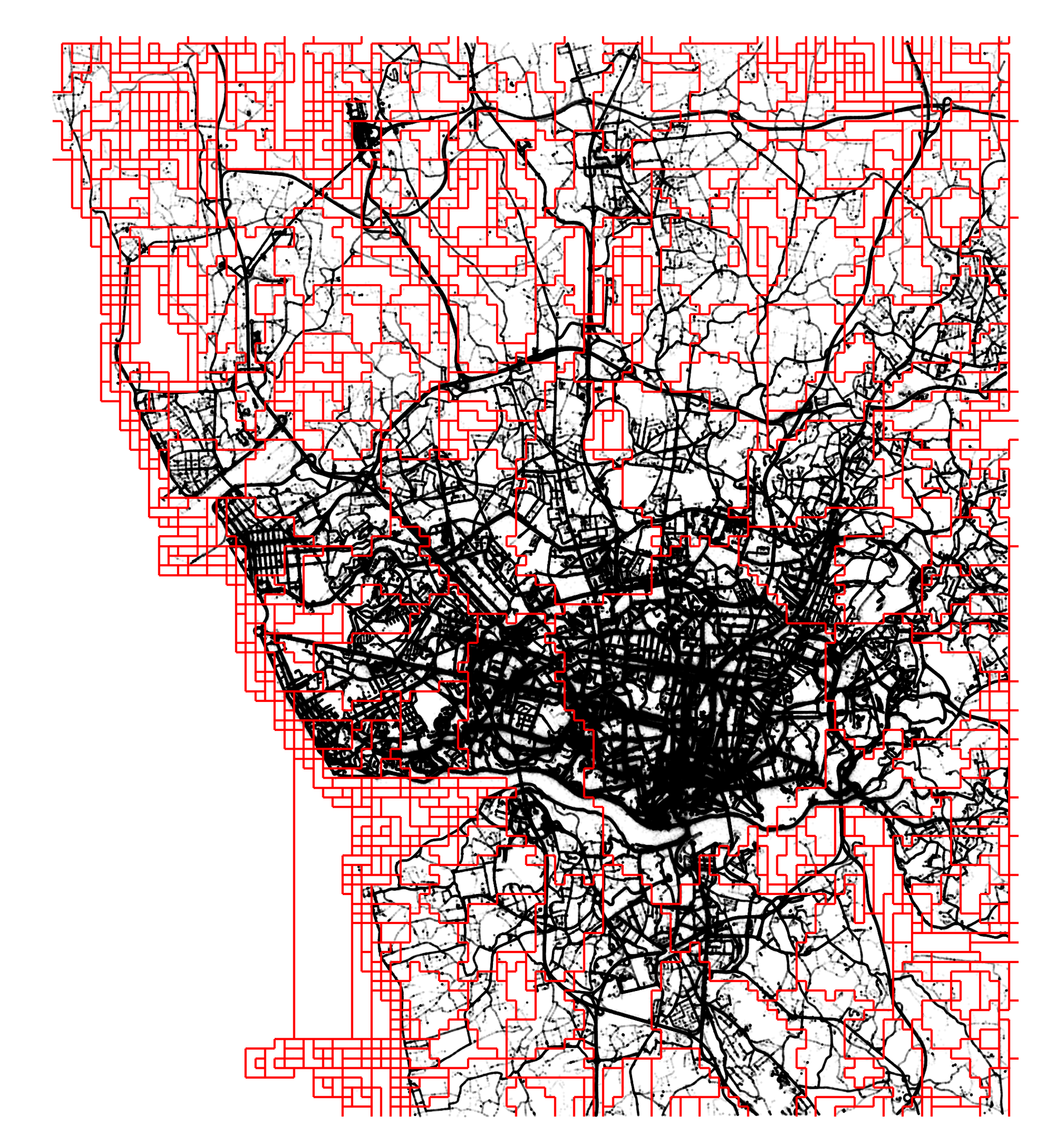}}
        \caption{$\alpha=0.5$ (default)}
    \end{subfigure}\hspace{4pt}%
    \begin{subfigure}[b]{\qpw}
        \centering
        \fbox{\includegraphics[width=\dimexpr\linewidth-2\fboxrule\relax]{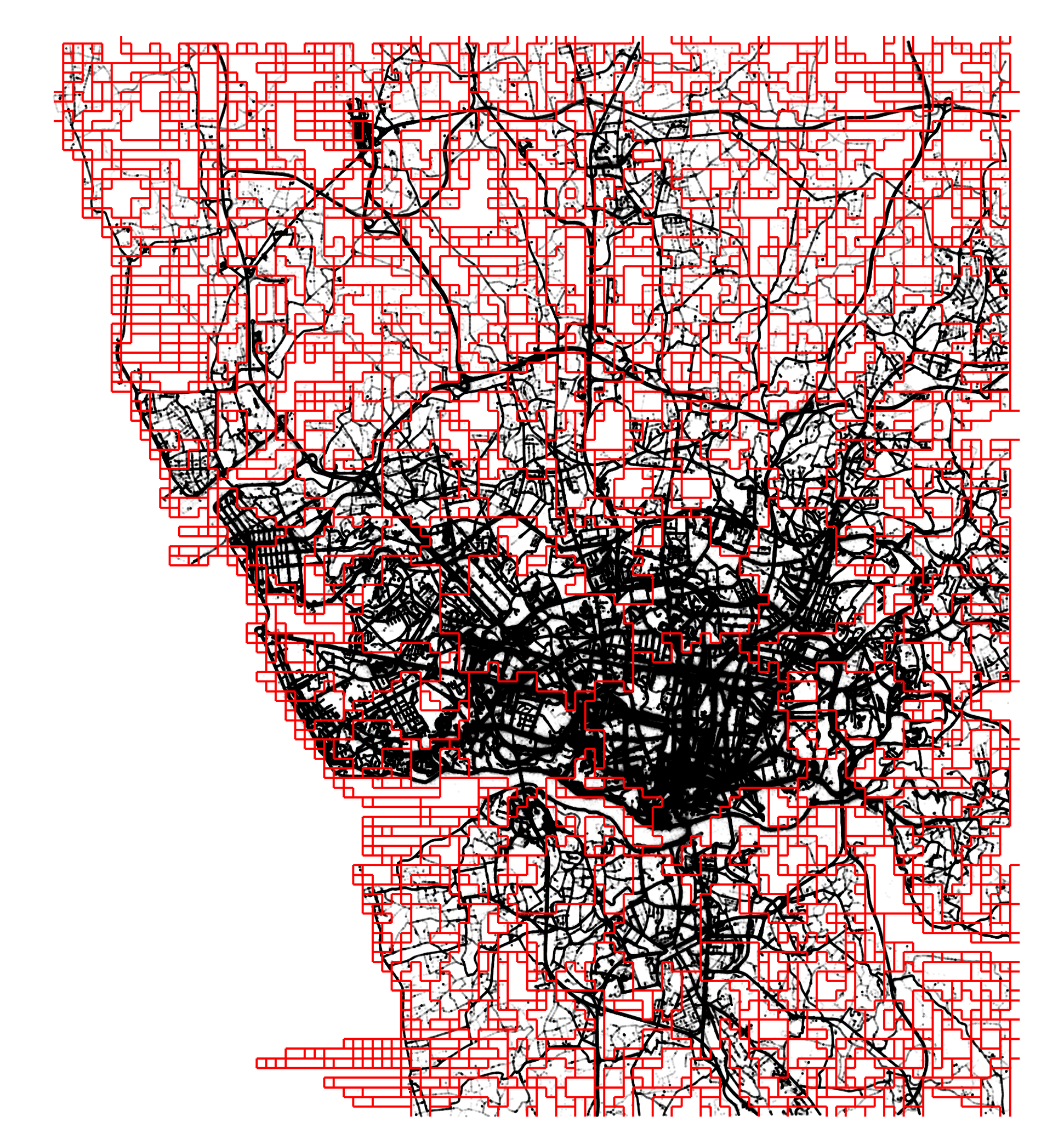}}
        \caption{$\alpha=1.0$}
    \end{subfigure}%
    }
    \caption{Density-adaptive partitions on \textit{Porto} under varying contrast balance parameter $\alpha$ at a fixed grid cell size $\Delta = (150, 150)$. (a)~$\alpha = 0.0$, (b)~$\alpha = 0.5$ (default), (c)~$\alpha = 1.0$. Red lines indicate bucket boundaries overlaid on the trajectory data; the number of buckets $|B|$ is $472$, $964$, and $1{,}704$ in (a)--(c), respectively.}
    \Description{Three maps of Porto with red bucket boundaries overlaid on the trajectory data, for contrast balance parameter alpha equal to 0.0, 0.5, and 1.0. Larger alpha yields more and finer buckets.}
    \label{fig:alpha-comparison}
\end{figure}

Having examined the effects of the CountGrid cell size $\Delta$ in~\cref{appendix:dense-grid}, we now analyze the two parameters governing density-adaptive bucket formation: the contrast balance $\alpha$ and threshold scaling $\beta$. Both operate within the bucket formation process (\cref{alg:density-adaptive_bucket_formation}), governing how local density contrasts are translated into partition boundaries. We examine how each parameter affects partition granularity and estimation accuracy across datasets and selectivity groups.

\begin{figure}[t]
    \centering
    \includegraphics[width=0.6\linewidth, trim=0 -10 0 0]{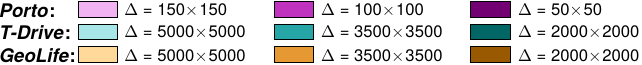}

    % both panels at the same scale, together 98% of the line width
    \ifdefined\qaw\else\newlength{\qaw}\fi
    \ifdefined\qbw\else\newlength{\qbw}\fi
    \settowidth{\qaw}{\includegraphics{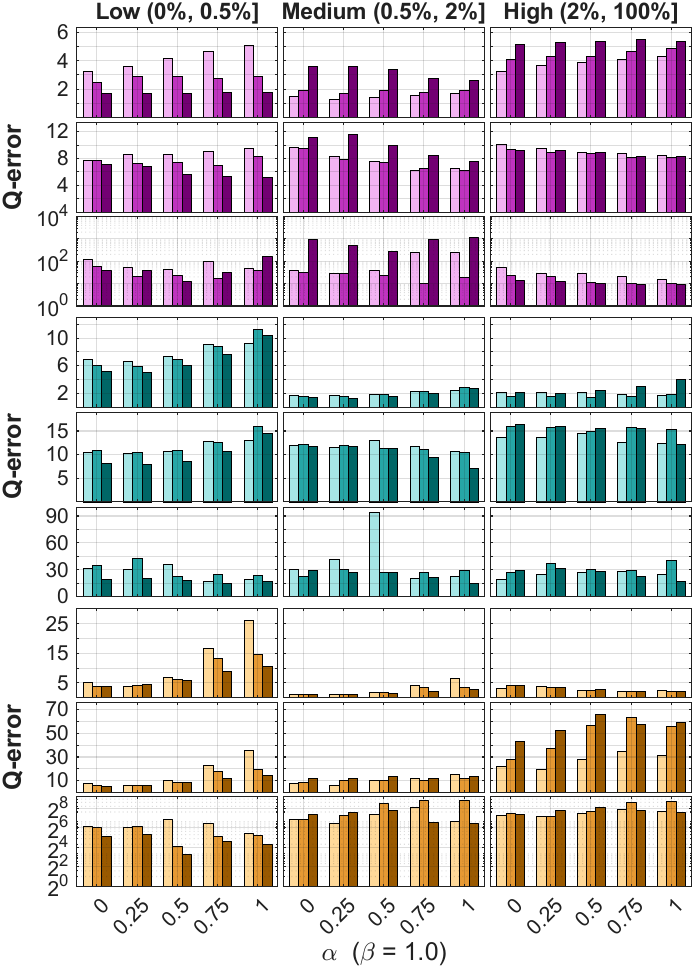}}%
    \settowidth{\qbw}{\includegraphics{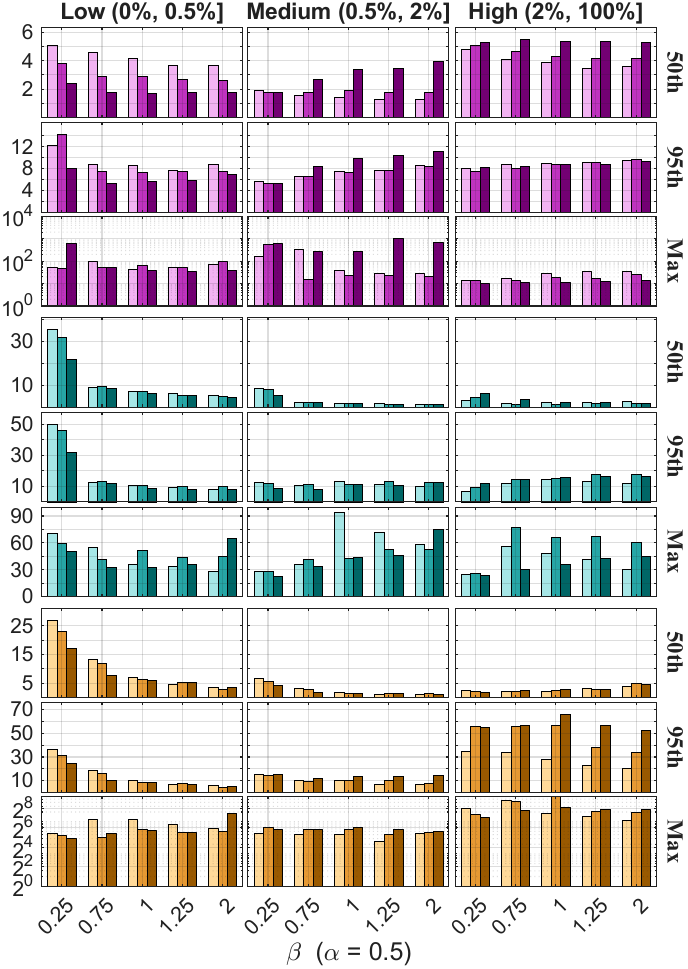}}%
    \setlength{\qaw}{\dimexpr 0.98\linewidth * \numexpr\qaw\relax / \numexpr\qaw+\qbw\relax\relax}%
    \setlength{\qbw}{\dimexpr 0.98\linewidth - \qaw\relax}%
    \begin{subfigure}[t]{\qaw}
        \centering
        \includegraphics[width=\linewidth]{figures/appendix/param_alpha_bar_all.pdf}
        \caption{Contrast balance $\alpha$ ($\beta=1.0$ fixed)}
        \label{fig:sensitivity-alpha}
    \end{subfigure}\hfill
    \begin{subfigure}[t]{\qbw}
        \centering
        \includegraphics[width=\linewidth]{figures/appendix/param_beta_bar_all.pdf}
        \caption{Threshold scaling $\beta$ ($\alpha=0.5$ fixed)}
        \label{fig:sensitivity-beta}
    \end{subfigure}
    \caption{Estimation accuracy under varying (a) contrast balance $\alpha$ and (b) threshold scaling $\beta$ on \textit{Porto}, \textit{T-Drive}, and \textit{GeoLife}. Median, 95th-percentile, and maximum Q-error are reported per selectivity group.}
    \Description{Two grids of grouped bar charts, one for the contrast balance parameter alpha and one for the threshold scaling parameter beta. Each grid has three rows per dataset (median, 95th percentile, and maximum Q-error) and three columns for low, medium, and high selectivity, with bars for three CountGrid cell sizes per dataset.}
    \label{fig:sensitivity-alpha-beta}
\end{figure}

\subsection{Contrast Balance $\alpha$}
The contrast balance parameter $\alpha\in[0,1]$ controls how local density differences are reflected in density-adaptive propagation (\cref{eq:local_cost}): larger $\alpha$ increases sensitivity to fine-grained density contrast, while smaller $\alpha$ favors the smoothed local density field and yields partitions more robust to noise. \cref{fig:alpha-comparison} illustrates this effect on \textit{Porto} at $\Delta=(150,150)$: as $\alpha$ increases, partitions become finer, with $|B|$ growing from 472 at $\alpha=0$ to 1,704 at $\alpha=1$; the default $\alpha=0.5$ yields an intermediate partition ($|B|=964$).
\par \cref{fig:sensitivity-alpha} reports estimation accuracy across $\alpha \in \{0, \allowbreak0.25, \allowbreak0.5, \allowbreak0.75, 1\}$ at three grid cell sizes on each dataset. Under low and medium selectivity, increasing $\alpha$ generally raises median Q-error, with the magnitude growing substantially from \textit{Porto} to \textit{T-Drive} to \textit{GeoLife}. On \textit{Porto}, the gain from improved density alignment is largely offset by the increase in fragmentation, resulting in only minor variation in median Q-error (e.g., $2.5$ to $2.9$ at $\Delta=(100,100)$, low selectivity). \textit{GeoLife}, in contrast, shows a stronger dependence: low selectivity median Q-error grows from $3.8$ to $14.7$ at $\Delta=(3500,3500)$ as $\alpha$ increases from $0$ to $1$. This reflects its heterogeneous sampling, making smaller $\alpha$ more effective for noise suppression. \textit{T-Drive} falls between these two regimes.
\par Under high selectivity, the response to $\alpha$ becomes more heterogeneous and dataset-dependent. \textit{Porto} preserves the same direction as the lower selectivity regimes, with Q-error increasing as $\alpha$ grows. \textit{GeoLife}, in contrast, shows the opposite pattern under high selectivity: median Q-error at $\alpha=1$ is consistently lower than at $\alpha=0$ across all evaluated $\Delta$, with the clearest trend appearing at $\Delta=(2000,2000)$ where it decreases monotonically from $4.1$ to $2.1$. \textit{T-Drive} shows a similar but weaker and $\Delta$-dependent reversal: at the coarsest $\Delta=(5000,5000)$, median Q-error decreases from $2.2$ to $1.7$, while at finer $\Delta=(2000,2000)$ larger $\alpha$ yields higher Q-error ($2.1$ to $4.1$). This behavior reflects the reduced impact of fragmentation under large query regions and the varying degree of density heterogeneity across datasets.
\par Overall, these observations indicate that $\alpha$ primarily controls the granularity of density-adaptive partitioning, while its impact on estimation accuracy depends on the selectivity regime and the density heterogeneity of the dataset. The 95th percentile Q-error follows the median trends under low and medium selectivity but diverges under high selectivity, whereas the maximum Q-error exhibits higher variability across configurations without a consistent trend in $\alpha$. Since low and medium selectivity dominate the challenging workloads emphasized in our main evaluation (\cref{subsec:cardinality-estimation-accuracy}), smaller to moderate $\alpha$ is generally preferable. The default $\alpha=0.5$ stays within a small factor of the configuration-wise optimum under low and medium selectivity (typically within $1.8\times$ in both median and 95th percentile), providing a robust and practical choice.

\begin{figure}[t]
    \centering
    \ifdefined\qpw\else\newlength{\qpw}\fi
    \setlength{\qpw}{0.22\linewidth}
    {%
    \setlength{\fboxsep}{0pt}
    \begin{subfigure}[b]{\qpw}
        \centering
        \fbox{\includegraphics[width=\dimexpr\linewidth-2\fboxrule\relax]{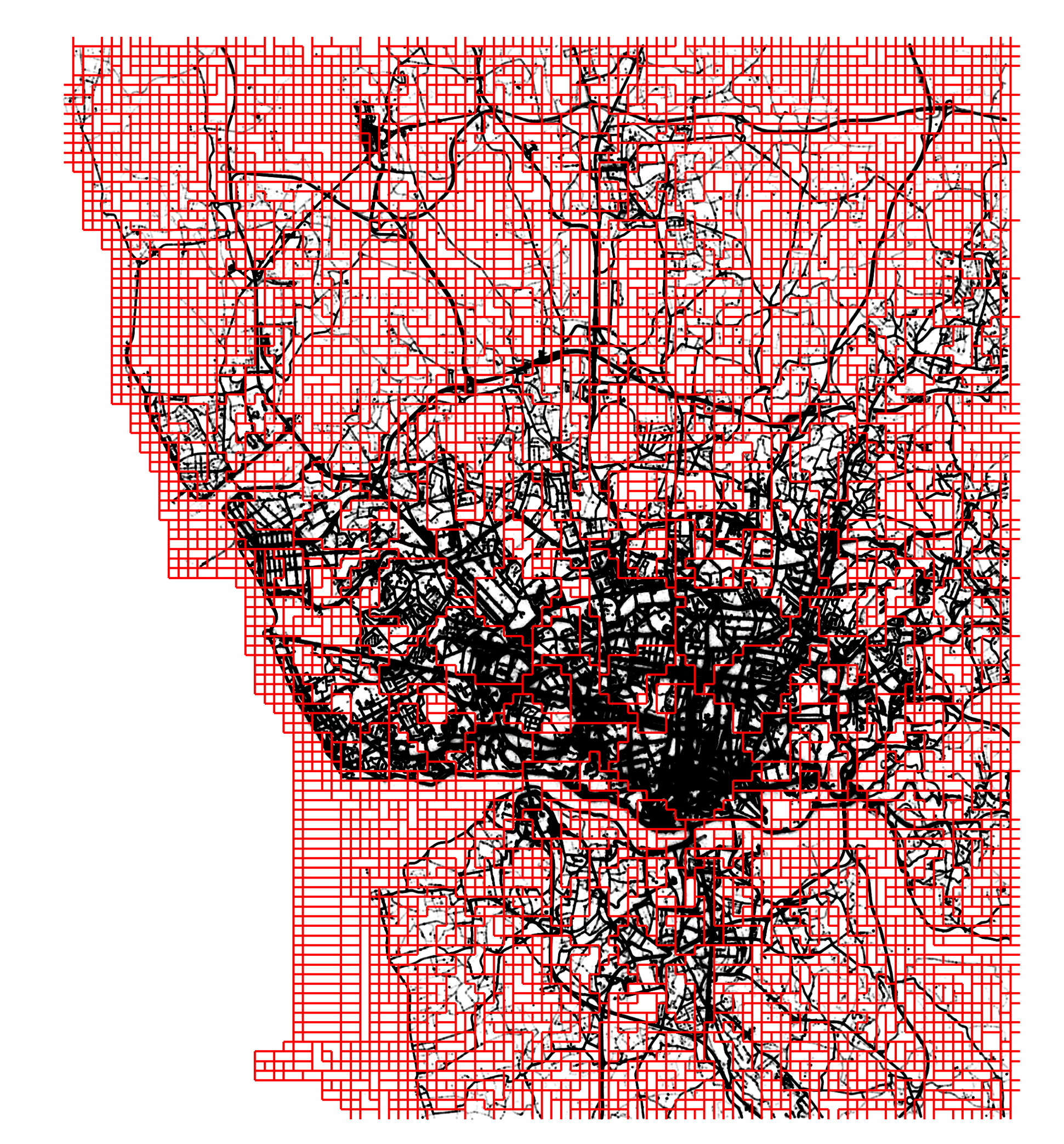}}
        \caption{$\beta=0.25$}
    \end{subfigure}\hspace{4pt}%
    \begin{subfigure}[b]{\qpw}
        \centering
        \fbox{\includegraphics[width=\dimexpr\linewidth-2\fboxrule\relax]{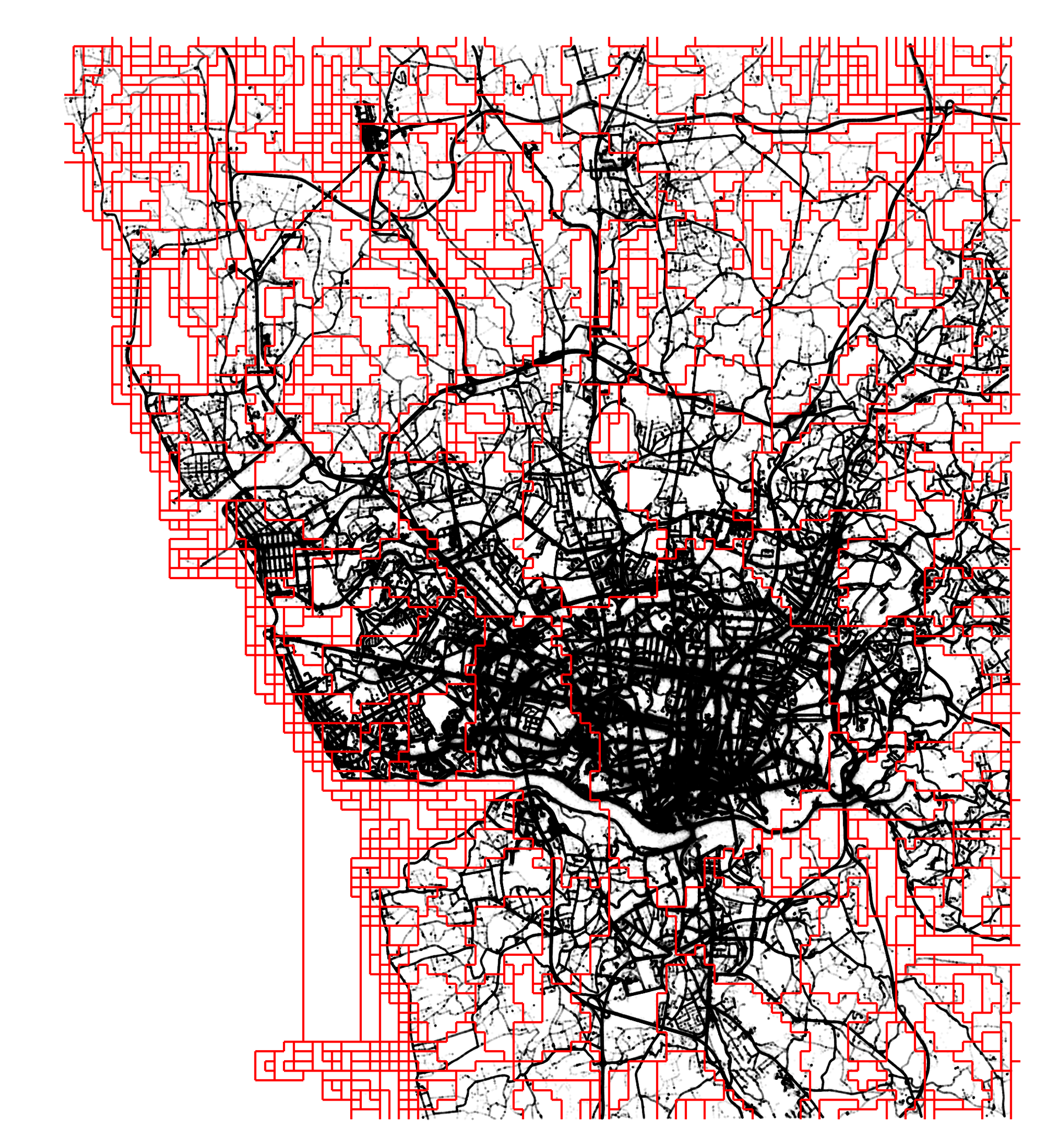}}
        \caption{$\beta=1.00$ (default)}
    \end{subfigure}\hspace{4pt}%
    \begin{subfigure}[b]{\qpw}
        \centering
        \fbox{\includegraphics[width=\dimexpr\linewidth-2\fboxrule\relax]{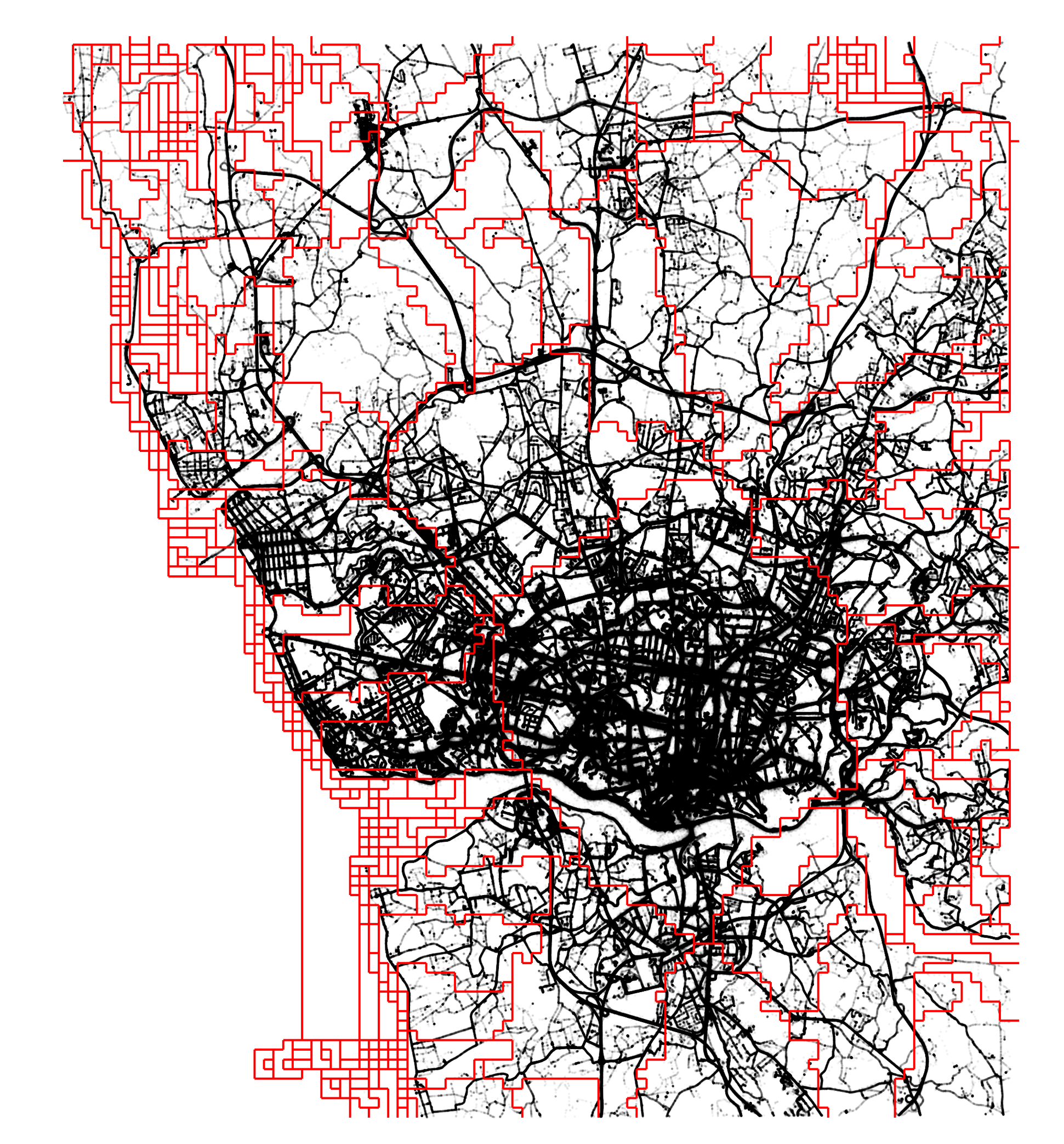}}
        \caption{$\beta=2.00$}
    \end{subfigure}%
    }
    \caption{Density-adaptive partitions on \textit{Porto} under varying threshold scaling parameter $\beta$ at a fixed grid cell size $\Delta = (150, 150)$. (a)~$\beta = 0.25$, (b)~$\beta = 1.00$ (default), (c)~$\beta = 2.00$. Red lines indicate bucket boundaries overlaid on the trajectory data; the number of buckets $|B|$ is $4{,}380$, $964$, and $380$ in (a)--(c), respectively.}
    \Description{Three maps of Porto with red bucket boundaries overlaid on the trajectory data, for threshold scaling parameter beta equal to 0.25, 1.00, and 2.00. Larger beta yields fewer and coarser buckets.}
    \label{fig:beta-comparison}
\end{figure}

\subsection{Threshold Scaling $\beta$}
The threshold scaling parameter $\beta$ controls the boundary formation criterion of density-adaptive propagation through the adaptive threshold $\tau(x) = \beta \cdot \mu(x)$ defined in~\cref{eq:threshold}. By scaling the local density level, $\beta$ determines the relative contrast required to trigger a boundary: smaller $\beta$ lowers the threshold and promotes finer partitions, while larger $\beta$ produces coarser partitions. \cref{fig:beta-comparison} illustrates this effect on \textit{Porto} at $\Delta=(150,150)$: as $\beta$ increases, partitions transition from highly fragmented to progressively coarser structures, with $|B|$ decreasing from $4{,}380$ at $\beta=0.25$ to $380$ at $\beta=2$; the default $\beta=1.0$ yields a balanced partition with $|B|=964$. 
\par \cref{fig:sensitivity-beta} summarizes estimation accuracy across $\beta \in \{0.25,\allowbreak 0.75, \allowbreak 1, 1.25, 2\}$ at three grid cell sizes on each dataset. Since $\beta$ scales the boundary formation threshold, it directly controls partition granularity, and this effect yields sharply different accuracy trends across selectivity groups.
\par Under low and medium selectivity, increasing $\beta$ improves median Q-error across all three datasets by suppressing excessive fragmentation in sparsely sampled regions, except in Porto’s densest partitioning regime, where partitions are already fine enough that the trend weakens or reverses (e.g., $\Delta=(50,50)$, medium selectivity). The effect is most pronounced on \textit{T-Drive} and \textit{GeoLife}, whose trajectory distributions exhibit substantial sparsity: on \textit{T-Drive} at $\Delta=(5000,5000)$, low selectivity median Q-error decreases from $35.5$ at $\beta=0.25$ to $5.2$ at $\beta=2$, and on \textit{GeoLife} at $\Delta=(3500,3500)$, from $23.1$ to $3.0$ over the same range. On \textit{Porto}, whose trajectories are more spatially concentrated, the improvement is milder (e.g., from $5.1$ to $3.7$ at $\Delta=(150,150)$, low selectivity).
\par Under high selectivity, the response diverges by dataset. On \textit{Porto} and \textit{T-Drive}, median Q-error tends to remain stable or decrease as $\beta$ grows, reflecting the reduced impact of fragmentation at large query regions. On \textit{GeoLife}, in contrast, the direction reverses: coarser partitions induced by larger $\beta$ aggregate heterogeneous density into single buckets, degrading accuracy. At $\Delta=(3500,3500)$, high selectivity median Q-error rises from $2.4$ at $\beta=0.25$ to $5.2$ at $\beta=2$, and analogous increases appear at other $\Delta$. This reversal reflects \textit{GeoLife}'s density distribution, where large buckets span regions with substantial variation.
\par Overall, $\beta$ controls partition granularity and introduces a dataset-dependent trade-off across selectivity regimes: larger $\beta$ reduces excessive fragmentation and benefits low and medium selectivity across all three datasets, while its effect under high selectivity depends on density heterogeneity---smaller $\beta$ benefits high selectivity accuracy on \textit{GeoLife}, whereas \textit{Porto} and \textit{T-Drive} show the same direction as the lower selectivity regimes, with median Q-error remaining stable or decreasing as $\beta$ increases. The 95th percentile generally follows the median trends under low and medium selectivity but diverges under high selectivity. The maximum Q-error exhibits higher variability without a consistent trend across configurations. The default $\beta=1.0$ lies near this trade-off boundary, staying within $2.2\times$ of the configuration-wise optimum for low and medium selectivity while avoiding the pronounced high selectivity degradation observed on \textit{GeoLife} at large $\beta$. This makes $\beta=1.0$ a balanced and practical default.

%%%%%%%%%%%%%%%%%%%%%%%%%%%%%Larger Budgets%%%%%%%%%%%%%%%%%%%%%%%%%%%%%%%%%

\setcounter{table}{0}
\setcounter{figure}{0}

\section{Extended Results for the Main Evaluation}\label{appendix:larger-budgets}

\begin{table}[t]
    \centering
    \caption{Q-error on \textit{Porto} under fixed memory budgets (1000 and 2000 KB), extending~\cref{tab:qerror-DD-Porto-memory}.}
    \label{tab:extend-qerror-DD-Porto-memory}
    {%
    \renewcommand{\arraystretch}{1}
    \setlength{\aboverulesep}{0pt}
    \setlength{\belowrulesep}{0pt}
    \setlength{\arrayrulewidth}{0.8pt}
    \setlength{\tabcolsep}{3pt}
    \newcommand{\topgap}{\rule{0pt}{2.7ex}}
    \newcommand{\botgap}{\rule[-1.5ex]{0pt}{0pt}}
    \ifdefined\qcolw\else\newlength{\qcolw}\fi
    \settowidth{\qcolw}{0000.0}
    \newcommand{\hd}[1]{\makebox[\qcolw][r]{#1}}
    \ifdefined\qhd\else\newlength{\qhd}\fi
    \settowidth{\qhd}{50th}
    \newcommand{\qlt}{\dimexpr\qcolw-\qhd\relax}
    \resizebox{\textwidth}{!}{%
    \begin{tabular}{@{}c@{\hspace{6pt}}l@{\hspace{6pt}}|@{\hspace{6pt}}rrrr@{\hspace{5pt}}c@{\hspace{5pt}}rrrr@{\hspace{5pt}}c@{\hspace{5pt}}rrrr@{\hspace{6pt}}}
        \toprule[0.8pt]
        \multirow{2}{*}{\begin{tabular}[c]{@{}c@{}}\textbf{Memory}\\\textbf{(KB)}\end{tabular}}
        & \multirow{2}{*}{\textbf{Est.}}
        & \multicolumn{4}{c}{\topgap\textbf{Low (0\%, 0.5\%]}}
        && \multicolumn{4}{c}{\textbf{Medium (0.5\%, 2\%]}}
        && \multicolumn{4}{c}{\textbf{High (2\%, 100\%]}} \\
        \cmidrule(l{\qlt}r{5pt}){3-6} \cmidrule(l{\qlt}r{5pt}){8-11} \cmidrule(l{\qlt}r{6pt}){13-16}
        & & \topgap\botgap\hd{50th} & \hd{95th} & \hd{99th} & \hd{Max}
        && \hd{50th} & \hd{95th} & \hd{99th} & \hd{Max}
        && \hd{50th} & \hd{95th} & \hd{99th} & \hd{Max} \\
        \midrule[0.8pt]
        \multirow{6}{*}{1000} & \topgap MEH & 39.2 & 1249.3 & 5770.1 & 5.9e4
            && 5.2 & 175.6 & 457.5 & 1.1e4
            && 6.4 & 10.7 & 11.9 & 22.8 \\
         & MHist & 67.5 & 1297.3 & 2911.3 & 3.5e4
            && 7.7 & 222.0 & 805.0 & 6354.2
            && 5.5 & 8.5 & 11.8 & 23.2 \\
         & LHist & 80.3 & 460.1 & 977.6 & 6.5e4
            && 31.2 & 411.2 & 998.6 & 4.6e4
            && 2.3 & 9.1 & 34.6 & 74.4 \\
         & DEH & \secondbest{8.5} & \secondbest{29.4} & \secondbest{43.0} & \best{170.3}
            && \best{2.5} & \best{8.0} & \best{11.5} & \best{25.3}
            && \best{1.2} & \best{1.6} & \best{2.2} & \best{4.4} \\
         & CASE+ & 40.7 & 87.8 & 115.1 & \secondbest{397.6}
            && 6.5 & 21.3 & 27.1 & \secondbest{51.7}
            && \secondbest{1.3} & \secondbest{2.3} & \secondbest{3.2} & \secondbest{5.1} \\
         & \botgap\textbf{TraHist} & \best{1.7} & \best{5.7} & \best{15.2} & 401.6
            && \secondbest{3.2} & \secondbest{9.2} & \secondbest{13.0} & 1.3e4
            && 5.4 & 9.1 & 10.9 & 14.2 \\
        \midrule
        \multirow{6}{*}{2000} & \topgap MEH & 34.0 & 877.8 & 5609.6 & 1.0e5
            && 4.9 & 137.1 & 536.4 & 2.1e4
            && 6.5 & 10.6 & 12.0 & 23.6 \\
         & MHist & 51.3 & 919.3 & 2547.2 & 2.5e4
            && 6.6 & 160.8 & 456.4 & 4558.9
            && 5.8 & 9.2 & 12.1 & 23.6 \\
         & LHist & 96.1 & 636.6 & 1381.1 & 1.3e5
            && 34.4 & 478.7 & 1163.7 & 8.7e4
            && 2.5 & 9.8 & 36.6 & 79.8 \\
         & DEH & \secondbest{7.8} & \secondbest{26.2} & \secondbest{37.8} & \secondbest{167.4}
            && \best{2.2} & \best{6.8} & \best{10.9} & \best{24.6}
            && \best{1.2} & \best{1.5} & \best{2.1} & \best{3.7} \\
         & CASE+ & 36.9 & 86.5 & 103.9 & 392.6
            && 5.3 & 20.6 & 24.9 & \secondbest{49.2}
            && \secondbest{1.3} & \secondbest{2.3} & \secondbest{3.1} & \secondbest{5.0} \\
         & \botgap\textbf{TraHist} & \best{1.8} & \best{6.6} & \best{14.7} & \best{48.5}
            && \secondbest{3.8} & \secondbest{13.3} & \secondbest{21.8} & 94.3
            && 5.5 & 8.4 & 9.6 & 11.4 \\
        \bottomrule[0.8pt]
    \end{tabular}%
    }
    }
\end{table}

\begin{table}[t]
    \centering
    \caption{Q-error on \textit{T-Drive} under fixed bucket budgets (1000 and 2000 buckets), extending~\cref{tab:qerror-DD-Tdrive-buckets}.}
    \label{tab:extend-qerror-DD-Tdrive-buckets}
    {%
    \renewcommand{\arraystretch}{1}
    \setlength{\aboverulesep}{0pt}
    \setlength{\belowrulesep}{0pt}
    \setlength{\arrayrulewidth}{0.8pt}
    \setlength{\tabcolsep}{3pt}
    \newcommand{\topgap}{\rule{0pt}{2.7ex}}
    \newcommand{\botgap}{\rule[-1.5ex]{0pt}{0pt}}
    \ifdefined\qcolw\else\newlength{\qcolw}\fi
    \settowidth{\qcolw}{0000.0}
    \newcommand{\hd}[1]{\makebox[\qcolw][r]{#1}}
    \ifdefined\qhd\else\newlength{\qhd}\fi
    \settowidth{\qhd}{50th}
    \newcommand{\qlt}{\dimexpr\qcolw-\qhd\relax}
    \resizebox{\textwidth}{!}{%
    \begin{tabular}{@{}c@{\hspace{6pt}}l@{\hspace{6pt}}|@{\hspace{6pt}}rrrr@{\hspace{5pt}}c@{\hspace{5pt}}rrrr@{\hspace{5pt}}c@{\hspace{5pt}}rrrr@{\hspace{6pt}}}
        \toprule[0.8pt]
        \multirow{2}{*}{\textbf{Buckets}}
        & \multirow{2}{*}{\textbf{Est.}}
        & \multicolumn{4}{c}{\topgap\textbf{Low (0\%, 0.5\%]}}
        && \multicolumn{4}{c}{\textbf{Medium (0.5\%, 2\%]}}
        && \multicolumn{4}{c}{\textbf{High (2\%, 100\%]}} \\
        \cmidrule(l{\qlt}r{5pt}){3-6} \cmidrule(l{\qlt}r{5pt}){8-11} \cmidrule(l{\qlt}r{6pt}){13-16}
        & & \topgap\botgap\hd{50th} & \hd{95th} & \hd{99th} & \hd{Max}
        && \hd{50th} & \hd{95th} & \hd{99th} & \hd{Max}
        && \hd{50th} & \hd{95th} & \hd{99th} & \hd{Max} \\
        \midrule[0.8pt]
        \multirow{6}{*}{1000} & \topgap MEH & 181.3 & 4938.7 & 1.8e4 & 1.8e5
            && 24.7 & 792.3 & 1774.0 & 2.6e4
            && 14.5 & 29.6 & 31.8 & 32.8 \\
         & MHist & 649.8 & 2.6e4 & 5.3e4 & 8.7e5
            && 84.6 & 5001.8 & 1.2e4 & 1.2e5
            && 3.6 & \secondbest{6.3} & \secondbest{6.8} & \best{7.2} \\
         & LHist & 111.9 & 734.6 & 1506.7 & 5365.5
            && 52.1 & 610.3 & 1000.6 & 5074.3
            && 4.4 & 24.6 & 88.0 & 127.9 \\
         & DEH & 231.5 & 659.6 & 765.7 & 851.3
            && 57.2 & 160.9 & 190.5 & 230.0
            && 3.8 & \secondbest{6.3} & 11.7 & 19.8 \\
         & CASE+ & \secondbest{88.4} & \secondbest{161.9} & \secondbest{180.9} & \secondbest{201.1}
            && \secondbest{21.3} & \secondbest{39.4} & \secondbest{45.1} & \secondbest{58.7}
            && \best{1.2} & \best{3.6} & \best{5.5} & \secondbest{8.9} \\
         & \botgap\textbf{TraHist} & \best{6.2} & \best{13.3} & \best{33.2} & \best{162.7}
            && \best{1.6} & \best{13.8} & \best{33.1} & \best{48.3}
            && \secondbest{1.6} & 18.2 & 33.8 & 69.4 \\
        \midrule
        \multirow{6}{*}{2000} & \topgap MEH & 134.3 & 3406.3 & 9573.2 & 1.2e5
            && 20.4 & 514.3 & 1366.1 & 1.7e4
            && 17.3 & 34.8 & 37.6 & 38.9 \\
         & MHist & 452.4 & 1.9e4 & 3.7e4 & 5.6e5
            && 68.7 & 3598.6 & 7789.6 & 7.9e4
            && 4.7 & 8.5 & 9.1 & \secondbest{9.5} \\
         & LHist & 147.9 & 1056.8 & 2125.6 & 7216.0
            && 53.5 & 825.1 & 1402.0 & 6457.2
            && 6.2 & 31.5 & 104.2 & 145.2 \\
         & DEH & 190.0 & 446.6 & 507.5 & 571.9
            && 48.0 & 108.4 & 126.7 & 139.3
            && 3.7 & \secondbest{5.3} & \secondbest{7.2} & 11.4 \\
         & CASE+ & \secondbest{81.9} & \secondbest{140.9} & \secondbest{156.4} & \secondbest{176.2}
            && \secondbest{20.0} & \secondbest{34.2} & \secondbest{39.0} & \best{42.9}
            && \best{1.1} & \best{2.9} & \best{4.5} & \best{9.0} \\
         & \botgap\textbf{TraHist} & \best{6.8} & \best{11.2} & \best{31.7} & \best{75.9}
            && \best{1.8} & \best{16.3} & \best{31.6} & \secondbest{55.8}
            && \secondbest{1.7} & 17.1 & 31.9 & 77.5 \\
        \bottomrule[0.8pt]
    \end{tabular}%
    }
    }
\end{table}

\begin{table}[t]
    \centering
    \caption{Extended query estimation latency (ms) under (a) fixed \textit{memory budgets} (\textit{Porto}: 1000 and 2000 KB, \textit{T-Drive}: 100 and 200 KB) and (b) fixed \textit{bucket budgets} (1000 and 2000 buckets), extending~\cref{tab:qtime}. Average and maximum latency are reported over 3,000 trajectory range queries per dataset across all selectivity groups.}
    \label{tab:extend-estimation-latency}
    {%
    \renewcommand{\arraystretch}{1}
    \setlength{\aboverulesep}{0pt}
    \setlength{\belowrulesep}{0pt}
    \setlength{\arrayrulewidth}{0.8pt}
    \setlength{\tabcolsep}{3pt}
    \newcommand{\topgap}{\rule{0pt}{2.7ex}}
    \newcommand{\botgap}{\rule[-1.5ex]{0pt}{0pt}}
    \ifdefined\qcolw\else\newlength{\qcolw}\fi
    \newcommand{\hd}[1]{\makebox[\qcolw][c]{#1}}
    \ifdefined\qhda\else\newlength{\qhda}\fi
    \settowidth{\qhda}{Avg}
    \ifdefined\qhdm\else\newlength{\qhdm}\fi
    \settowidth{\qhdm}{Max}
    \newcommand{\qlt}{\dimexpr(\qcolw-\qhda)/2\relax}
    \newcommand{\qrta}{\dimexpr5pt+(\qcolw-\qhdm)/2\relax}
    \newcommand{\qrtb}{\dimexpr6.25pt+(\qcolw-\qhdm)/2\relax}
    \newcommand{\qrtc}{\dimexpr\qend+(\qcolw-\qhdm)/2\relax}
    \ifdefined\qvt\else\newlength{\qvt}\fi
    \settowidth{\qvt}{000.000}
    \newcommand{\vt}[1]{\makebox[\qvt][r]{#1}}
    \ifdefined\qvs\else\newlength{\qvs}\fi
    \settowidth{\qvs}{00.000}
    \newcommand{\vs}[1]{\makebox[\qvs][r]{#1}}
    \ifdefined\qend\else\newlength{\qend}\fi
    \setlength{\qend}{\dimexpr6.35pt-(\qvt-\qvs)/2\relax}
    \resizebox{\textwidth}{!}{%
    \setlength{\qcolw}{\qvt}%
    \begin{tabular}[t]{@{}c@{\hspace{6pt}}l@{\hspace{6pt}}|@{\hspace{6pt}}cc@{\hspace{5pt}}c@{\hspace{5pt}}cc@{\hspace{\qend}}}
        \multicolumn{7}{c}{\textbf{(a) Fixed memory budgets}} \\[2pt]
        \toprule[0.8pt]
        \multirow{2}{*}{\begin{tabular}[c]{@{}c@{}}\textbf{Memory}\\\textbf{(KB)}\end{tabular}}
        & \multirow{2}{*}{\textbf{Est.}}
        & \multicolumn{2}{c}{\topgap\textbf{\textit{Porto}}}
        && \multicolumn{2}{c}{\textbf{\textit{T-Drive}}} \\
        \cmidrule(l{\qlt}r{\qrta}){3-4} \cmidrule(l{\qlt}r{\qrtc}){6-7}
        & & \topgap\botgap\hd{Avg} & \hd{Max} && \hd{Avg} & \hd{Max} \\
        \midrule[0.8pt]
        \multirow{6}{*}{1000 / 100} & \topgap MEH & \vt{2.173} & \vt{5.501} && \vs{0.676} & \vs{1.820} \\
         & MHist & \vt{277.641} & \vt{357.605} && \vs{55.265} & \vs{62.270} \\
         & LHist & \vt{3.834} & \vt{9.132} && \vs{1.653} & \vs{3.496} \\
         & DEH & \vt{\best{0.547}} & \vt{\best{0.564}} && \vs{\best{0.537}} & \vs{\best{0.557}} \\
         & CASE+ & \vt{\secondbest{0.560}} & \vt{\secondbest{0.586}} && \vs{\secondbest{0.548}} & \vs{\secondbest{0.582}} \\
         & \botgap\textbf{TraHist} & \vt{0.591} & \vt{0.623} && \vs{0.577} & \vs{0.593} \\
        \midrule
        \multirow{6}{*}{2000 / 200} & \topgap MEH & \vt{3.902} & \vt{10.035} && \vs{1.337} & \vs{3.680} \\
         & MHist & \vt{705.594} & \vt{745.284} && \vs{39.958} & \vs{40.621} \\
         & LHist & \vt{5.369} & \vt{13.068} && \vs{2.114} & \vs{4.713} \\
         & DEH & \vt{\best{0.542}} & \vt{\best{0.571}} && \vs{\best{0.538}} & \vs{\best{0.557}} \\
         & CASE+ & \vt{\secondbest{0.555}} & \vt{\secondbest{0.587}} && \vs{\secondbest{0.550}} & \vs{\secondbest{0.580}} \\
         & \botgap\textbf{TraHist} & \vt{0.588} & \vt{0.633} && \vs{0.574} & \vs{0.590} \\
        \bottomrule[0.8pt]
    \end{tabular}%
    \hspace{14pt}%
    \setlength{\qcolw}{\qvs}%
    \begin{tabular}[t]{@{}c@{\hspace{6pt}}l@{\hspace{6pt}}|@{\hspace{6pt}}cc@{\hspace{5pt}}c@{\hspace{5pt}}cc@{\hspace{6.25pt}}}
        \multicolumn{7}{c}{\textbf{(b) Fixed bucket budgets}} \\[2pt]
        \toprule[0.8pt]
        \multirow{2}{*}{\textbf{Buckets}}
        & \multirow{2}{*}{\textbf{Est.}}
        & \multicolumn{2}{c}{\topgap\textbf{\textit{Porto}}}
        && \multicolumn{2}{c}{\textbf{\textit{T-Drive}}} \\
        \cmidrule(l{\qlt}r{\qrta}){3-4} \cmidrule(l{\qlt}r{\qrtb}){6-7}
        & & \topgap\botgap\hd{Avg} & \hd{Max} && \hd{Avg} & \hd{Max} \\
        \midrule[0.8pt]
        \multirow{6}{*}{1000} & \topgap MEH & \vs{\best{0.265}} & \vs{\secondbest{0.570}} && \vs{\best{0.343}} & \vs{0.899} \\
         & MHist & \vs{19.379} & \vs{19.387} && \vs{18.893} & \vs{18.981} \\
         & LHist & \vs{9.823} & \vs{24.557} && \vs{10.758} & \vs{27.064} \\
         & DEH & \vs{\secondbest{0.538}} & \vs{\best{0.557}} && \vs{\secondbest{0.541}} & \vs{\best{0.559}} \\
         & CASE+ & \vs{0.554} & \vs{0.580} && \vs{0.550} & \vs{\secondbest{0.582}} \\
         & \botgap\textbf{TraHist} & \vs{0.566} & \vs{0.575} && \vs{0.567} & \vs{0.583} \\
        \midrule
        \multirow{6}{*}{2000} & \topgap MEH & \vs{\best{0.246}} & \vs{\best{0.552}} && \vs{0.617} & \vs{1.663} \\
         & MHist & \vs{16.181} & \vs{16.338} && \vs{33.862} & \vs{33.955} \\
         & LHist & \vs{18.433} & \vs{46.737} && \vs{20.018} & \vs{50.996} \\
         & DEH & \vs{\secondbest{0.540}} & \vs{\secondbest{0.559}} && \vs{\best{0.542}} & \vs{\best{0.559}} \\
         & CASE+ & \vs{0.556} & \vs{0.586} && \vs{\secondbest{0.555}} & \vs{0.582} \\
         & \botgap\textbf{TraHist} & \vs{0.577} & \vs{0.593} && \vs{0.565} & \vs{\secondbest{0.580}} \\
        \bottomrule[0.8pt]
    \end{tabular}%
    }
    }
\end{table}
This appendix extends the budget ranges reported in the main evaluation for \textit{Porto} and \textit{T-Drive}. \Cref{tab:extend-qerror-DD-Porto-memory} and \cref{tab:extend-qerror-DD-Tdrive-buckets} extend the accuracy results in \cref{tab:qerror-DD-Porto-memory} and \cref{tab:qerror-DD-Tdrive-buckets} to 2000\,KB and 2000 buckets, respectively, while \cref{tab:extend-estimation-latency} extends the query estimation latency results in \cref{tab:qtime} to the corresponding larger memory and bucket budgets. The overall trends remain consistent with the main evaluation. Under low selectivity, TraHist achieves the lowest median Q-error on both datasets. The advantage continues under medium selectivity on \textit{T-Drive}, whereas DEH achieves the lowest median Q-error on \textit{Porto}. Under high selectivity, DEH and CASE+ generally achieve lower error, while the accuracy gap among estimators narrows. The latency result (\cref{tab:extend-estimation-latency}) shows the same efficiency trends as in the main evaluation: DEH, CASE+, and TraHist maintain stable sub-millisecond latency, while MHist and LHist become substantially slower as the budgets increase, with latency reaching up to three orders of magnitude higher.

%%%%%%%%%%%%%%%%%%%%%%%%%%%%%%%%%%%%%%%%%%%%%%%%%%%%%%%%%%%%%%%%%

\end{document}